\documentclass[preprintnumbers,nofootinbib,noshowpacs,eqsecnum,prd,superscriptaddress,letterpaper]{revtex4}

\usepackage[utf8]{inputenc}
\usepackage{graphicx}
\usepackage{amsmath,amssymb}
\usepackage{feynmp}
\usepackage{url}
\usepackage{ulem}
\usepackage{multirow}
\usepackage{hyperref,color}

\newcommand\one{\leavevmode\hbox{\small1\normalsize\kern-.33em1}}

\newcommand{\lag}{\mathcal{L}}

\newcommand{\ope}{\mathcal{O}}
\newcommand{\qqquad}{\qquad \qquad}
\newcommand{\qqqquad}{\qquad \qquad \qquad}

\newcommand{\gev}{\text{GeV}}
\newcommand{\tev}{\text{TeV}}

\newcommand{\pb}{\text{pb}}
\newcommand{\br}{\text{BR}}

\newcommand{\ifb}{\text{fb}^{-1}}

\newcommand{\itevx}{\text{TeV}^{-2}}

\def\slashchar#1{\setbox0=\hbox{$#1$}           
   \dimen0=\wd0                                 
   \setbox1=\hbox{/} \dimen1=\wd1               
   \ifdim\dimen0>\dimen1                        
      \rlap{\hbox to \dimen0{\hfil/\hfil}}      
      #1                                        
   \else                                        
      \rlap{\hbox to \dimen1{\hfil$#1$\hfil}}   
      /                                         
   \fi}

\newcommand{\eg}{\textsl{e.g.}\;}
\newcommand{\ie}{\textsl{i.e.}\;}

\newcommand{\be}{\begin{eqnarray*}}
\newcommand{\ee}{\end{eqnarray*}}

\newcommand{\bee}{\begin{eqnarray}}
\newcommand{\eee}{\end{eqnarray}}
\newcommand{\beeq}{\begin{equation}}
\newcommand{\eeeq}{\end{equation}}

\begin{document}

\title{The Higgs Legacy of the LHC Run I}

\author{Tyler Corbett}
\affiliation{C.N. Yang Institute for Theoretical Physics, SUNY, Stony Brook, USA}
\author{Oscar J. P. \'Eboli}
\affiliation{Instituto de Fisica, Universidade de Sao Paulo, Sao Paulo, Brazil}
\author{Dorival Gon\c{c}alves}
\affiliation{Institute for Particle Physics Phenomenology, Department of Physics,
Durham University, UK}
\author{J.~Gonzalez--Fraile}
\affiliation{Institut f\"ur Theoretische Physik, Universit\"at Heidelberg, Germany}
\author{Tilman Plehn}
\affiliation{Institut f\"ur Theoretische Physik, Universit\"at Heidelberg, Germany}
\author{Michael Rauch}
\affiliation{Institute for Theoretical Physics, Karlsruhe Institute of Technology, Germany}

\begin{abstract} 
Based on Run I data we present a comprehensive analysis of Higgs
couplings. For the first time this SFitter analysis includes
independent tests of the Higgs-gluon and top Yukawa couplings, Higgs
decays to invisible particles, and off-shell Higgs measurements. The
observed Higgs boson is fully consistent with the Standard Model, both
in terms of coupling modifications and effective field theory. Based
only on Higgs total rates the results using both approaches are
essentially equivalent, with the exception of strong correlations in
the parameter space induced by effective operators. These correlations
can be controlled through additional experimental input, namely
kinematic distributions. Including kinematic distributions the typical
Run I reach for weakly interacting new physics now reaches 300 to 500 GeV.
\end{abstract}

\maketitle
\tableofcontents
\newpage

\section{Introduction}
\label{sec:intro}

The discovery of a light narrow Higgs boson~\cite{higgs} in
2012~\cite{discovery} was a triumph of particle physics. While its
statistical significance was largely driven by the peak in the $\gamma
\gamma$ invariant mass spectrum, already Run~I of the LHC allowed
ATLAS and CMS to perform a large number of tests of the nature of the
observed resonance. One of these tests is the analysis of Higgs
couplings relative to their Standard Model values. No significant
deviations from the Standard Model properties were observed in the
Higgs production and decay rates. However, we need to keep in mind
that these constraints are at a numerical level where in typical
weakly interacting models for new physics~\cite{bsm_review} we would
not expect significant deviations, either.\bigskip

One of the main physics programs defining the upcoming LHC runs will
be a comprehensive precision analysis of Higgs properties. It will
eventually utilize up to $3000~\ifb$ of integrated luminosity,
covering a wide range of production and decay channels and
observables. Technically, the Higgs coupling analysis from Run~I is
based on comparably simple total cross sections and branching ratios;
this simple structure of underlying measurements allows us to limit
the interpretation to an independent variation of all Higgs couplings
in the Standard Model without missing much of the experimental
information.

Planning for the corresponding Run~II analysis we need to implement
major extensions to the Higgs coupling analysis. The reason is that
the expected wealth of measurements, including kinematic
distributions, will probe modifications of the Standard Model
Lagrangian which are not captured by a simple shift in Higgs
couplings.\footnote{This does not mean that the Higgs coupling
  analysis should be abandoned entirely. For reasons discussed in this
  paper we will keep it in the \textsc{SFitter} framework to search
  for physics beyond the Standard Model which merely shifts Higgs
  couplings.} Including kinematic distributions in the analysis of the
Higgs couplings suggests to describe the theoretical basis in terms of
an effective field theory. It also implies that the interface between
experiment and theory has to be re-defined. Eventually, at least three
questions need to be addressed by the Higgs measurements:
\begin{enumerate}
\item do any of the observed Higgs couplings show deviations from their
  Standard Model predictions?
\item can possible deviations be linked to a mass scale through an
  effective theory of the electroweak Lagrangian?
\item what does the Higgs sector tell us about specific new physics models?
\end{enumerate}
These questions are closely tied to each other, but the complex
structure of correlations between the experimental measurements and
between theoretical predictions render a single answer to all three
questions unrealistic. In the past, \textsc{SFitter} has been one of
the driving forces to answer the
first~\cite{before_sfitter,sfitter_orig,sfitter_delta,higgs_couplings} and
third~\cite{sfitter_bsm} questions. The second question has 
been tackled by Higgs specialized tools, for example presented in
Refs.~\cite{barca,eft_fits}. A shift from the analysis of Higgs
couplings based on total rates to an effective field theory of the
Higgs sector is the appropriate step to include kinematic
distributions in the Higgs fit.\bigskip

In this paper we first update the Higgs coupling analysis to include
the full Run~I data set. One focus of \textsc{SFitter} is the
appropriate treatment of theoretical uncertainties, which will also
play a major role in this paper.  Second, we present a first
\textsc{SFitter} study of the electroweak effective field theory,
which will in the future allow us to treat the three above questions
in the same framework.  With that purpose we include in the analysis
several differential distributions and we study their implementation
and impact.  The Run~I constraints in terms of the accessible new
physics scale stay below the TeV range, which is consistent with a
$20\%$ precision on coupling analyses and supports the statement that
within known theoretical frameworks we do not expect to see deviations
in the Higgs couplings during Run~I. Finally, we include off-shell
measurements of Higgs production~\cite{off_shell_th} in the
\textsc{SFitter} analysis and estimate their current impact. While it
has been established that off-shell measurements will not serve as
model-independent measurements of the Higgs width~\cite{off_shell_no},
we can establish their role in hypothesis-based Higgs studies.

\subsection{Experimental input}
\label{sec:intro_ex}

In all \textsc{SFitter} Higgs coupling analyses the experimental
input is not the published set of Higgs signal strengths, but the
number of signal and background events for each analysis. This allows
us to independently study statistical, systematic, and theoretical
uncertainties and include them in a profile likelihood analysis
largely independent of the ATLAS and CMS assumptions. As a matter of
fact, this independence is our motivation to maintain the
\textsc{SFitter} effort in spite of more and more advanced
experimental Higgs coupling analyses.\bigskip

As experimental input we use the following Higgs searches and
measurements as published by the ATLAS and CMS collaborations:
\begin{center}
\begin{tabular}{l|ll}
production/decay mode & ATLAS & CMS \\ \hline
$H \to WW$             & Ref.~\cite{1412.2641}           & Ref.~\cite{1312.1129} \\
$H \to ZZ$             & Ref.~\cite{1408.5191}           & Ref.~\cite{1312.5353} \\
$H \to \gamma\gamma$   & Ref.~\cite{1408.7084}           & Ref.~\cite{1407.0558} \\
$H \to \tau\bar{\tau}$ & Ref.~\cite{1501.04943}          & Ref.~\cite{1401.5041} \\
$H \to b\bar{b}$       & Ref.~\cite{1409.6212}           & Ref.~\cite{1310.3687} \\
$H \to Z\gamma$        & Ref.~\cite{ATLAS-CONF-2013-009} & Ref.~\cite{1307.5515} \\
$H \to \text{invisible}$& Ref.~\cite{1402.3244,ATLAS_inv2,ATLAS_inv3,ATLAS_inv4}
& Ref.~\cite{1404.1344,CMS_inv2} \\
\hline
$t\bar{t}H$ production & Ref.~\cite{1408.7084,1409.3122}
& Ref.~\cite{1407.0558,1408.1682,1502.02485} \\ \hline
kinematic distributions & Ref.~\cite{1409.6212,1407.4222} & \\ \hline
off-shell rate & Ref.~\cite{off_shell_atlas} & Ref.~\cite{off_shell_cms} \\
\end{tabular}
\end{center}
From all these analyses we extract the number of observed, signal and
background events after appropriate cuts.  The several categorizations
in the experimental searches listed in the above table lead to the 159
measurements that we include in the rate based analyses. In
Sec.~\ref{sec:eff_full} we add 14 extra measurements from kinematic
distributions. Finally, the off-shell distributions considered in
Sec.~\ref{sec:off-shell} contribute with 37 additional
measurements.\bigskip

We will show Higgs coupling analyses in the \textsc{SFitter} framework
starting with one universal modification, moving to five tree level
SM-like couplings, and then allowing for additional contributions to
the Higgs-photon and Higgs-gluon loop-induced couplings.  A new
channel which we did not include in previous Higgs coupling analyses
is the direct measurement of a modified top-quark Yukawa coupling in
$t\bar{t}H$ production. This is necessary to identify new physics
contributing to the effective Higgs-gluon coupling. While there is not
yet enough sensitivity to properly observe the $t\bar{t}H$ production
channel with SM coupling strength, the current searches do provide
upper bounds on the production and decay rates.  Both experiments have
looked for this channel as part of their di-photon final-state
analysis~\cite{1408.7084,1409.3122,1407.0558}; moreover, CMS has
published additional searches based on multi-lepton final states
arising from Higgs decays into $WW^*$, $ZZ^*$ and
$\tau\bar{\tau}$~\cite{1408.1682} as well as a dedicated $b\bar{b}$
analysis~\cite{1502.02485}. These $t\bar{t}H$ measurements allow us to
separate an extra new contribution to the Higgs-gluon coupling,
leading to the analysis with seven independent coupling modifications.
After setting the two Higgs-weak-boson coupling modifications equal, these seven
modifications correspond to the relevant parameters in the non-linear
effective Lagrangian expansion~\cite{non-linear}, if we restrict them to the
leading terms following Ref.~\cite{non-linear2}.  In the last step we
will also include Higgs searches to invisible particles, \ie
generating missing transverse momentum. This will allow us to add an
extra modification to the analysis, accounting for possible Higgs
invisible decays. All coupling analyses in the traditional
\textsc{SFitter} framework rely on the same measurements listed
above.\bigskip

For the dimension-6 operators, which not only change the coupling
strengths but also the momentum dependence of the vertices leading to
modifications of kinematic distributions, we make use of several
differential distributions. These are the transverse momentum
distributions of the gauge boson in $VH$, $H\rightarrow b\bar{b}$
production~\cite{1409.6212} for all 0, 1 and 2-lepton final states,
and the $\Delta \phi_{jj}$ distribution in $H$+2 jets with di-photon
decays of the Higgs boson~\cite{1407.4222}. In
Sec.~\ref{sec:eff_distri} we will discuss in detail how we include the
crucial information from kinematic distributions in addition to total
rates.\bigskip

Finally, we will for the first time include off-shell Higgs production
rates~\cite{off_shell_atlas,off_shell_cms}  in a global fit of Higgs properties.
We will analyze their effect to the usual coupling determination and
quantify their potential to constrain the Higgs total decay width.

\subsection{Fit setup}
\label{sec:intro_setup}

The most relevant shortcoming of Higgs production at the LHC is that
there is no method which allows us to directly access the Higgs width
if we only consider on-shell data. For the traditional Higgs couplings
analysis this means that we will be sensitive to all observed Higgs
decays, including decays to invisible particles passing through the
detector without leaving a trace.  Because the total width entering
the coupling measurement is a sum of partial widths, we can easily
construct a lower limit to the total width by summing the partial
widths induced by the observed Higgs couplings.

Once we add off-shell Higgs data we will start being sensitive for
example to Higgs decays to hadrons or more than two states. The reason
is that off-shell Higgs rate measurements add a possible upper limit
on the Higgs width, as discussed in detail in
Sec.~\ref{sec:off-shell}. This limit is currently weak, so for the
present section we identify the sum of observed partial widths with
the total width, as usually done in \textsc{SFitter}.  The one
additional assumption we need to make is generation universality,
meaning that the relation between the mass and the Yukawa coupling for
up-type quarks, down-type quarks, and charged leptons is universal for
the second and third generations each. For example, the charm Yukawa
coupling is shifted by $g_c = g_c^\text{SM} (1 + \Delta_t)$ in the usual
\textsc{SFitter} conventions which will be reviewed in
Eq.\eqref{eq:def_delta}.

With these assumptions we know what would happen if there was a
sizeable unobserved contribution to the total width: in the ideal case
of all measurements otherwise in agreement with the Standard Model and
equally constraining, all measured Higgs couplings based on the
underestimate of the total width will appear to be too small by a
universal factor. The experimental challenge would be to distinguish
such a scenario for example from a Higgs portal where the SM Higgs
state mixes with a heavy additional scalar~\cite{lhc_portal}. In this
situation the off-shell Higgs rate measurement will be extremely
useful.\bigskip

For all experimental uncertainties we assume a Poisson distribution for 
the statistical error of the rate measurements and Gaussian
distributions of associated nuisance parameters. For the systematic
uncertainty distribution this is justified as long as it is resolved
with help of some kind of measurement outside the analysis we are
considering.  For theoretical uncertainties a Gaussian modeling might
be technically convenient, but it needs to be established that
it is at least conservative~\cite{kyle}. We will give a
detailed discussion of this issue in Sec.~\ref{sec:coup_th}.

The usual determination of model parameters in \textsc{SFitter}
proceeds in two steps. First, we construct Markov chains which probe
the multi-dimensional space of Higgs couplings. These chains cover
typical values of $\Delta_x \in [-5,7]$.  The error treatment includes
all experimental and theoretical error bars with full correlations.  From
these chains we for example compute the 2-dimensional correlations
shown later in this work. In a second step we can focus on the SM-like
solution, defined as $\Delta_x > -1$ for all couplings, 
and compute the 68\% and 95\% confidence-level (CL) interval 
of the log-likelihood for each individual Higgs coupling.

In the Higgs analysis presented in this paper we
already derive 1-dimensional profile likelihoods and the corresponding
68\% CL interval from the Markov chains. Unlike for the 2-dimensional
correlations we limit our 68\% CL analysis in the individual couplings
to the SM-like solution. The interval is defined as the range of
coupling modifications or Wilson coefficients covering 68\% of the
integrated profile log-likelihood. Because in particular for non-Gaussian
distributions this definition allows for a simultaneous shift of both
limits and is hence not uniquely defined, we also require the value of
the profile log-likelihood to coincide on both sides of the error
band.

\section{Higgs couplings}
\label{sec:coup}

One question we can ask in the Higgs sector is: \textsl{how well does
  the Standard Model describe all available Higgs data at the LHC?}
There are (at least) two ways of answering this question: first, we
can compute an over-all confidence level of the Standard Model given
all available Higgs data, possibly combined with electroweak precision
measurements etc. The answer to this question is statistically well
defined, but to give us useful information we need to carefully
analyze the pulls of different measurements.

Second, we can measure the parameters in the renormalizable Higgs
Lagrangian: we start from the quadratic and quartic terms in the Higgs
potential, which can be exchanged for the Higgs mass and the vacuum 
expectation value $v$. This
relation can eventually be tested in measurements of the Higgs
self-coupling. In addition, we can measure the Higgs coupling to each
Standard Model particle.  The Lagrangian underlying this measurement
consists of the Standard Model operators with free couplings. These
couplings can be extracted from LHC rate measurements,
\begin{alignat}{6}
 g_x &= g_x^\text{SM} \; (1 + \Delta_x) \notag \\
 g_\gamma &
= g_\gamma^\text{SM} \; (1 + \Delta_\gamma^\text{SM} + \Delta_\gamma ) 
\equiv g_\gamma^\text{SM} \; (1 + \Delta_\gamma^\text{SM+NP} ) \notag \\
 g_g &= g_g^\text{SM} \; (1 + \Delta_g^\text{SM} + \Delta_g ) 
\equiv g_g^\text{SM} \; (1 + \Delta_g^\text{SM+NP} ) \; ,
\label{eq:def_delta}
\end{alignat} 
where the modifications of the tree-level couplings appearing in the
Standard Model loops are encoded into $\Delta_{\gamma,g}^\text{SM}$
while extra possible new physics contributions are included in
$\Delta_{\gamma,g}$.  In general, the loop-induced couplings have a
non-trivial momentum dependence; for the Higgs couplings measurement we
assume that all three external momenta are fixed, reducing the
coupling to a single number. In terms of a Lagrangian we can write our
hypothesis as
\begin{alignat}{5}
\lag 
= \lag_\text{SM} 
&+ \Delta_W \; g m_W H \; W^\mu W_\mu
+ \Delta_Z \; \frac{g}{2 c_w} m_Z H \; Z^\mu Z_\mu
- \sum_{\tau,b,t} \Delta_f \; 
\frac{m_f}{v} H \left( \bar{f}_R f_L + \text{h.c.} \right) \notag \\
&+  \Delta_g F_G \; \frac{H}{v} \; G_{\mu\nu}G^{\mu\nu}
+  \Delta_\gamma F_A \; \frac{H}{v} \; A_{\mu\nu}A^{\mu\nu} 
+ \text{invisible decays} \; ,
\label{eq:lag_delta}
\end{alignat} 
with $c_w$ denoting the cosine of the weak mixing angle.
A possible Higgs decay to invisible states could be described by a
wide variety of Lagrangian terms, but as long as we only search for an
invisible branching ratio there is no need for further
specifications.  The contributions to the higher-dimensional
Higgs-gluon and Higgs-photon couplings are normalized to their
Standard Model values $F_G$ and $F_A$. We use their values for finite
loop masses, while their normalization is illustrated by the limit
$F_G^{(\infty)} \to \alpha_s/(12 \pi)$ for a heavy top mass. The
invisible decay width can for example be generated in a Higgs portal
model~\cite{lhc_portal} and will be quoted in terms of an invisible
branching ratio.\bigskip

The form $(1+\Delta)$ of the coupling deviations suggests that we will
focus on scenarios not too different from the Standard Model. Large
deviations from the Standard Model should be taken with a grain of
salt, because in such a situation it is not clear if the kinematic
distributions of the `Higgs' signal and the associated detector
efficiencies are well controlled.  Values around $\Delta = -2$
indicate a switch in the sign of the coupling, which should be checked
individually. In some figures we show the full range of $\Delta$ values for
illustration, but we will limit our interpretation to small values of
$|\Delta| \lesssim 0.5$.

The coupling measurement according to Eqs.\eqref{eq:def_delta}
and~\eqref{eq:lag_delta} is
clearly well defined in the sense that it corresponds to a weak scale
Lagrangian which we compare to experimental results. Higher-order QCD
corrections can be included, because renormalizability with respect to
the strong coupling is not affected by changes in the Higgs
couplings. Electroweak corrections cannot be computed in a model with
free Higgs couplings, but their impact on LHC rates can safely be
neglected for Run~I data.\bigskip

In the spirit of an effective field theory the free couplings ansatz
of Eq.\eqref{eq:def_delta} can be linked to extended Higgs
sectors~\cite{sfitter_bsm}: in such models the Higgs coupling
measurement will search for modifications to the couplings of the
SM-like Higgs boson.  Using a well defined ultraviolet completion,
possible shifts can also be interpreted for example using the full set
of model parameters in an aligned two-Higgs-doublet model. If we
include $\Delta_t$ as well as $\Delta_g$ in the coupling analysis we
need to supplement the extended Higgs sector for example with an
additional strongly interacting fermion. In addition, the non-diagonal
photon--$Z$--Higgs is missing, but can be trivially added to this
ansatz as we will discuss. We have shown that a \textsc{SFitter}
analysis of the aligned two-Higgs-doublet model --- with coupling
deviations $\Delta$ computed from the underlying parameters --- and
the weak scale Higgs coupling analysis indeed give identical
results~\cite{sfitter_bsm}. Furthermore, neglecting the invisible
decays and setting $\Delta_W=\Delta_Z$ the Lagrangian in
Eq.\eqref{eq:lag_delta} corresponds to the non-linear effective
Lagrangian~\cite{non-linear} of the Higgs sector, but restricted to
the leading terms of the expansion defined in Ref.~\cite{non-linear2}.

Our couplings approach only
tracks a deviation in the leading coupling of the Higgs boson to each
SM particle. We only consider the dimension-4 Lagrangian plus
non-decoupling dimension-6 operators coupling the Higgs to photons or
gluons. Strictly speaking, in the linear effective Lagrangian approach
this kind of dimension-6 operators also modify the Higgs couplings to
electroweak gauge bosons. As long as the
experimental analysis is limited to total rates, we can expect
loop-induced corrections for the $HWW$ coupling to be suppressed with
respect to shifts in the tree-level coupling $g_W$.\bigskip

A shortcoming of the approach is that it does not include effects of the
additional new particles on the SM Lagrangian outside the Higgs
sector. For example, it does not link deviations in Higgs couplings to
anomalous triple gauge couplings. For a comprehensive analysis of all
effects of such new states we have to extend the free Higgs couplings
to an effective field theory based on a non-linear sigma model in the
broken phase of the electroweak symmetry, see for instance
Ref.~\cite{non-linear}.

Going beyond the original \textsc{SFitter} ansatz indeed means
re-writing and extending the Lagrangian by additional operators which
couple Standard Model fields to the Higgs boson. These operators can
be classified by their dimension. For example for the Higgs coupling
to $W$ and $Z$ bosons this question has been studied independently of
the coupling strength~\cite{barca,eft_fits,non-linear}.

\subsection{Standard Model couplings}
\label{sec:coup_sm}

\begin{figure}[t]
  \centering
  \includegraphics[width=0.45\textwidth]{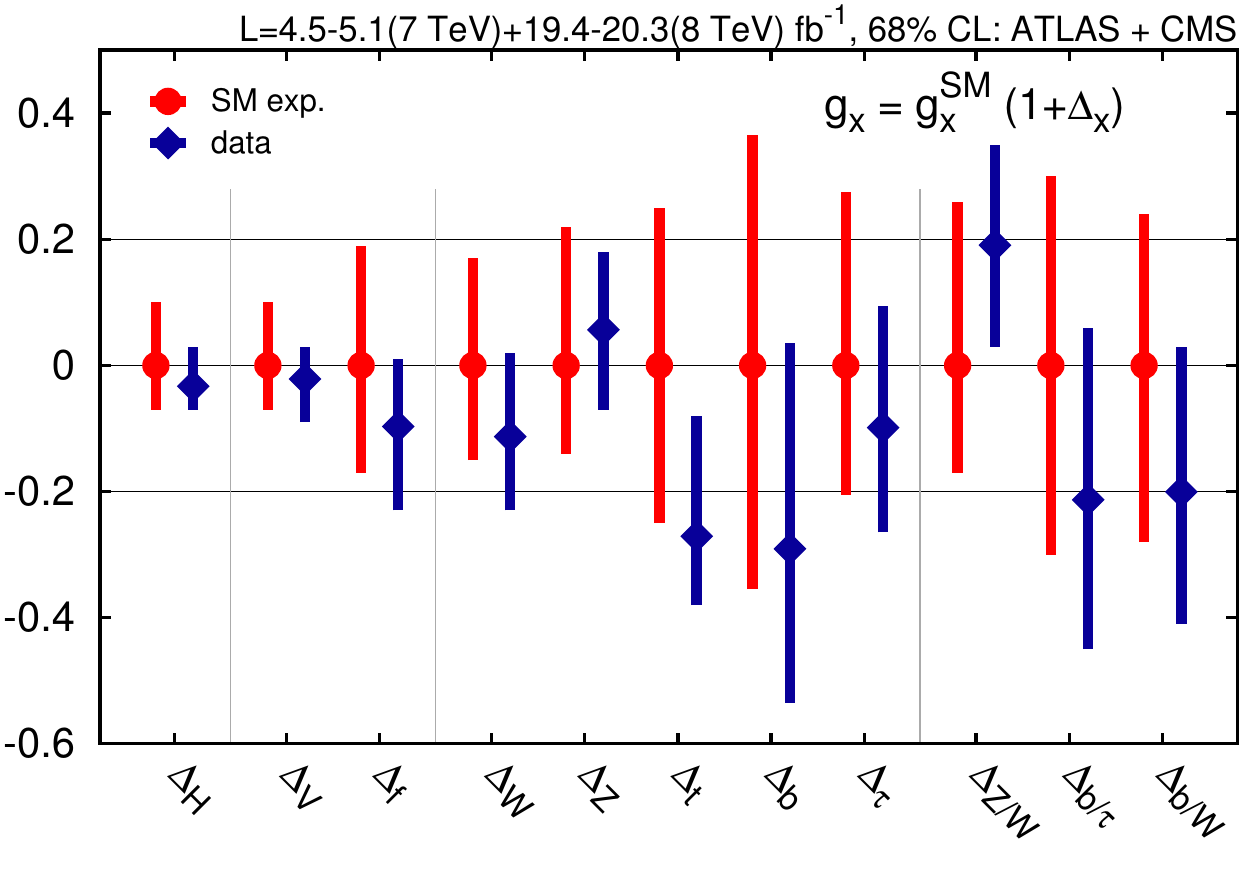}
  \caption{68\% CL error bars on the deviations $\Delta_x$ from all Standard
    Model couplings of the observed Higgs boson. In this fit we do not
    allow for new particles in the effective Higgs couplings to
    photons and gluons, $\Delta_\gamma = 0 = \Delta_g$. The results 
    labelled `SM exp' assume central values on the Standard Model
    expectation, but the current data error bars.}
\label{fig:delta}
\end{figure}

In the first step we can fit the five tree level Higgs couplings to all Standard
Model particles relevant for the LHC observations. The result is shown
in Fig.~\ref{fig:delta}. The red bars labelled `SM exp' show results where we have
injected a Standard Model Higgs signal on top of the background, \ie the measured
rate in each channel is exactly the SM expectation, but leave everything else
unchanged. They indicate that the observed errors are
slightly smaller than expected. This is a universal effect of the
theoretical uncertainties which we will discuss in detail in
Sec.~\ref{sec:coup_th}.\bigskip

The simplest model, motivated for example by
a Higgs portal~\cite{lhc_portal} or a single form factor from a
strongly interacting Higgs sector~\cite{mchm}, consists of a
universal coupling modification $\Delta_H$. Such a coupling
modification is constrained to around $3\%$ at 68\% CL, in agreement with the
Standard Model or $\Delta_H=0$. Translated into a mixing angle
$\alpha$ from the Higgs portal whose preferred range at 68\% CL is
\begin{alignat}{5}
\cos \alpha &= 1 + \Delta_H \in [0.93,1.03] \; .
\end{alignat}

The second simplest model is a universal coupling modification for the
Higgs interaction with the gauge bosons $\Delta_V$, and one
modification for the coupling with the fermions $\Delta_f$. 
Possible ultraviolet completions are given by models with additional
Higgs multiplets beyond singlets or doublets, where certain
combinations allow us to circumvent the otherwise strong limits by
electroweak precision data~\cite{higgsmultiplets}.  In this case,
$\Delta_V$ is still constrained to around $\pm 6\%$ at 68\% CL, while the
fermionic coupling shows a reduced precision of around $\pm 12\%$
at 68\% CL, all consistent with the Standard Model.\bigskip

An independent variation of five Higgs couplings is also in complete
agreement with the Standard Model. Again, the actual error bars on the
$\Delta_x$ are slightly smaller than what we would expect from exact
Standard Model values, an effect we will discuss in Sec.~\ref{sec:coup_th}.
The 15\% measurement of the top Yukawa coupling
is driven by the Higgs couplings to photons and gluons under the
assumption that no new particles contribute to these loop-induced
couplings. The measurement of the bottom Yukawa benefits from the
normalization of all rates, because in the Standard Model the total
width is largely driven by the partial decay width $H \to b\bar{b}$.

In addition to the individual couplings we also show the deviations of
ratios of couplings. Such ratios are useful to remove systematic and
theoretical uncertainties. Indeed, we see that the ratio 
\begin{equation}
\frac{g_b}{g_W} = 
\frac{g_b^\text{SM}}{g_W^\text{SM}} \, \left( 1 + \Delta_{b/W} \right)
\end{equation}
shows a smaller variation than $\Delta_b$ alone. The corresponding
positive correlation of $\Delta_b$ and $\Delta_W$ arises from the
total width in the denominator of the predicted event
numbers.\bigskip

\begin{figure}[t]
  \centering
  \includegraphics[width=0.45\textwidth]{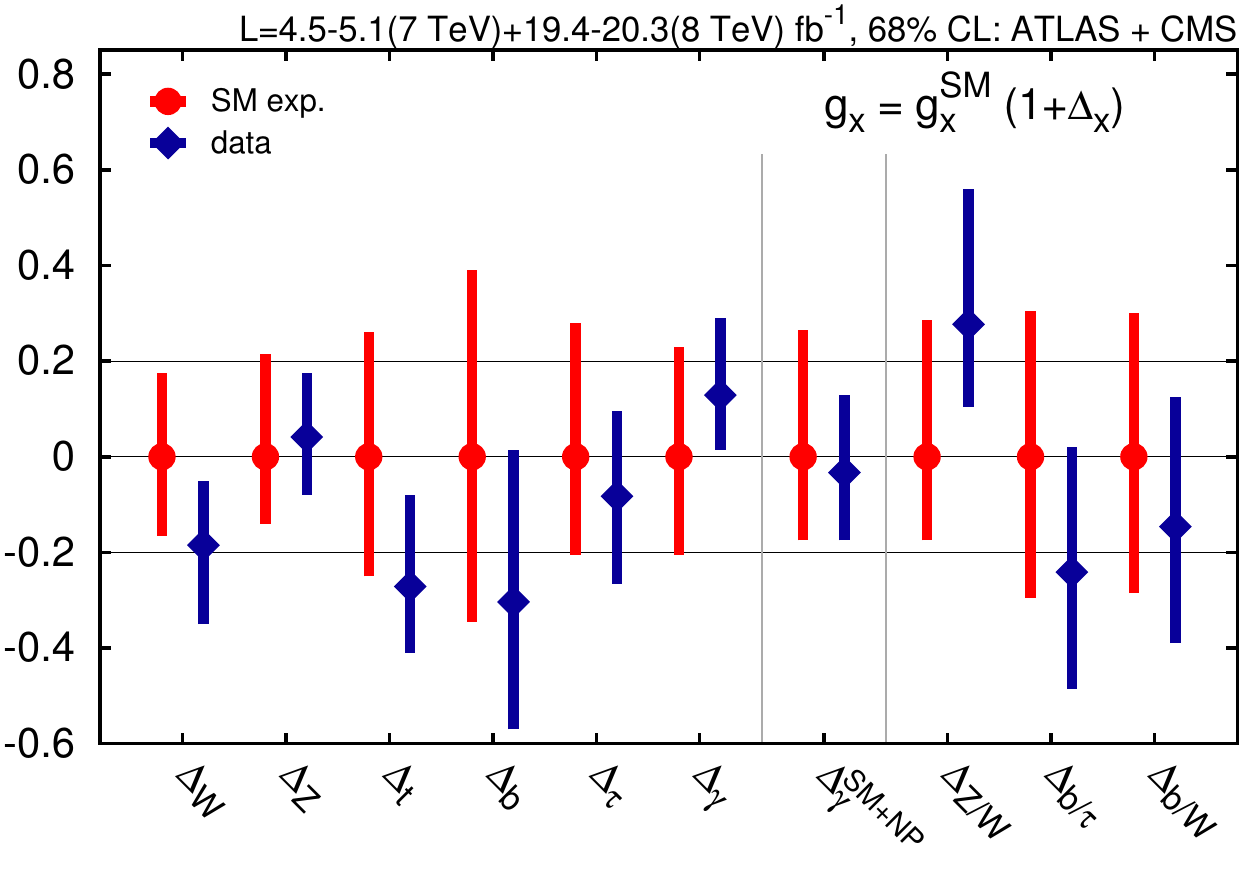}
  \hspace*{0.03\textwidth}
  \includegraphics[width=0.45\textwidth]{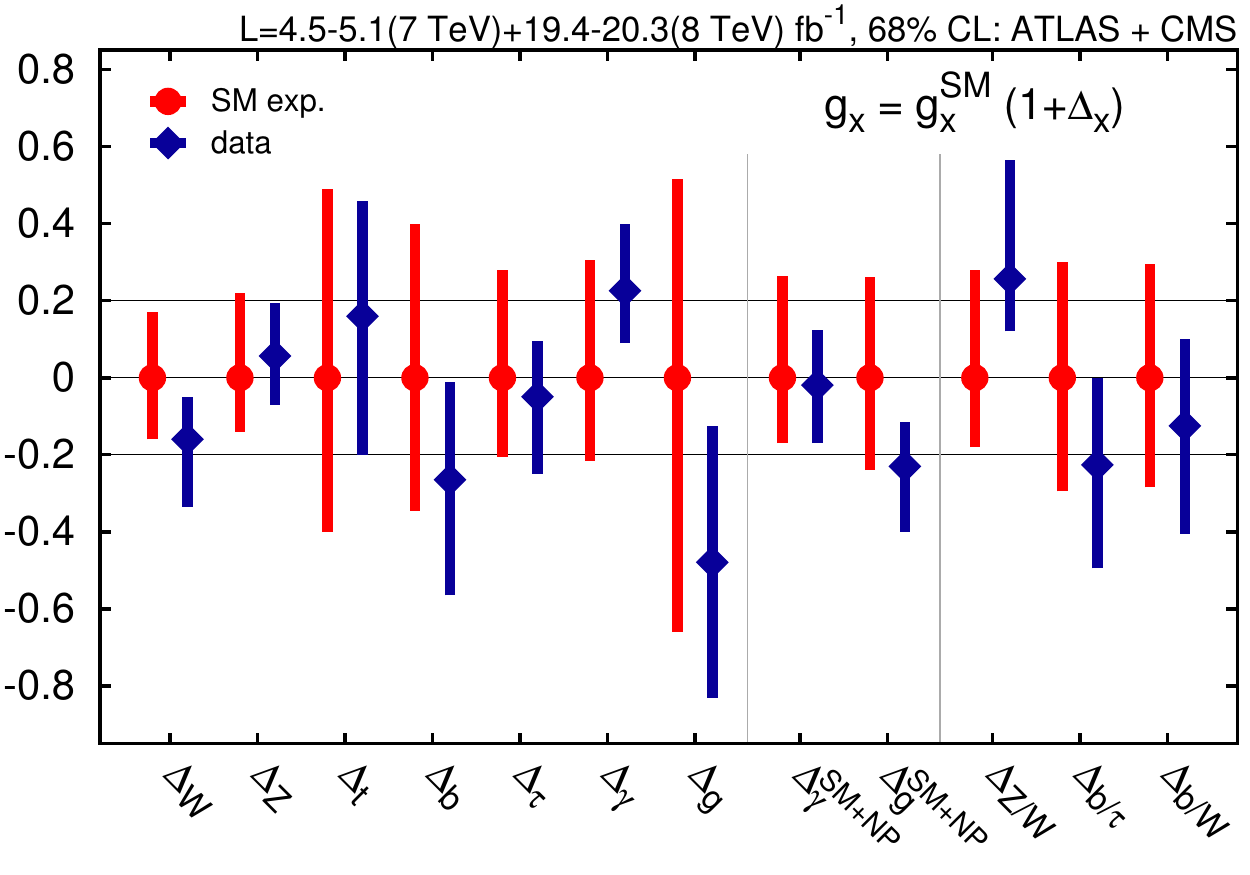}
  \caption{68\% CL error bars on the deviations $\Delta_x$ from all Standard
    Model couplings of the observed Higgs boson. For the loop-induced
    couplings we allow new contributions to the $H\gamma\gamma$
    coupling only ($\Delta_g=0$, left) and to the $H\gamma\gamma$ and
    $Hgg$ couplings (right). The results labelled `SM exp' assume
    central values on the Standard Model expectation, but the current data
    error bars.}
\label{fig:delta_g}
\end{figure}

Because the Higgs decay $H \to \gamma \gamma$ has been precisely
measured at the LHC we can extend the coupling fit by a new physics
contribution to this loop-induced coupling. Of course, the variations
of the Standard Model couplings $\Delta_{b,t}$ and $\Delta_W$ are
consistently reflected in the full $\Delta^\text{SM+NP}_\gamma$. Following
Eq.\eqref{eq:def_delta} this deviation consists of two terms, the
parametric shift from the Standard Model loops and an additional shift
from new physics. The latter is shown as part of the coupling
measurement in the left panel of Fig.~\ref{fig:delta_g}, where we present
the results of the six parameter analysis.

The only modification with respect to the fit with $\Delta_\gamma = 0$
is a slight downward shift of the central value of $\Delta_W$. It
decreases the contribution from the dominant $W$-loop and therefore
has to be compensated by a small positive new physics contribution
$\Delta_\gamma \sim 0.13$. This also leads to a very slight increase
in the uncertainty on $\Delta_W$ and $\Delta_t$.  While the new
physics contribution $\Delta_\gamma$ based on all available ATLAS and
CMS analyses has a one-sigma preference for an additional
contribution, the combination $\Delta_\gamma^\text{SM+NP}$ is in
perfect agreement with the Standard Model.  The error bars for
$\Delta_\gamma$ and $\Delta_\gamma^\text{SM+NP}$ have the same size,
which means that the interference structure between $\Delta_W$ and
$\Delta_t$ breaks any strong correlation with $\Delta_\gamma$ in this
fit.\bigskip

Finally, we show in the right panel of Fig.~\ref{fig:delta_g} the
first \textsc{SFitter} Higgs couplings fit including a new physics
contribution to the effective Higgs-gluon coupling, $\Delta_g \ne 0$.
As argued in Sec.~\ref{sec:coup} and suggested by our notation, in
this seven parameter analysis we focus on the SM-like solutions, \ie
small values of the $|\Delta_x|$.  As expected, the increase in the
error bar of $\Delta_t$ is dramatic. The central value of $\Delta_t$
increases by one standard deviation of the new measurement, while
$\Delta_g$ resides about one standard deviation below the Standard
Model expectation, keeping the Higgs production rate close to the
Standard Model prediction. Larger deviations of $\Delta_t$ are
forbidden by constraints on the $t\bar{t}H$ channels.  For example,
the combined CMS analysis of $t\bar{t}H$ channels reports a signal
strength of $2.8^{+1.0}_{-0.9}$~\cite{1408.1682}, and a specific CMS
analysis of the $t\bar{t}H \to t\bar{t} b\bar{b}$ channel arrives at a
signal strength of $1.2^{+1.6}_{-1.5}$~\cite{1502.02485}. Typical
uncertainties around 100\% on the cross section translate into a 30\%
uncertainty on the top Yukawa coupling.

\begin{figure}[b!]
  \centering
  \includegraphics[width=0.29\textwidth]{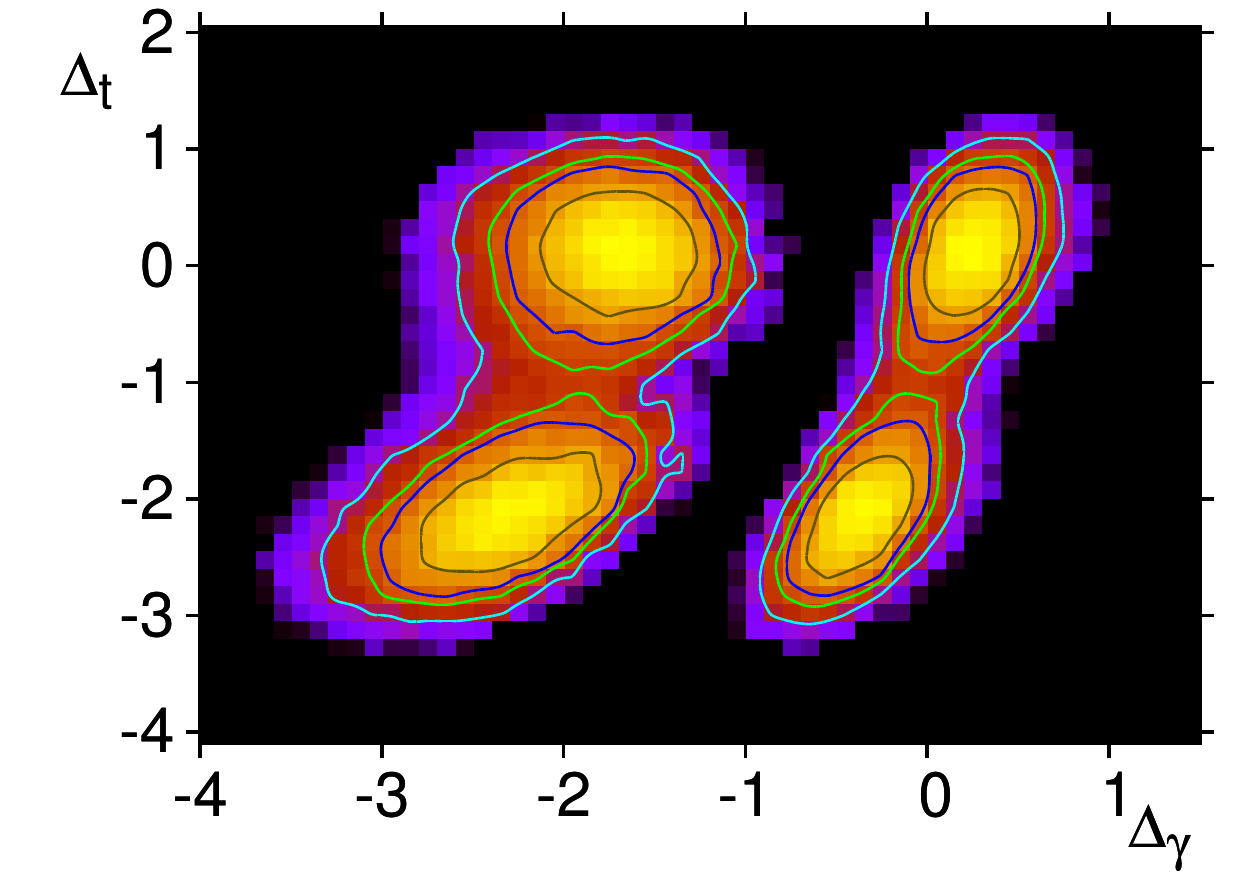}
  \hspace*{1ex}
  \includegraphics[width=0.29\textwidth]{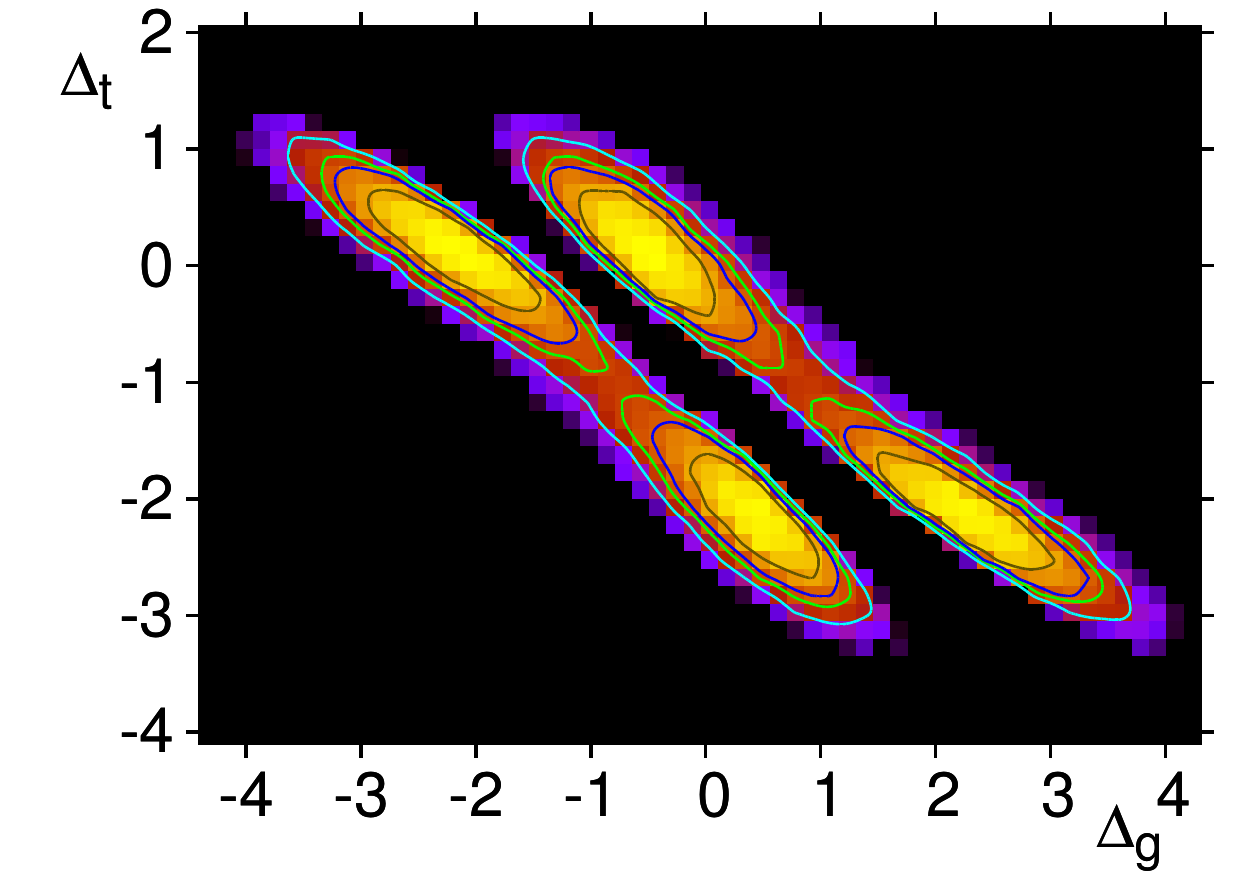}
  \hspace*{1ex}
  \includegraphics[width=0.29\textwidth]{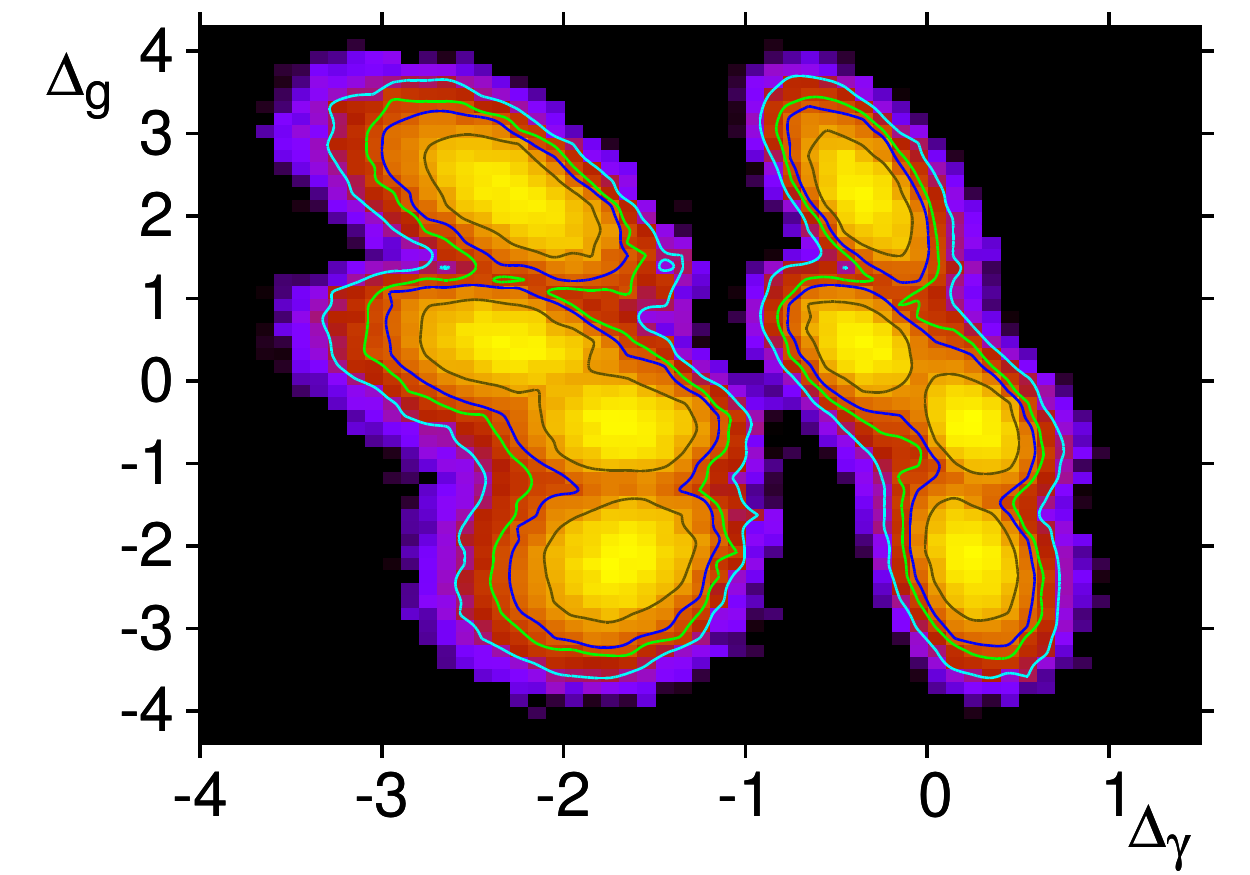}
  \hspace*{1ex}
  \raisebox{3pt}{\includegraphics[width=0.0545\textwidth]{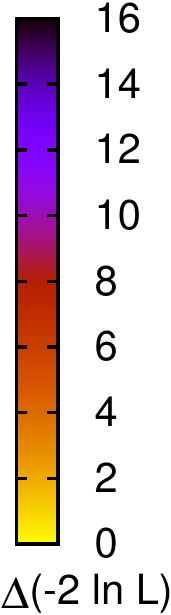}}
  \caption{Correlations between different coupling modifications for
    the fit including $\Delta_\gamma$ as well as $\Delta_g$. The
    1-dimensional profile likelihoods correspond to the results
    shown as the blue bars in the right panel of
    Fig.~\ref{fig:delta_g}.}
\label{fig:delta_gcorr}
\end{figure}

One reason to consider the $t\bar{t}H$ measurements with care is that
their significance hardly adds to an independent evidence for this
production channel. For example, the combination of
Ref.~\cite{1408.1682} rules out the Standard Model at two standard
deviations, just slightly less significantly than it establishes the
$t\bar{t}H$ production process. An appropriate hypothetical question
to ask is if these results would have been published the same way if
the signature had been a sign of physics beyond the Standard Model
instead of a very much expected signal.

Aside from the large error bars these measurements in the $t\bar{t}H$
channel have less obvious control of the signal kinematics than other
Higgs channels; for example, they might or might not include a clear
Higgs mass reconstruction, which is crucial for the unambiguous
interpretation of the rate measurement but poses a well known
combinatorics problem~\cite{tth}. Such a global analysis does lead to
a valid upper limit on the $t\bar{t}H$ cross section, but for a lower
limit we need to assume that $t\bar{t}H$ production is the only source
of relevant events.

As expected, the individual error bars for $\Delta_g$ and $\Delta_t$
are around three times as large as the error bar for the combination
$\Delta_g^\text{SM+NP}$, where the latter is known to better than
20\%. The remaining Higgs couplings are again hardly affected by the
additional parameter $\Delta_g$. The error bar of $\Delta_\gamma$ is
slightly increased because of the enlarged error bar on $\Delta_t$.
Unlike for $\Delta_\gamma$ this is a signal for a very strong
correlation between $\Delta_t$ and $\Delta_g$ in the 2-dimensional
profile likelihood.

In Fig.~\ref{fig:delta_gcorr} we show some relevant 2-dimensional
correlations of coupling modifications as obtained for the
  discussed analysis spanning the seven coupling
  modifications. First, we see that in the
$\Delta_t$ vs $\Delta_\gamma$ plane there are four solutions
corresponding to a sign flip in each of the two couplings. We fix the
global sign of all Higgs couplings to $\Delta_W >
-1$~\cite{sfitter_orig}. As long as we limit our analysis to total
rates each individual coupling modification at tree level will show a
perfect degeneracy between $\Delta_x = 0$ and $\Delta_x = -2$. The
loop-induced Higgs-gluon coupling is dominated by the top loop, with a
small contribution from the bottom quark, so it will not lift this
degeneracy. In contrast, the Higgs-photon coupling is strongly
sensitive to the relative sign of the top and $W$-contributions. 

The moderate positive correlation in the SM-like solution reflects the
fact that an increase of the top Yukawa coupling leads to a decrease
in the $H\gamma\gamma$ coupling and hence has to be compensated by a
positive value of $\Delta_\gamma$. As shown in the central panel the correlation between $\Delta_t$
and $\Delta_g$ is the strongest correlation in the Higgs couplings
analysis.  It reflects the fact that the gluon fusion Higgs cross
section constrains the sum of the two with a slight re-weighting from
the top mass dependence of the loop-induced Higgs--gluon
coupling~\cite{h_jets}. We will come back to this aspect when
discussing the effective theory analysis in Sec.~\ref{sec:eff} and top
mass effects in Sec.~\ref{sec:off-shell}.  The resulting
correlation of $\Delta_\gamma$ and $\Delta_g$ first of all features
eight solutions, arising from the indirect combination through
$\Delta_t \sim -2,0$. They are clearly separated into the two regimes
$\Delta_\gamma = -2,0$, while in $\Delta_g$ they are merged through
the strong correlation with $\Delta_t$. For example in the SM-like
regime the correlation between the two loop-induced couplings is at a
similarly weak level as the correlation between $\Delta_\gamma$ and
$\Delta_t$.

Without showing any detailed results we can also take advantage of the
first studies of the Higgs interaction with a photon and a $Z$
boson~\cite{ATLAS-CONF-2013-009,1307.5515}. We include a new physics
contribution to the loop-induced vertex in the Standard Model, in
complete analogy to the modifications $\Delta_\gamma$ and $\Delta_g$
in Eqs.\eqref{eq:def_delta} and~\eqref{eq:lag_delta}.  The
corresponding 68\% CL allowed region on $\Delta_{Z\gamma}$ bounds
$\Delta_{Z\gamma}<0.7$ (1.8 at 95\% CL), without any visible effect on the rest of
studied parameters shown in Fig.~\ref{fig:delta_g}.

\subsection{Invisible decays}
\label{sec:coup_inv}

\begin{figure}[t]
  \centering
  \includegraphics[width=0.45\textwidth]{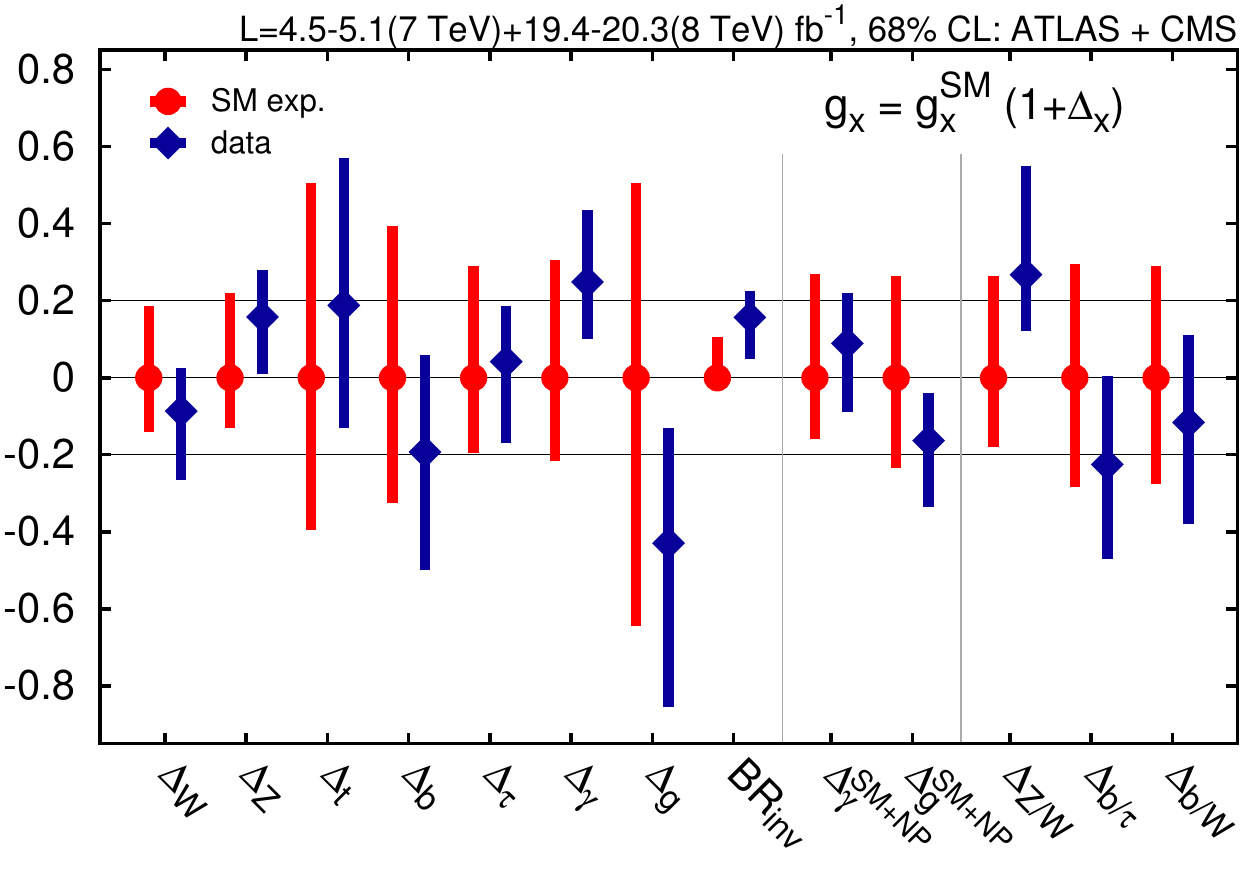}
  \caption{68\% CL error bars on the deviations $\Delta_x$ from all Standard
    Model couplings of the observed Higgs boson. In addition to all
    couplings predicted by the Standard Model we include a Higgs decay
    to invisible particles. The results 
    labelled `SM exp' assume central values on the Standard Model
    expectation, but the current data error bars.}
\label{fig:delta_gi}
\end{figure}

Higgs decays to invisible particles can only be observed in Higgs
production channels with a measurable recoil system. Examples are weak
boson fusion~\cite{eboli_zeppenfeld} and $ZH$
production~\cite{zh_inv}, where the more sensitive weak boson fusion (WBF)
channel might be able to probe invisible branching ratios to $2-3\%$ with an
ultimate integrated luminosity of
$3000~\ifb$~\cite{eboli_zeppenfeld}. To date there are ATLAS and CMS analyses
available in these two channels~\cite{1402.3244,ATLAS_inv2,ATLAS_inv3,ATLAS_inv4,1404.1344,CMS_inv2}.\bigskip

In Fig.~\ref{fig:delta_gi} we show the status for the full set of SM
Higgs couplings and a hypothetical Higgs coupling to invisible
states. Unlike for the other couplings we do not define a coupling
deviation $\Delta_\text{inv}$, but directly refer to the invisible
branching ratio $\br_\text{inv}$. In the Standard Model this invisible
branching ratio is generated by the decay $H \to ZZ^* \to 4\nu$. It
only reaches around $1\%$ and is therefore unlikely to ever be
observed at the LHC. The current limit on invisible Higgs decays in
the full Higgs couplings analysis is around 10\%. Obviously, for a
dedicated analysis with a more constraining model assumption the
limits will be stronger~\cite{higgs_couplings}.

Both relevant production processes responsible for invisible Higgs
decay searches are mediated by the $ZZH$ and $WWH$ interactions, where
$\Delta_{W,Z}$ are the best measured couplings in the analysis
described in Sec.~\ref{sec:coup_sm}. The additional searches for
invisible decays will not add any new information on the determination
on $\Delta_{W,Z}$, so we expect the invisible branching ratio to be
orthogonal to the other Higgs coupling measurements, \ie uncorrelated
with all other channels. A slight correlation of the invisible
contribution to the total Higgs width leads to a minor upwards shift
of all other couplings.

In Fig.~\ref{fig:delta_gicorr} we show the 2-dimensional profile
likelihoods for BR$_\text{inv}$ versus the $\Delta_W$ and $\Delta_Z$
appearing in the production processes and with $\Delta_b$ dominating
the total width. None of them show a significant correlation. In the
absence of strong correlations with any other model parameters in the
Lagrangian of Eq.\eqref{eq:lag_delta}, our best fit value and 68\% CL
limits on an invisible Higgs branching ratio $\br_\text{inv}=
0.16^{+0.07}_{-0.11}$ will hardly depend on the assumptions for
example made about the loop-induced Higgs couplings.

\begin{figure}[b!]
  \centering
  \includegraphics[width=0.29\textwidth]{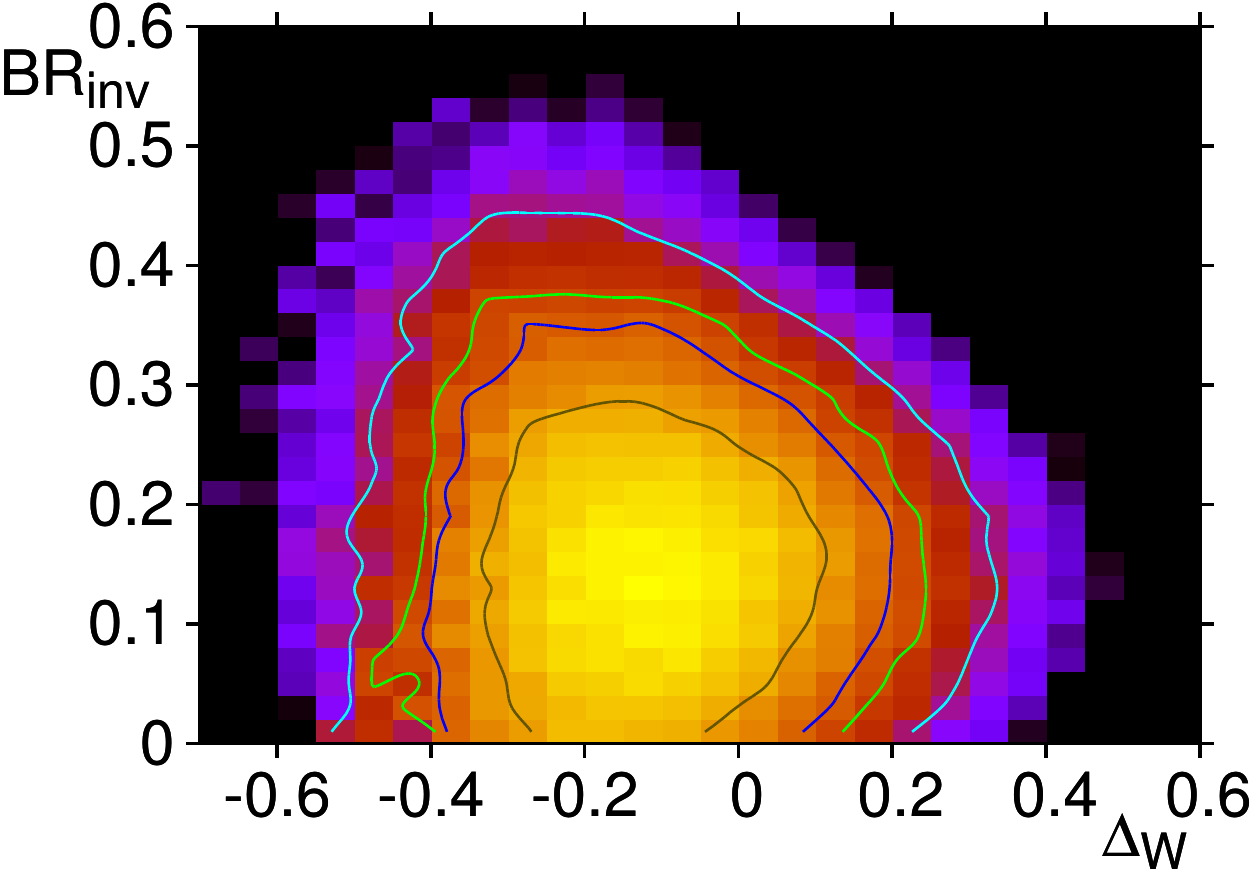}
  \hspace*{1ex}
  \includegraphics[width=0.29\textwidth]{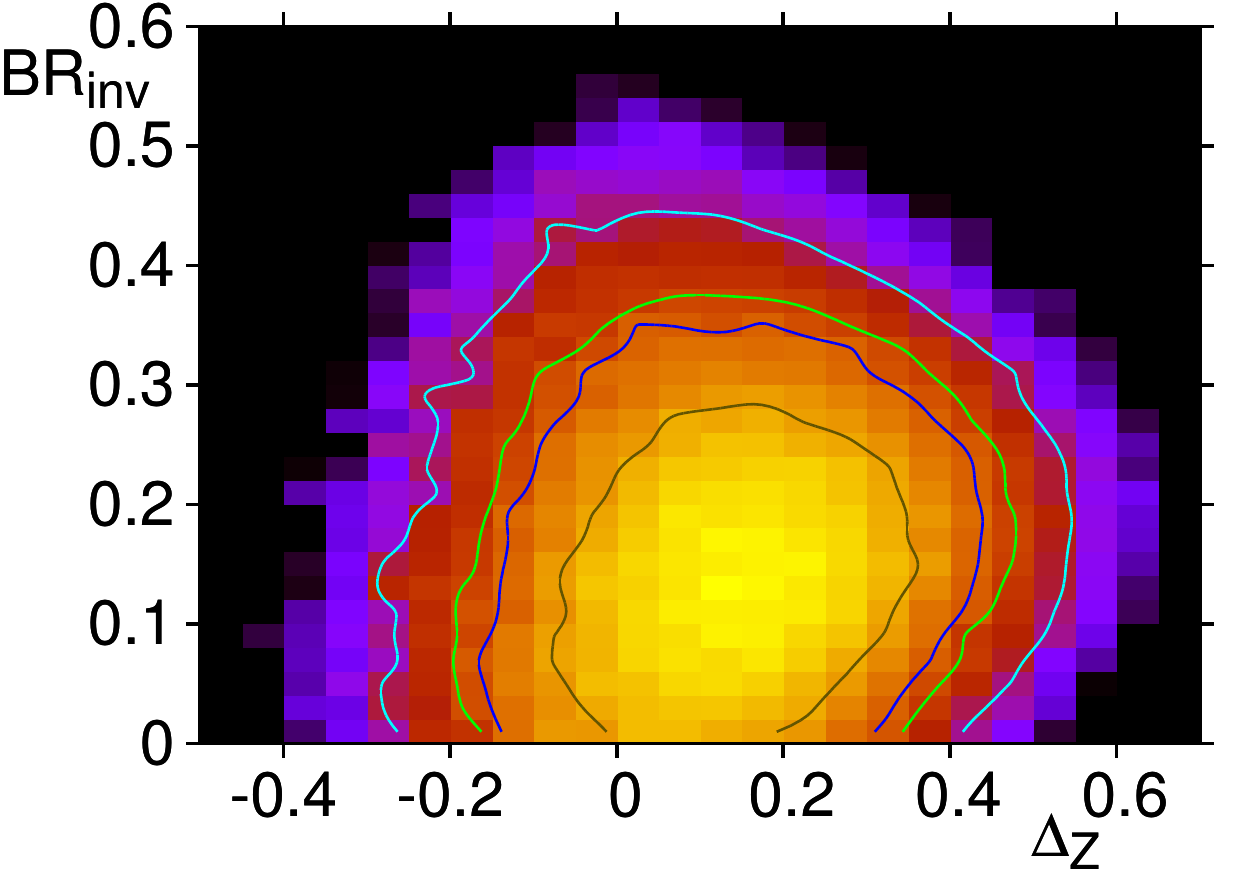}
  \hspace*{1ex}
  \includegraphics[width=0.29\textwidth]{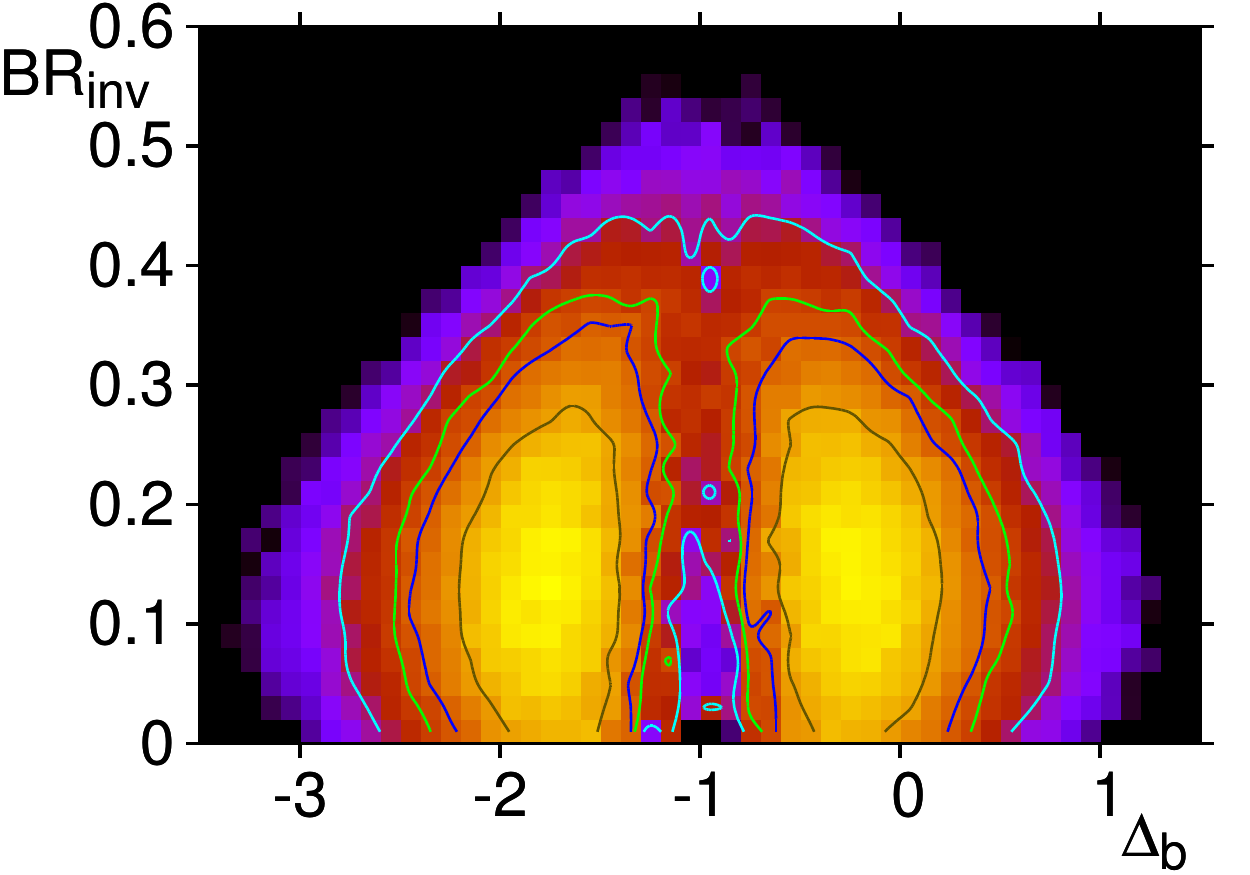}
  \hspace*{1ex}
  \raisebox{3pt}{\includegraphics[width=0.0545\textwidth]{figs/colorbox}}
  \caption{Correlations between different coupling modifications to SM
    particles and the invisible branching ratio. The corresponding 1-dimensional
    profile likelihoods are shown as the blue bars in Fig.~\ref{fig:delta_gi}.}
\label{fig:delta_gicorr}
\end{figure}

\subsection{Theoretical uncertainties}
\label{sec:coup_th}

In this last part of the Higgs couplings analysis we highlight open
questions related to the treatment of theoretical
uncertainties. Unlike experimental uncertainties, the estimate of
for example higher-order contributions missing in the calculation of
an LHC cross section do not offer a frequentist interpretation. It is
unclear what kind of likelihood distribution of the associated
nuisance parameters we need to assume. On the other hand, once we
define a likelihood distribution for these nuisance parameters the rest
of the likelihood analysis is completely defined.\footnote{Because of
  this lack of uniqueness in the definition of theoretical uncertainties we
  advocate for not including them in the experimental analyses or (if
  unavoidable) for factoring them out to allow for a flexible
  analysis~\cite{kyle}.}

All we can say from a theory perspective is that a certain deviation
from the best available cross section or rate prediction is in some
kind of agreement with the Standard Model or beyond the level where we
are willing to consider such an interpretation. This problem is
independent of the way we determine the uncertainty range on an
observable. Varying the unphysical factorization and renormalization
scales is only one method, and there might be many others.  If we
assume a flat distribution for the theoretical uncertainty to remove
any bias between different predictions within the allowed range, the
\textsc{RFit} scheme~\cite{rfit} is uniquely defined as the profile
likelihood combination with the experimental uncertainties.
Uncertainties on the parton densities are treated in complete analogy
to the theoretical uncertainties from unknown higher orders, including any
assumption on their correlations.

Note that a similar problem of choosing a prior for the theoretical
uncertainty arises in the Bayesian approach~\cite{cacciari}. On the
one hand the Bayesian approach allows for a choice of priors in
general, including the theoretical uncertainty, without having to ask
for a statistical interpretation. On the other hand, this renders one
assumption on the prior as ad-hoc as any other. This leaves us with
the crucial task to carefully check the prior dependence of our
result.\bigskip

\begin{figure}[b!]
  \centering
  \includegraphics[width=0.345\textwidth]{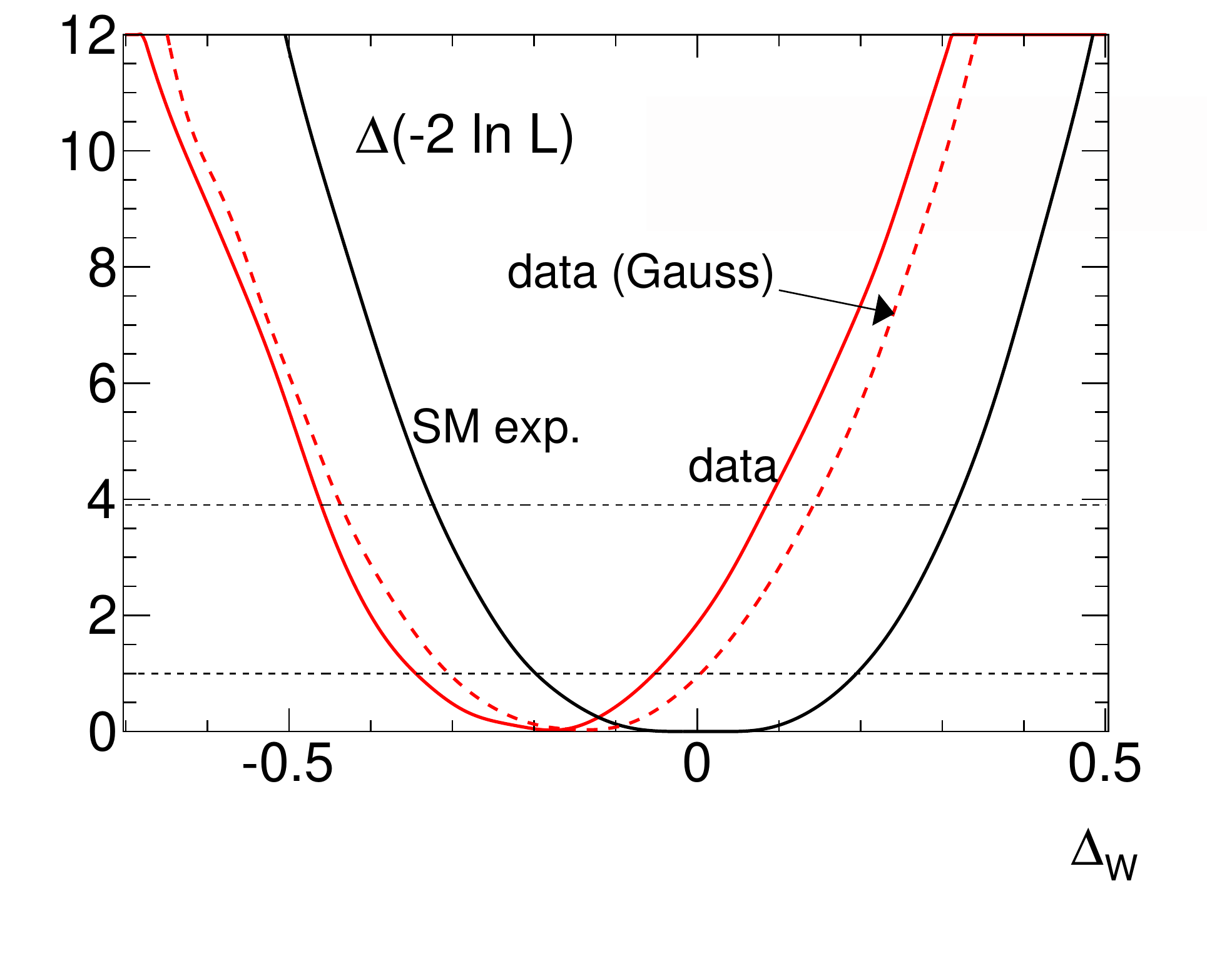}
  \hspace*{-0.03\textwidth}
  \includegraphics[width=0.345\textwidth]{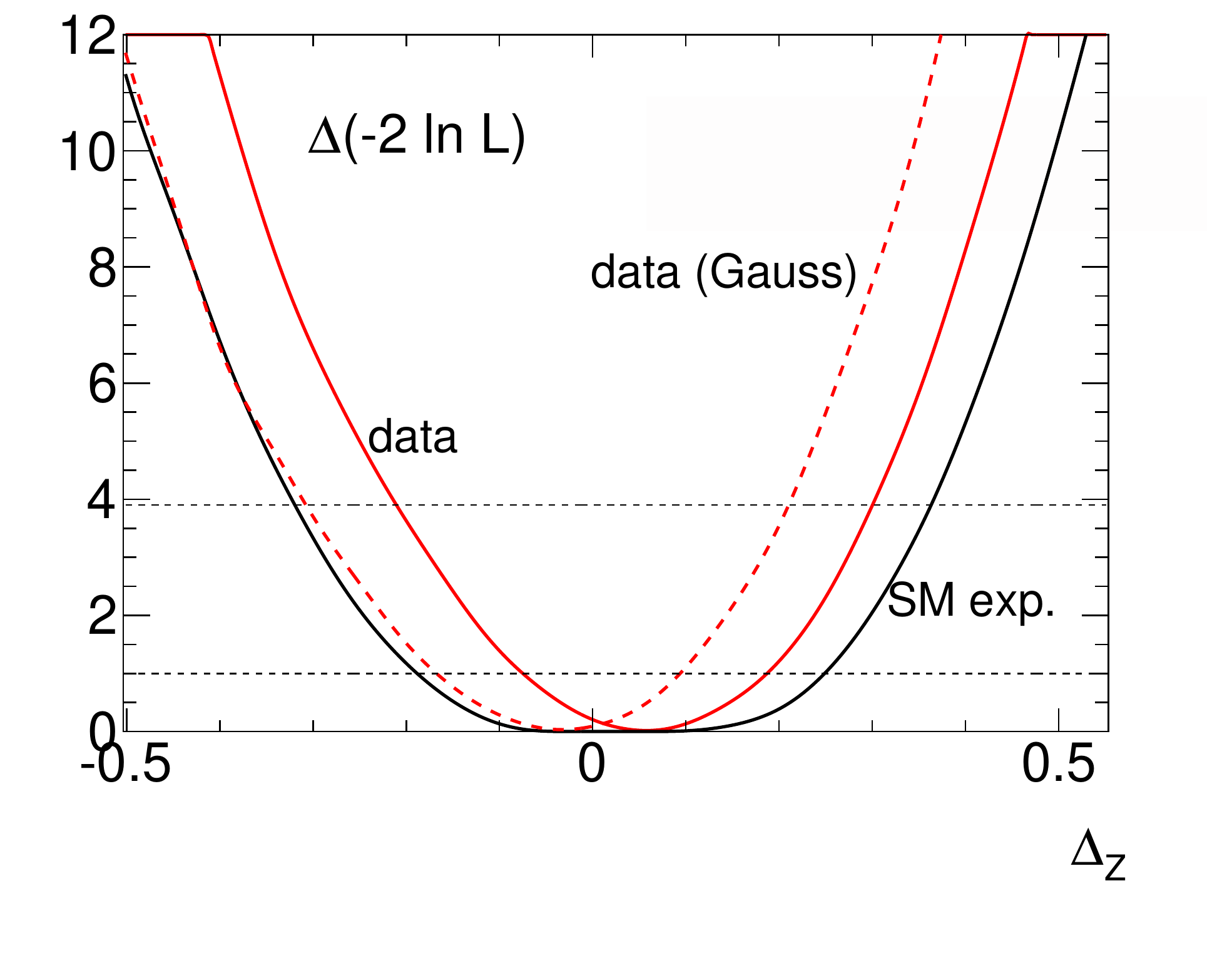}
  \hspace*{-0.03\textwidth}
  \includegraphics[width=0.345\textwidth]{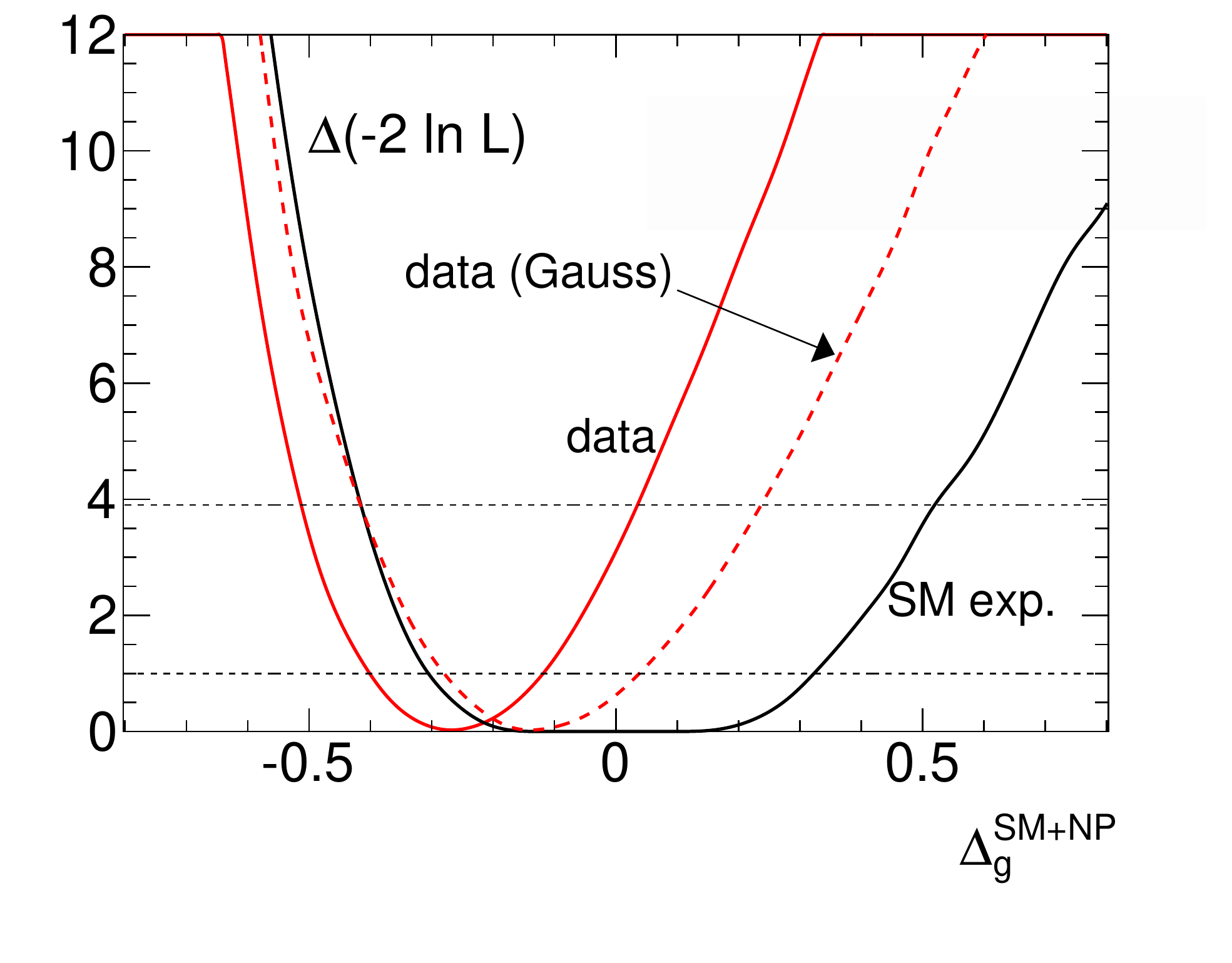}
  \caption{1-dimensional profile likelihoods for $\Delta_W$,
    $\Delta_Z$, and $\Delta_g^{\text{SM+NP}}$. We assume all
    measurements on the Standard Model values with flat theoretical
    uncertainties (black), observed rates with flat (solid red) and
    Gaussian (dashed red) theoretical uncertainties.}
\label{fig:delta_uncertainties}
\end{figure}

No matter what approach we follow, it is important to recognize that
we should employ a conservative estimate of theoretical uncertainties in
addition to a flexible framework which allows us to test different
assumptions as efficiently as possible~\cite{kyle}. This includes the
shape of the likelihood associated with the theoretical error bars as well
as the size of the theoretical error bars. Before we discuss the modelling
of the likelihood including a theoretical uncertainty we need to understand
an effect which we observe all through Sec.~\ref{sec:coup_sm}
and~\ref{sec:coup_inv}: the expected size of the error bars is
consistently larger than the observed errors. The reason is linked to
the behavior of flat theoretical uncertainties once the measurements start
developing a pull. In Fig.~\ref{fig:delta_uncertainties} we show a set
of 1-dimensional profile likelihoods. For the expected limits, \ie
assuming that all rate measurements agree perfectly with the
Standard Model, we clearly see the flat central range, induced by the
theoretical uncertainties.

However, once we allow for a statistical distribution of the
measurements all 1-dimensional profile likelihoods lose the flat
central regions and instead follow the Gaussian shape of the dominant
experimental uncertainties.  Note that this does not have to be the
case based on first principles: if all uncertainties were flat in the
rates, the resulting profile likelihood would keep its box shape, and
the errors would be added linearly~\cite{rfit,lecture}.  The central
limit theorem does not guarantee a Gaussian distribution, because the
profile likelihood does not involve a convolution. The resulting
Gaussians in Fig.~\ref{fig:delta_uncertainties} instead reflect the
fact that theoretical uncertainties are smaller than their
experimental counter parts, and the Gaussian features of the latter
dominate the final distribution once we allow for a spread of
measurements. 
The curves in Fig.~\ref{fig:delta_uncertainties} illustrate the
general observation, that actual 1-dimensional error bands with their
Gaussian behavior are smaller than the expected errors with their flat
central range, once we include real data.\bigskip

\begin{figure}[t]
  \centering
  \includegraphics[width=0.45\textwidth]{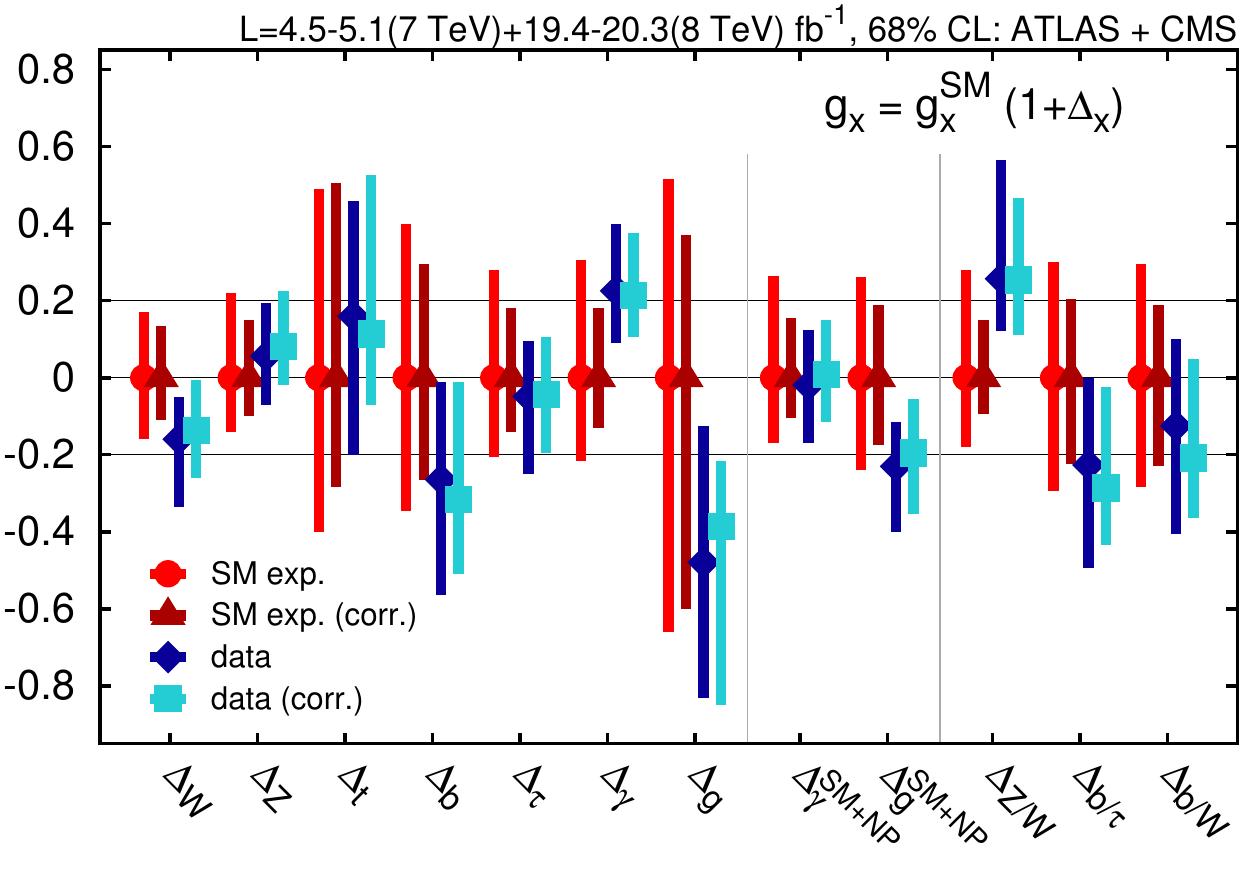}
  \hspace*{0.03\textwidth}
  \includegraphics[width=0.45\textwidth]{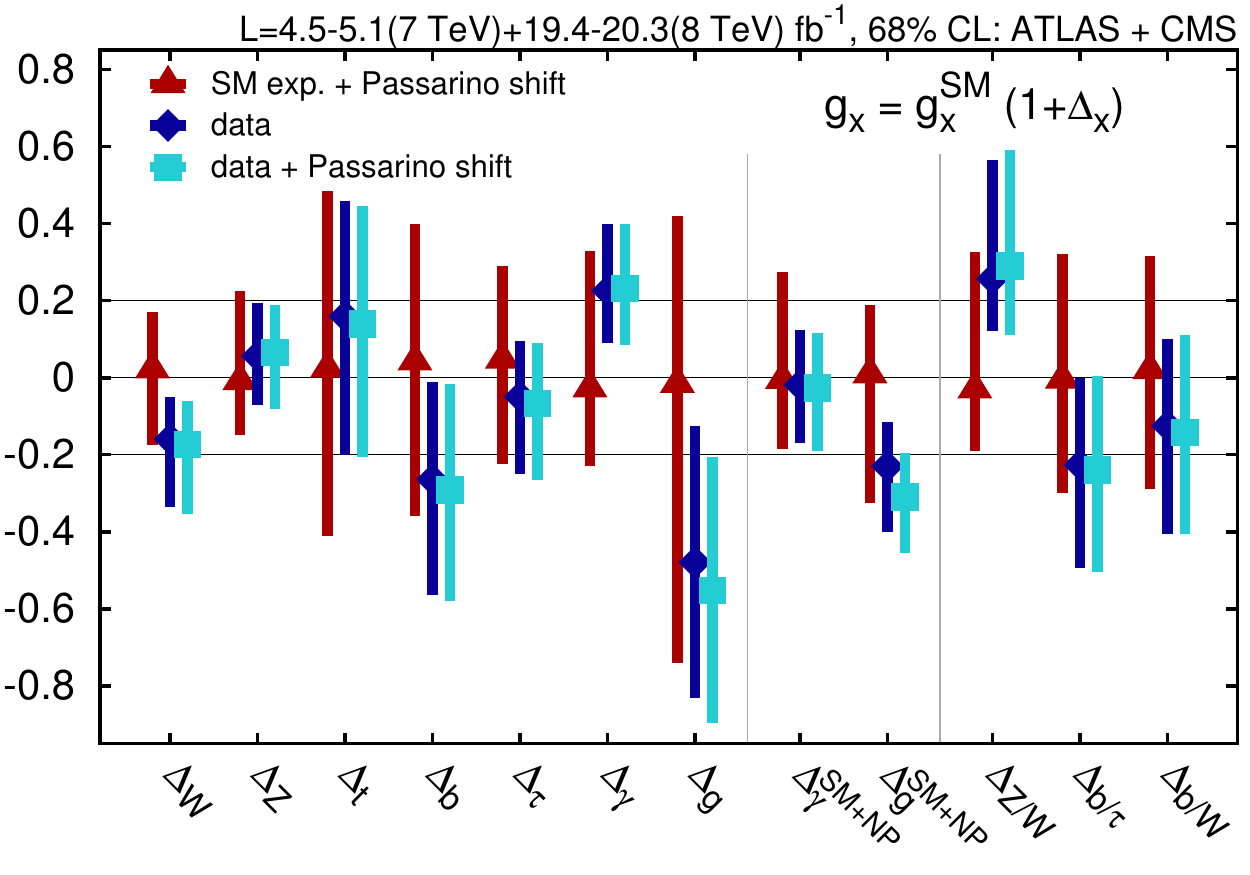}
  \includegraphics[width=0.45\textwidth]{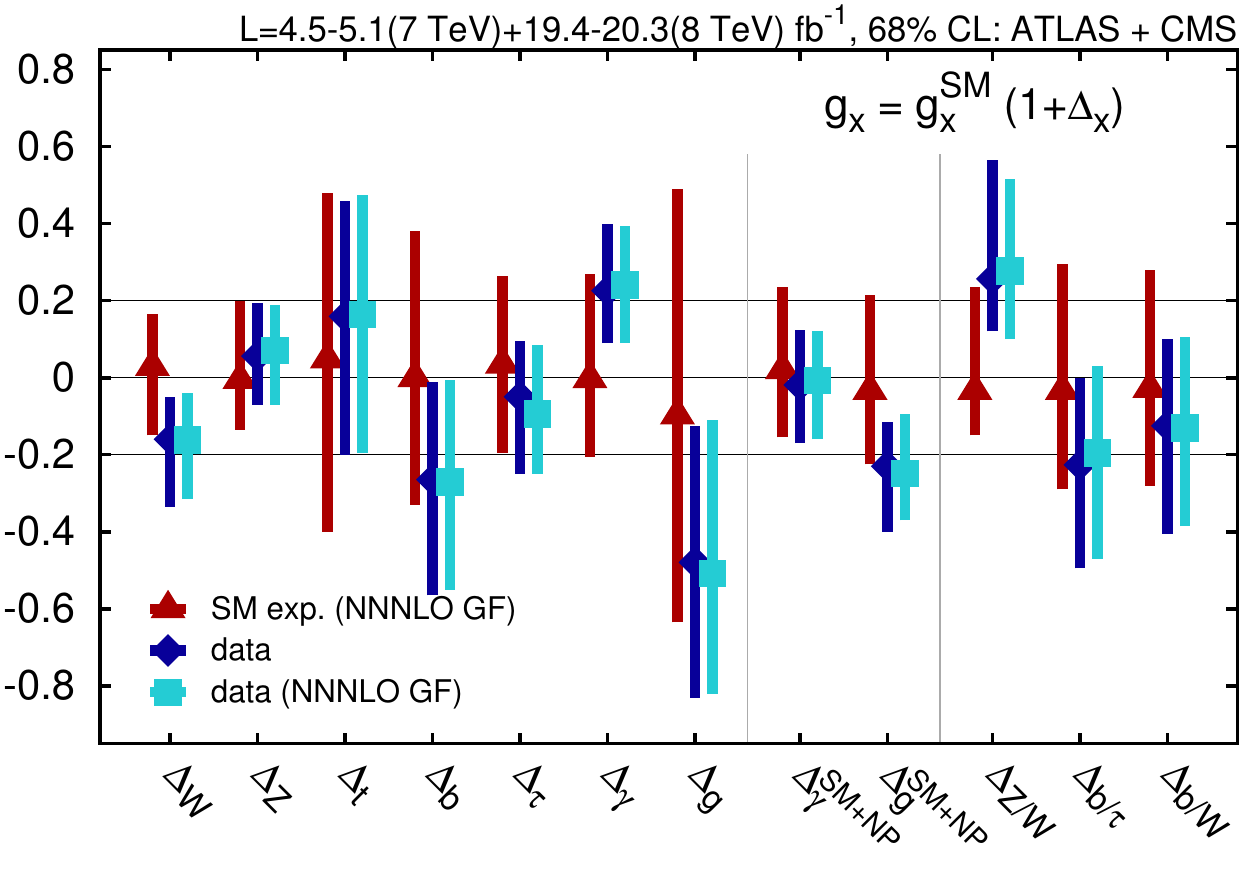}
  \hspace*{0.03\textwidth}
  \includegraphics[width=0.45\textwidth]{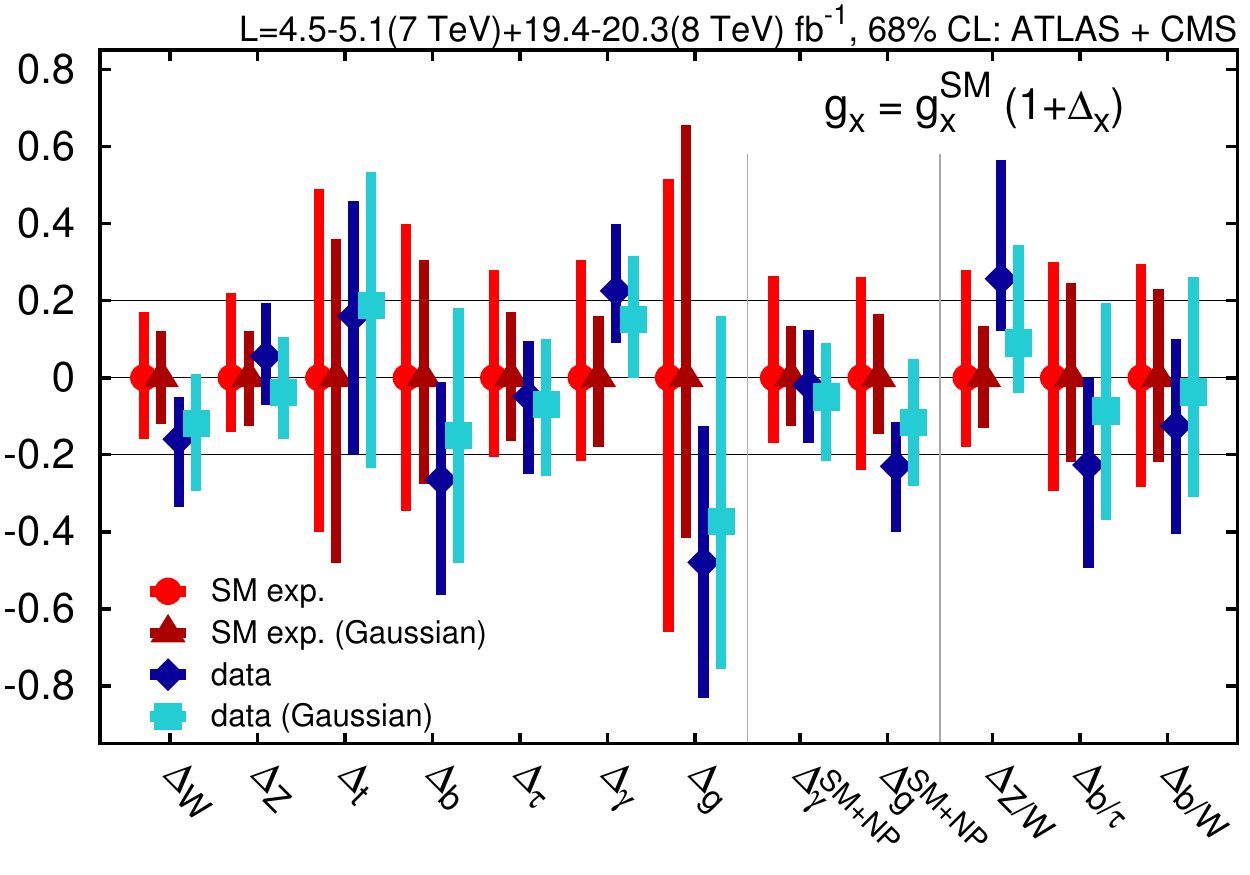}
  \caption{68\% CL error bars on the deviations $\Delta_x$ from all Standard
    Model couplings of the observed Higgs boson. First, we show the
    effect of fully correlated theoretical uncertainties on the
    different production processes (upper left); next, we show the
    results with the modified theoretical uncertainties proposed in
    Ref.~\cite{passarino} (upper right); then, we illustrate the
    effect of the N$^3$LO calculation of the Higgs cross
    section~\cite{n3lo} (lower left); finally, we illustrate what
    happens when we simulate the theoretical uncertainties with a
    Gaussian distribution (lower right). The results labelled `SM exp'
    assume central values on the Standard Model expectation, but the
    current data error bars.}
\label{fig:delta_err}
\end{figure}

With this observation in mind we show a set of results from our
systematic study of the appropriate treatment of theoretical
uncertainties in Fig.~\ref{fig:delta_err}. The \textsc{SFitter}
standard approach to theoretical uncertainties is based on
\begin{itemize}
\item uncorrelated uncertainties for the production, to account for
  very different kinematic selections;
\item an allowed cross section range given by the scale dependence
  of the best available prediction;
\item a flat likelihood distribution of the associated nuisance
  parameter for the cross sections and the decays.
\end{itemize}
These three assumptions we check one-by-one. In the upper left panel
of Fig.~\ref{fig:delta_err} we show an alternative \textsc{SFitter}
analysis with fully correlated uncertainties for the production rates,
including the error bar from the parton densities.  Because of the
strong correlations between the different production and decay
processes expressed in terms of Higgs couplings such a shift in the
assumed correlations for the theoretical uncertainties could have a
significant effect. However, we see that for full correlations the
size of the error bars is only slightly reduced, in spite of the fact
that the central values for example for $\Delta_b$ shifts by half a
standard deviation. The truth lies somewhere in between the fully
correlated and the fully uncorrelated theoretical uncertainties, where for
the upcoming Run~II there will be a tendency towards less correlation
because of the more specific analysis strategies. On the other hand,
the difference between fully uncorrelated and fully correlated errors
is not worrisome and we stay on the conservative, uncorrelated
side.\bigskip

In the upper right panel of Fig.~\ref{fig:delta_err} we show the
change in the extracted Higgs couplings when we modify the assumed
theoretical uncertainties following Ref.~\cite{passarino}.  For a
collider energy of 8~TeV and a Higgs mass of 125~GeV the default
prediction of the gluon fusion production cross section is
\begin{equation}
\sigma_{pp \to H}
= 19.52~\pb \pm 7.5\%_\text{pert} \pm 7.2\%_\text{pdf}  \; .
\end{equation}
Using the modified definition of Ref.~\cite{passarino} it becomes
\begin{equation}
\sigma_{pp \to H}
= 22.49~\pb \pm 13\%_\text{pert} \pm 7.2\%_\text{pdf}  \; .
\end{equation}
The change in central values as well as in the size of the error bars
is hardly observable, in spite of the sizeable change in the size of
the theoretical uncertainties. This confirms the earlier observation
that theoretical uncertainties are subleading for the Run~I analysis.

Also related to the theoretical uncertainty we include the recent
computation of the N$^3$LO corrections to the Higgs production rate at
the LHC~\cite{n3lo}. The corresponding results are shown in the lower
left panel of Fig.~\ref{fig:delta_err}.  Now the cross section
prediction reads
\begin{equation}
\sigma_{pp \to H}
= 19.95~\pb \pm 1.5\%_\text{pert} \pm 7.2\%_\text{pdf}  \; ,
\end{equation}
with a significantly more optimistic error based on the central scale
choice $\mu_{R,F} = m_H/2$. Again, the now strongly reduced
theoretical uncertainty hardly affects the Run~I results.\bigskip

Finally, we compare the precision of the Higgs couplings determination
with flat theoretical uncertainties with a Gaussian nuisance
parameter. The main differences between the frequentist \textsc{RFit}
treatment and Gaussian theoretical uncertainties are not related to
the shape of the final distribution, but to the size of the combined
theoretical uncertainties. First, combining two flat theoretical
uncertainties, for example from unknown higher orders and the parton
densities, will lead to a linear combination of the two error bars in
the frequentist \textsc{RFit} scheme~\cite{xswg,rfit,lecture}. In the
Gaussian approach they are added in quadrature. Second, it is not
clear with which Gaussian significance we should identify the ends of
the box-shaped distribution. For example, computing the standard
deviation of a flat data set stays well below the size of the
box. This means that if we compare the range of one standard deviation
for the \textsc{RFit} scheme with one standard deviation of the
Gaussian, the error on the flat distribution appears smaller. 

Per se, it is not clear which of the two effects will dominate in a
given fit.  In this situation we could choose the flat and Gaussian
theoretical uncertainties without a clear preference. We stick to the
former because we assume that it will be the conservative approach
once theoretical errors actually affect LHC results with larger data
sets.

\section{Higgs operators}
\label{sec:eff}

Going beyond a measurement of all couplings predicted by the Standard
Model we can ask a different question: \textsl{Which consistent
  Lagrangian describes all LHC measurements best?}  A standard
approach is defined by effective field theory~\cite{effective}, where
we categorize a Lagrangian with the appropriate symmetries in terms of
the expansion parameter. While the results of the previous section can
be interpreted in the framework of a non-linear effective Lagrangian
approach as we have explained, in this section we focus on the linear
case. In the linear sigma model we construct a $SU(2)_L\times
U(1)_Y$-symmetric Higgs Lagrangian based on the doublet $\Phi$ and
order it according to the inverse powers of the cutoff scale,
$1/\Lambda$~\cite{effective-linear,higgsmultiplets,kaoru,Grzadkowski:2010es}.
The Lagrangian, here restricting to all dimension-6 operators
\begin{alignat}{5}
\lag = \sum_x \frac{f_x}{\Lambda^2} \; \ope_x 
\label{eq:def_f}
\end{alignat}
is gauge invariant, but not fully renormalizable or unitary. 

Strictly speaking, in the SM Higgs sector we should separate two
sources of dimension-6 operators. Yukawa couplings or gauge boson
couplings from spontaneous symmetry breaking violate the
Appelquist--Carazzone decoupling theorem~\cite{appcar}, which means
that the Higgs couplings to photons and gluons are only suppressed by
$1/v$.  New physics generally gives rise to dimension-6 operators
suppressed by $1/\Lambda^2$, leading to Higgs coupling strengths to
photons and gluons scaling like $v/\Lambda^2$. This distinction will
be reflected in the normalization of the respective operators below.

\subsection{Dimension-6 operator basis}

Before we present the result of the LHC analysis we need to define our
basis of dimension-6 operators. The minimum independent set of
dimension-6 operators with the SM particle content (including the
Higgs boson as an $SU(2)_L$ doublet) and compatible with the SM gauge
symmetries as well as baryon number conservation contains 59
operators, up to flavor and Hermitian conjugation~\cite{Grzadkowski:2010es}.
To present our choice of operator basis~\cite{barca}, we start by
imposing $C$ and $P$ invariance and employing for the bosonic sector
the classical non-minimal set of dimension-6 operators in the
HISZ basis~\cite{kaoru}, with the following
operators contributing to the Higgs interactions with gauge bosons:
\begin{alignat}{9}
 \ope_{GG} &= \Phi^\dagger \Phi \; G_{\mu\nu}^a G^{a\mu\nu}  
& \ope_{WW} &= \Phi^{\dagger} \hat{W}_{\mu \nu} \hat{W}^{\mu \nu} \Phi  
& \ope_{BB} &= \Phi^{\dagger} \hat{B}_{\mu \nu} \hat{B}^{\mu \nu} \Phi 
\notag \\
 \ope_{BW} &=  \Phi^{\dagger} \hat{B}_{\mu \nu} \hat{W}^{\mu \nu} \Phi 
& \ope_W &= (D_{\mu} \Phi)^{\dagger}  \hat{W}^{\mu \nu}  (D_{\nu} \Phi) 
& \ope_B &=  (D_{\mu} \Phi)^{\dagger}  \hat{B}^{\mu \nu}  (D_{\nu} \Phi)
\notag \\
 \ope_{\Phi,1} &=  \left ( D_\mu \Phi \right)^\dagger \Phi\  \Phi^\dagger
                  \left ( D^\mu \Phi \right ) \qquad 
& \ope_{\Phi,2} &= \frac{1}{2} \partial^\mu\left ( \Phi^\dagger \Phi \right)
                            \partial_\mu\left ( \Phi^\dagger \Phi \right) \qquad
& \ope_{\Phi,4} &= \left ( D_\mu \Phi \right)^\dagger \left(D^\mu\Phi \right)
                 \left(\Phi^\dagger\Phi \right ) \; .
\label{eq:eff}  
\end{alignat}
Here the Higgs doublet covariant derivative is
$D_\mu\Phi= \left(\partial_\mu+ i g' B_\mu/2 + i g
\sigma_a W^a_\mu/2 \right)\Phi $, the hatted field strengths
are $\hat{B}_{\mu \nu} = i g' B_{\mu \nu}/2$ and
$\hat{W}_{\mu\nu} = i g\sigma^a W^a_{\mu\nu}/2$, with the
Pauli matrices written as $\sigma^a$. The $SU(2)_L$ and $U(1)_Y$ gauge
couplings are $g$ and $g^\prime$. The additional operator
$\ope_{\Phi,3} = (\Phi^\dag \Phi)^3$ is crucial for the structure of
the Higgs potential and for a theoretical interpretation of the
measurement of the Higgs self-coupling, but we can safely omit it for
the LHC Run~I analysis.

The final choice of structures for our global Higgs analysis follows
Ref.~\cite{barca}, relying on operators contributing to existing
data. We first use the equations of motion (including all necessary
fermionic operators~\cite{Grzadkowski:2010es} omitted in this brief
introduction) to rotate to a basis where there are not blind
directions linked to electroweak precision data. We then neglect all
operators contributing to the bulk of electroweak precision data at
tree level; their coefficients will be too constrained to lead to
observable deviations in LHC Higgs measurements. After using the
remaining equation of motion to remove redundancy, we finally neglect
all operators that we know will not be constrained by LHC Higgs
measurements. We are left with the final set of nine operators that
parametrize the Higgs interactions at the LHC.  For the gauge boson
interactions they are
\begin{alignat}{5}
\lag_\text{eff}^{HVV} = &
- \frac{\alpha_s }{8 \pi} \frac{f_{GG}}{\Lambda^2} \ope_{GG}  
+ \frac{f_{BB}}{\Lambda^2} \ope_{BB} 
+ \frac{f_{WW}}{\Lambda^2} \ope_{WW} 
+ \frac{f_B}{\Lambda^2} \ope_B 
+ \frac{f_W}{\Lambda^2} \ope_W 
+ \frac{f_{\Phi,2}}{\Lambda^2} \ope_{\Phi,2} \; .
\label{eq:ourleff}
\end{alignat}
The operator $\ope_{\Phi,2}$ appears in the gauge and fermionic
Lagrangians, because it leads to a finite renormalization of the Higgs
field and hence a universal shift of all Higgs couplings to Standard Model
fields~\cite{barca}. This set of dimension-6 effective operators gives rise to the
following Higgs interactions with SM gauge boson pairs,
\begin{alignat}{5}
\lag^{HVV} 
&= g_{Hgg} \; H G^a_{\mu\nu} G^{a\mu\nu} 
+  g_{H \gamma \gamma} \; H A_{\mu \nu} A^{\mu \nu}
+ g^{(1)}_{H Z \gamma} \; A_{\mu \nu} Z^{\mu} \partial^{\nu} H 
+  g^{(2)}_{H Z \gamma} \; H A_{\mu \nu} Z^{\mu \nu} \notag \\
&+ g^{(1)}_{H Z Z}  \; Z_{\mu \nu} Z^{\mu} \partial^{\nu} H 
+  g^{(2)}_{H Z Z}  \; H Z_{\mu \nu} Z^{\mu \nu} 
+  g^{(3)}_{H Z Z}  \; H Z_\mu Z^\mu \notag \\
&+ g^{(1)}_{H W W}  \; \left (W^+_{\mu \nu} W^{- \, \mu} \partial^{\nu} H 
                            +\text{h.c.} \right) 
+  g^{(2)}_{H W W}  \; H W^+_{\mu \nu} W^{- \, \mu \nu} 
+  g^{(3)}_{H W W}  \; H W^+_{\mu} W^{- \, \mu} \; ,
\label{eq:lhvv}
\end{alignat}
where $V_{\mu \nu} = \partial_\mu V_\nu - \partial_\nu V_\mu$,
with $V=A,Z,W,G$. These effective couplings
are related to the coefficients in Eq.\eqref{eq:ourleff} through
\begin{alignat}{5}
g_{Hgg} &=
         -\frac{\alpha_s}{8 \pi} \frac{f_{GG} v}{\Lambda^2} 
& g^{(1)}_{H Z \gamma} &= \frac{g^2 v}{2 \Lambda^2} \; \frac{s_w (f_W - f_B) }{2 c_w} \notag \\
g_{H \gamma \gamma} &= - \frac{g^2 v s_w^2}{ 2\Lambda^2} \; \frac{f_{BB} + f_{WW}}{2} 
& g^{(2)}_{H Z \gamma} &= \frac{g^2 v}{2 \Lambda^2} \; \frac{s_w (2 s_w^2 f_{BB} - 2 c_w^2 f_{WW} )}{2 c_w} \notag  \\
g^{(1)}_{H Z Z} &= \frac{g^2 v}{ 2\Lambda^2} \; \frac{c_w^2 f_W + s_w^2 f_B}{2 c_w^2} 
&g^{(1)}_{H W W} &= \frac{g^2 v}{2\Lambda^2} \; \frac{f_W}{2} \notag \\
g^{(2)}_{H Z Z} &= - \frac{g^2 v}{2\Lambda^2} \; \frac{s_w^4 f_{BB} +c_w^4 f_{WW}}{2 c_w^2} 
&g^{(2)}_{H W W} &= - \frac{g^2 v }{2\Lambda^2} \; f_{WW} \notag \\
g^{(3)}_{H Z Z} &= m_Z^2 (\sqrt{2} G_F)^{1/2} \left(1-\frac{v^2}{2\Lambda^2}f_{\Phi,2}\right) \qqqquad 
&g^{(3)}_{H W W} &= m_W^2(\sqrt{2} G_F)^{1/2} \left(1-\frac{v^2}{2 \Lambda^2} f_{\Phi,2} \right) \; ,
\label{eq:g} 
\end{alignat}
where $s_w$ and $c_w$ stands for the sine and cosine of
  the weak mixing angle.\bigskip

We finally focus on the huge set of dimension-6 operators contributing
to the Higgs interactions with fermion pairs~\cite{Grzadkowski:2010es}.
Because of a lack of appropriate observables in the LHC Higgs measurements,
from the fermionic operators left in the final basis
we limit ourselves to the flavor-diagonal Yukawa structures
\begin{equation}
\ope_{e\Phi,33}=(\Phi^\dagger\Phi)(\bar L_3 \Phi e_{R,3}) 
\qquad 
\ope_{u\Phi,33}=(\Phi^\dagger\Phi)(\bar Q_3 \tilde \Phi u_{R,3})
\qquad 
\ope_{d\Phi,33}=(\Phi^\dagger\Phi)(\bar Q_3 \Phi d_{R,3}) \; ,
\label{eq:hffop}
\end{equation}
with $\tilde \Phi=i\sigma_2\Phi^*$, and where the conventions for the
fermion fields are $L$ for the lepton doublet, $Q$ for the quark
doublet, and $f_R$ for the $SU(2)_L$ singlet fermions.  The
corresponding effective Lagrangian for the fermionic interactions
reads
\begin{alignat}{5}
\lag_\text{eff}^{Hff} = &
& \frac{f_\tau m_\tau}{v \Lambda^2} \ope_{e\Phi,33} 
+ \frac{f_b m_b}{v \Lambda^2} \ope_{d\Phi,33} 
+ \frac{f_t m_t}{v \Lambda^2} \ope_{u\Phi,33} 
+ \frac{f_{\Phi,2}}{\Lambda^2} \ope_{\Phi,2} \; .
\label{eq:ourleff2}
\end{alignat}
As mentioned above, $\ope_{\Phi,2}$ affects the Higgs couplings universally. In
analogy to the Higgs-gluon coupling we scale the fermionic $f_x$ by a
factor $m/v$ to reflect the chiral nature of the Higgs coupling
operator~\cite{d6_review}. For the Higgs couplings to SM fermions
this implies
\begin{equation}
 \lag^{Hff} = g_f H \bar f_{L} f_{R} + \text{h.c.} 
\qquad \text{with} \quad 
g_f  =  - \frac{m_f}{v} \left( 1-\frac{v^2}{2\Lambda^2}f_{\Phi,2} - \frac{v^2}{\sqrt{2}\Lambda^2}  f_f \right) \; ,
\label{eq:g2}
\end{equation}
where we define the physical masses and fermions in the mass basis ($f = \tau,b,t$). 

\subsection{Rate-based analysis}
\label{sec:eff_fit}

As a first step we update the global analysis of dimension-6
operators based on the complete Run~I data in the \textsc{SFitter}
framework.  The main difference to the analysis of Ref.~\cite{barca}
is the variable top--Yukawa operator, which can now be constrained by
$t\bar{t}H$ production as well as the Higgs production via gluon fusion.

The contributions of the dimension-6 operators to the production rates
and decay widths are calculated using
\textsc{MadGraph5}~\cite{Alwall:2011uj} and
\textsc{FeynRules}~\cite{Christensen:2008py}. We check our results
with \textsc{Comphep}~\cite{Pukhov:1999gg,Boos:2004kh} and
\text{VBFNLO}~\cite{VBFNLO}. We approximately include
higher-order corrections through $K$-factors computed for the Standard
Model processes~\cite{xswg}. Similarly, for this
rate-based analysis we assume that all detector
efficiencies are identical for both the SM Higgs processes and the
corresponding dimension-6 contributions. The results of this
9--parameter global analysis are shown in Fig.~\ref{fig:dim6_corr}
and Fig.~\ref{fig:dim6}, after performing a statistical
analysis as described in Sec.~\ref{sec:intro_setup}.
For the present case we show the multiple degenerate
solutions.\bigskip

\begin{figure}[b!]
  \includegraphics[width=0.29\textwidth]{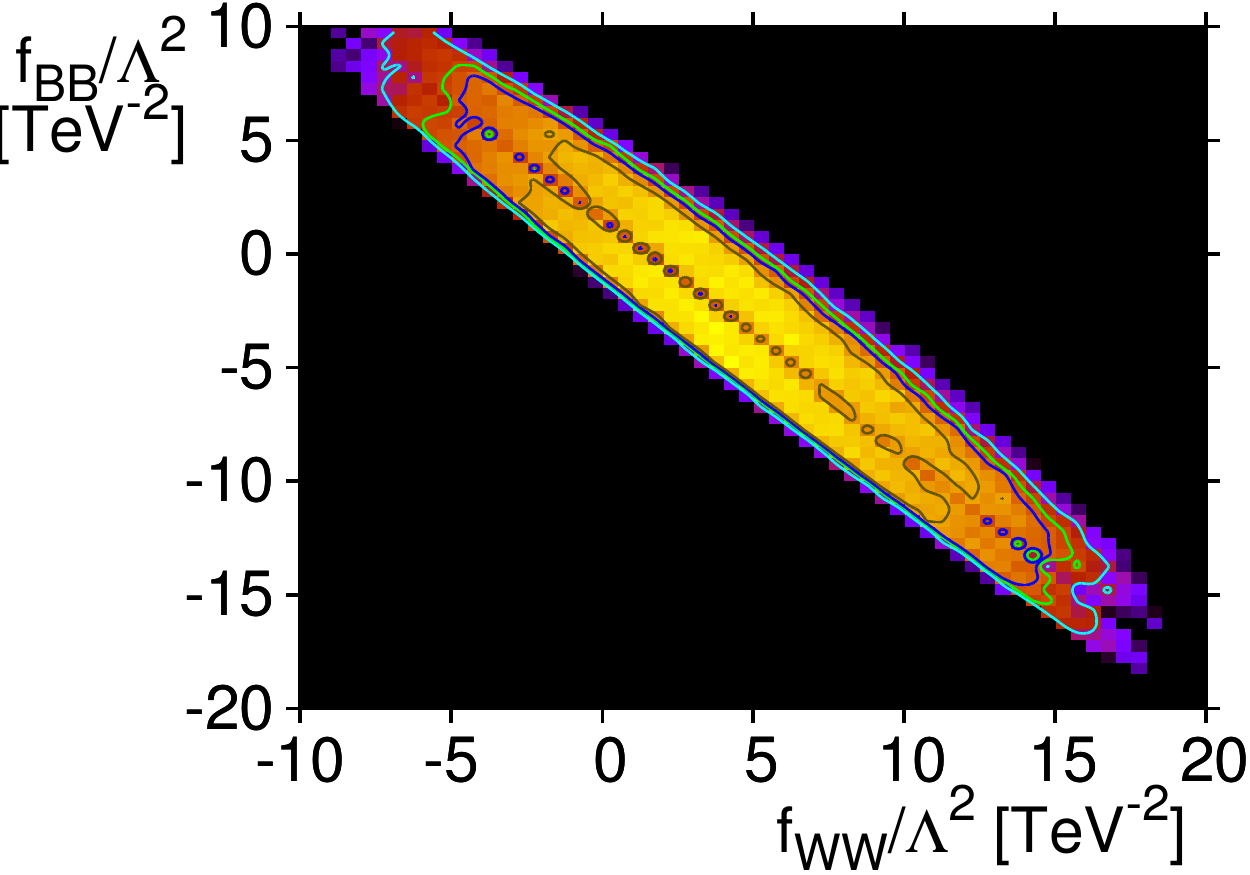}
  \hspace*{1ex}
  \includegraphics[width=0.29\textwidth]{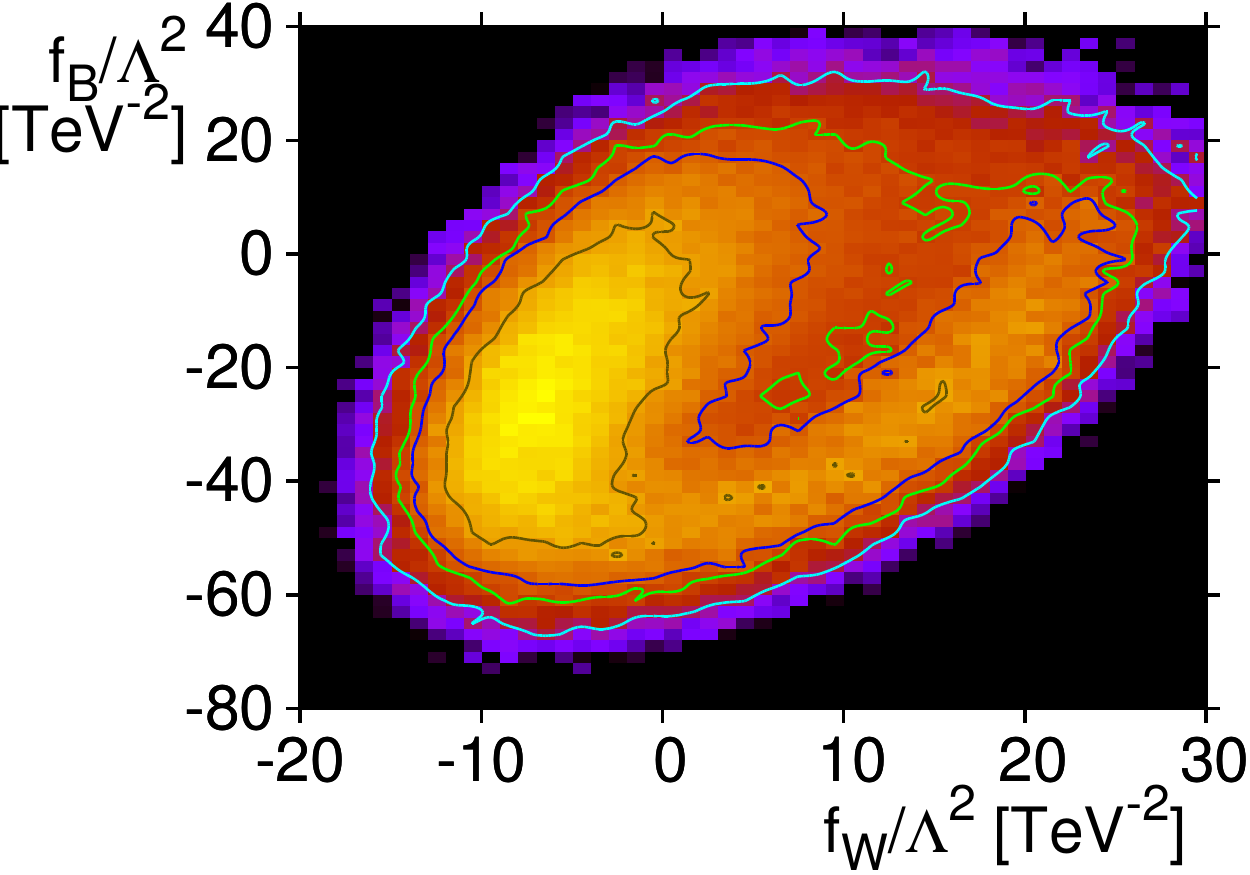}
  \hspace*{1ex}
  \includegraphics[width=0.29\textwidth]{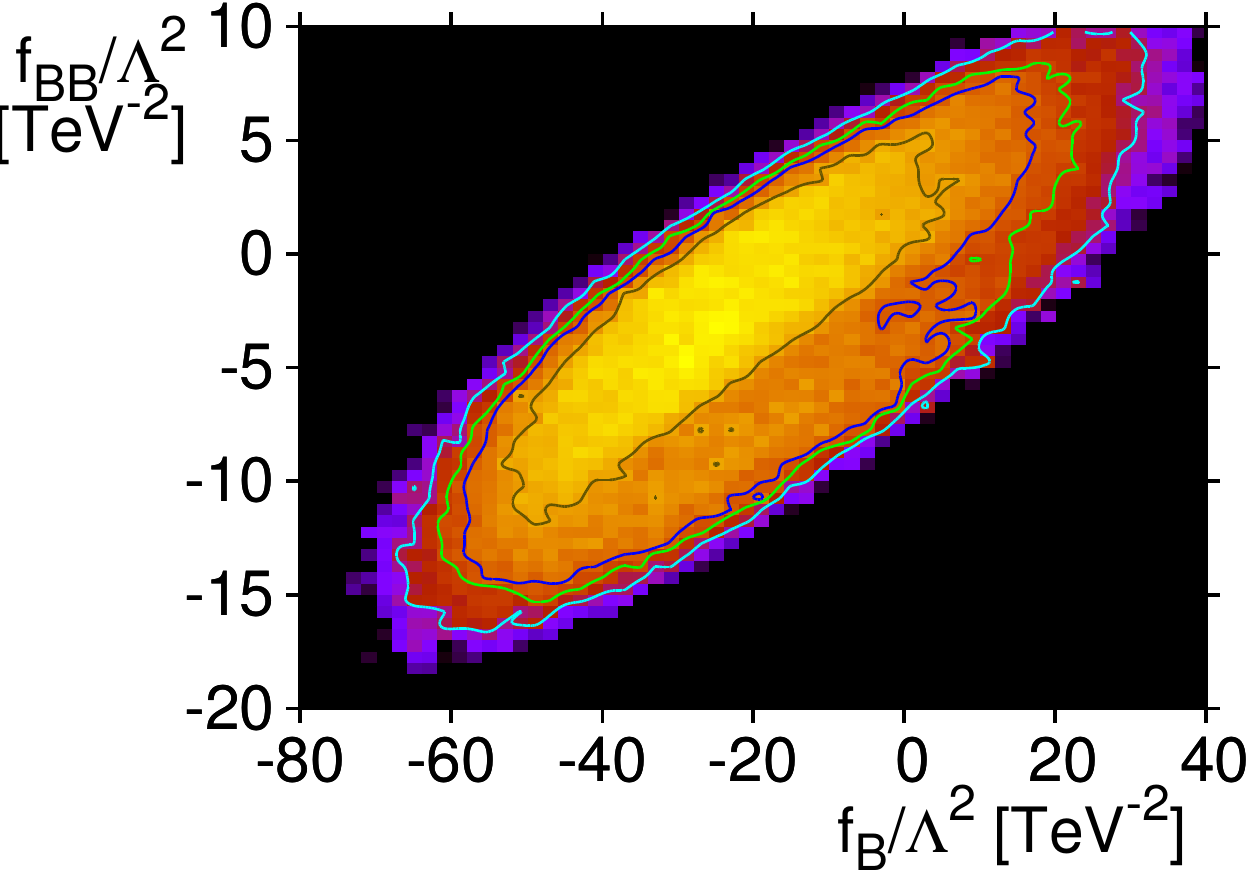}
  \hspace*{1ex}
  \raisebox{3pt}{\includegraphics[width=0.0545\textwidth]{figs/colorbox}}
  \caption{Correlations between different coefficients $f_x/\Lambda^2$,
    measured in $\itevx$. The 1-dimensional profile
    likelihoods corresponding to these results are shown as the blue bars in
    Fig.~\ref{fig:dim6}.}
\label{fig:dim6_corr}
\end{figure}

\begin{figure}[t]
  \centering
  \includegraphics[width=0.65\textwidth]{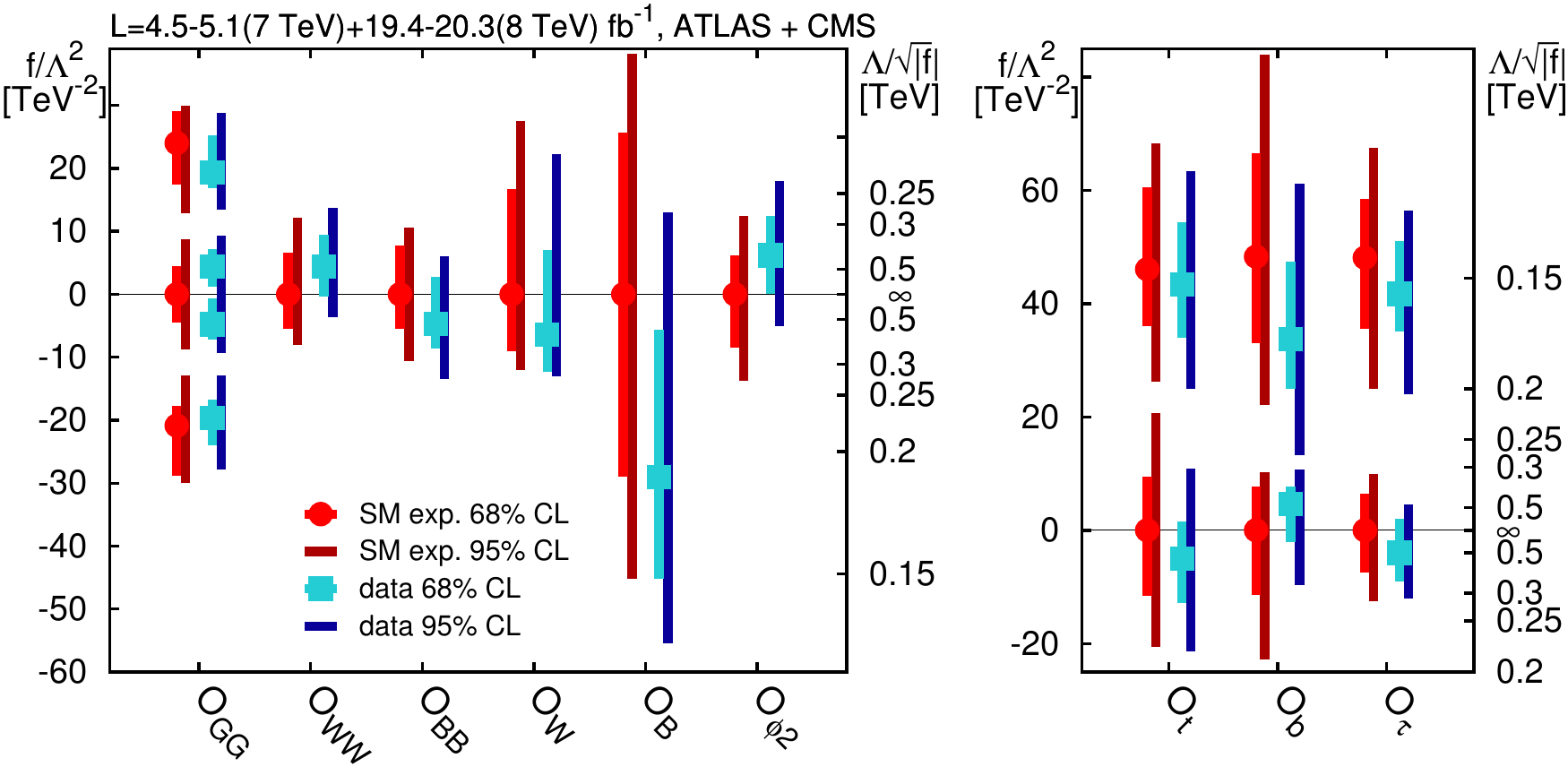}
  \caption{Error bars on the coefficients $f_x/\Lambda^2$ for the
    dimension-6 operators defined in Eq.\eqref{eq:ourleff} (left
    panel) and Eq.\eqref{eq:ourleff2} (right panel). We only include
    total rate information and show 68\%~CL as well as 95\%~CL
    contours. Unlike for the other 1-dimensional profile likelihoods
    we keep track of the secondary minima in this one figure.  The
    results labelled `SM~exp' assume central values on the Standard
    Model expectation, but the current data error bars.}
\label{fig:dim6}
\end{figure}

In Fig.~\ref{fig:dim6_corr} we depict a selection of interesting
correlations between the dimension-6 operators.
In addition to the correlations discussed
in the previous section, \eg $f_{GG}$ vs $f_t$ shown in
Fig.~\ref{fig:delta_gcorr}, the dimension-6 operators introduce a rich
structure of correlations related to the Higgs interactions with
electroweak gauge bosons. As long as the
analysis is only based on rate measurements in the Higgs sector,
these correlations are the main difference compared to the
$\Delta$-framework. The strongest of
these correlations is due to the di-photon channel, as it is measured with the
highest precision. Therefore, the tree-level contributions from $f_{WW}$ and
$f_{BB}$ to the Higgs coupling to photon pairs generate the strong
correlation in the left panel of Fig.~\ref{fig:dim6_corr}; see Eq.\eqref{eq:lhvv}.
The two, slightly separated, allowed regions at 68\%~CL are
due to the interference between the dimension-6 amplitudes and the
Standard Model ones. The fact that both $f_{WW}$ and $f_{BB}$ receive
their strongest constraints from the di-photon channel, reflects that
their contribution in the rate based analysis is very similar to the
addition of $\Delta_\gamma$ in the previous section.
While this strong correlation is partially broken by their smaller
contribution to the other channels in the analysis, we will see in the
following section that the addition of kinematic distributions
will increase the sensitive to $f_{WW}$ and $f_{BB}$ stemming from VBF
and Higgs associate production channels.

In the central panel of Fig.~\ref{fig:dim6_corr} we show the
correlation between $f_B$ and $f_W$. The Wilson coefficient $f_W$ is
much more strongly constrained than $f_B$, because of the large
contributions of the former to the $HVV$ vertices ($V=Z,W^\pm$)
while $f_B$ only contributes to $HZZ$ with a weak mixing angle
suppression. The mild impact of $f_B$ will eventually be compensated
by measurements of $H \to Z\gamma$ decays. Moreover, the
  contributions to $HVV$ also correlate $f_B$ to $f_{WW}$ and
$f_{BB}$, as displayed in the right panel of Fig.~\ref{fig:dim6_corr}.

The universal contribution of $f_{\Phi,2}$ to all Higgs couplings
strongly correlates this operator with the rest of dimension-6
structures, both in the bosonic and in the fermionic sectors.  This
way, $f_{\Phi,2}$ in principle lifts the degeneracy between the two
allowed regions for $f_b$ and $f_{\tau}$, which is due to the
interference between the SM amplitudes and the higher-dimensional
operators. The actual likelihood values for the two minima are still
equivalent though. In the $\Delta$-framework these regions are almost
entirely degenerate, allowing us to focus on the SM-like solution in
that case.

Starting with the assumption that to first approximation the
rate-based analysis of dimension-6 operators is physically equivalent to the
Higgs coupling analysis described in Sec.~\ref{sec:coup_sm} the strong
correlations shown in Fig.~\ref{fig:dim6_corr} still pose a technical
problem. The Higgs coupling modifications $\Delta_x$ are by definition
well aligned with the experimental measurements, which means that the
profile likelihood construction down to 1-dimensional likelihoods is
straightforward. For example the correlation between $f_B$ and $f_W$
makes it obvious that a profile likelihood either in $f_B$ or $f_W$
will have to deal with strongly non-Gaussian distributions, including
secondary minima.\bigskip

With this technical caveat in mind we show in Fig.~\ref{fig:dim6} the
best fit points and the corresponding 1-dimensional 68\% and 95\% CL
regions for each effective operator. We follow the procedure described
in Sec.~\ref{sec:intro_setup}, in this case keeping all possible
solutions for $\ope_{t,b,\tau}$ and $\ope_{GG}$. As we have discussed
the strongest constraints apply to $f_{WW}$ and $f_{BB}$. Next are
$f_{W}$ and $f_{\phi,2}$, and finally the weaker constraint $f_{B}$,
as discussed above. Just like for $\Delta_t$ the free value of $f_t$
enlarges the error bars for $f_{GG}$ and splits the allowed parameter
range into more or less distinct regions, like those shown in
Figs.~\ref{fig:delta_gcorr} and~\ref{fig:dim6}.

In the right panel we observe the expected secondary solutions for all
three $f_{t,b,\tau}$. To compare the errors on the couplings to
fermions we should keep in mind that in Eq.\eqref{eq:ourleff2} the
chiral factor is taken out of the definition of the operator and its
associated scale $\Lambda$. As we can see, at the 68\% CL the
secondary solutions appear as clear additional structures, while at
95\% CL the SM-like and secondary solutions barely separate for $f_b$.
This allows us to cleanly separate SM-like solutions from those with
merely switched signs of the Yukawa couplings. Note that the latter
correspond to a new physics scale $\Lambda \sim 150$~GeV in the
presence of a chiral symmetry factor, shedding some doubt on the
effective theory treatment as a whole.

From a statistical point of view it is not clear how one would deal
with such alternative solutions; in our case we show the solutions
with flipped signs of the Yukawa couplings in Fig.~\ref{fig:dim6}, but
will omit them in the 1-dimensional profile likelihood for the rest of
the present section.  In the Markov chain analysis they will be of
course still included.  We will revisit this issue in
Sec.~\ref{sec:off-shell} for the case of the top Yukawa coupling.

\subsection{Kinematic distributions}
\label{sec:eff_distri}

Based exclusively on total event rates, the results from the
previous section do not take full advantage of the available
information. In Eqs.\eqref{eq:lhvv} and~\eqref{eq:g2} we observe that
$\ope_{\Phi,2}$, $\ope_b$, $\ope_\tau$, and $\ope_t$ merely modify the
SM coupling strengths, but the other dimension-6 operators do generate
new Lorentz structures. These anomalous Lorentz structures are best
visible in Higgs production rather than decays, because the momentum
flow is not limited by the Higgs mass. Their study is indeed one of the
most interesting aspects of our effective field theory analysis.

To establish a framework for an implementation of kinematic
distributions into the Higgs operator analysis we first focus on $VH$
production and weak boson fusion. Adding kinematics to our global
analysis faces a considerable challenge, because we are limited to
fully documented distributions. When multi-variate analysis techniques
are applied, the documented distributions are usually not optimized.
However for two test cases we will show how we can consistently combine
information from rates with kinematic distributions
without weakening the analysis.

Finally, the effective Higgs Lagrangian does not define a UV-complete
theory if we only include dimension-6 operators.  The cutoff scale
$\Lambda$ is encoded in the ansatz, and at least for a weakly
interacting theory the experimental sensitivity offers a consistency
test.  There exist several ways to define a model which we can
consistently compare to data:
\begin{enumerate}
\item take the alternative model at face value, unless a prediction
  actually violates unitarity. This approach maximizes the
  distinguishing power of the measurement, but it only rules out the
  ultraviolet completion with the least SM-like behavior.
\item attach momentum--dependent form factors to soften the
  ultraviolet behavior. The main problem is that after going through a
  lot of trouble of defining an effective field theory hypothesis, we
  spoil it by introducing ad-hoc non-local interactions in the
  position--space Lagrangian.
\item only use data in phase space regions which are not sensitive to
  the ultraviolet completions, for example requiring $p_T < 100$~GeV
  for the tagging jets in weak boson fusion~\cite{higgs_spin} or an
  upper bound on $p_T^V$ in $VH$ production. The obvious disadvantage
  of this approach is that we lose experimental information and
  produce worse bounds, as we will see.
\end{enumerate}
In this analysis we will attempt to include as much of the kinematic
information as possible, but carefully check how much of the
distinguishing power comes from phase space regions not obviously
consistently described by the effective field theory.\bigskip

 \begin{figure}[t]
   \includegraphics[width=0.4\textwidth]{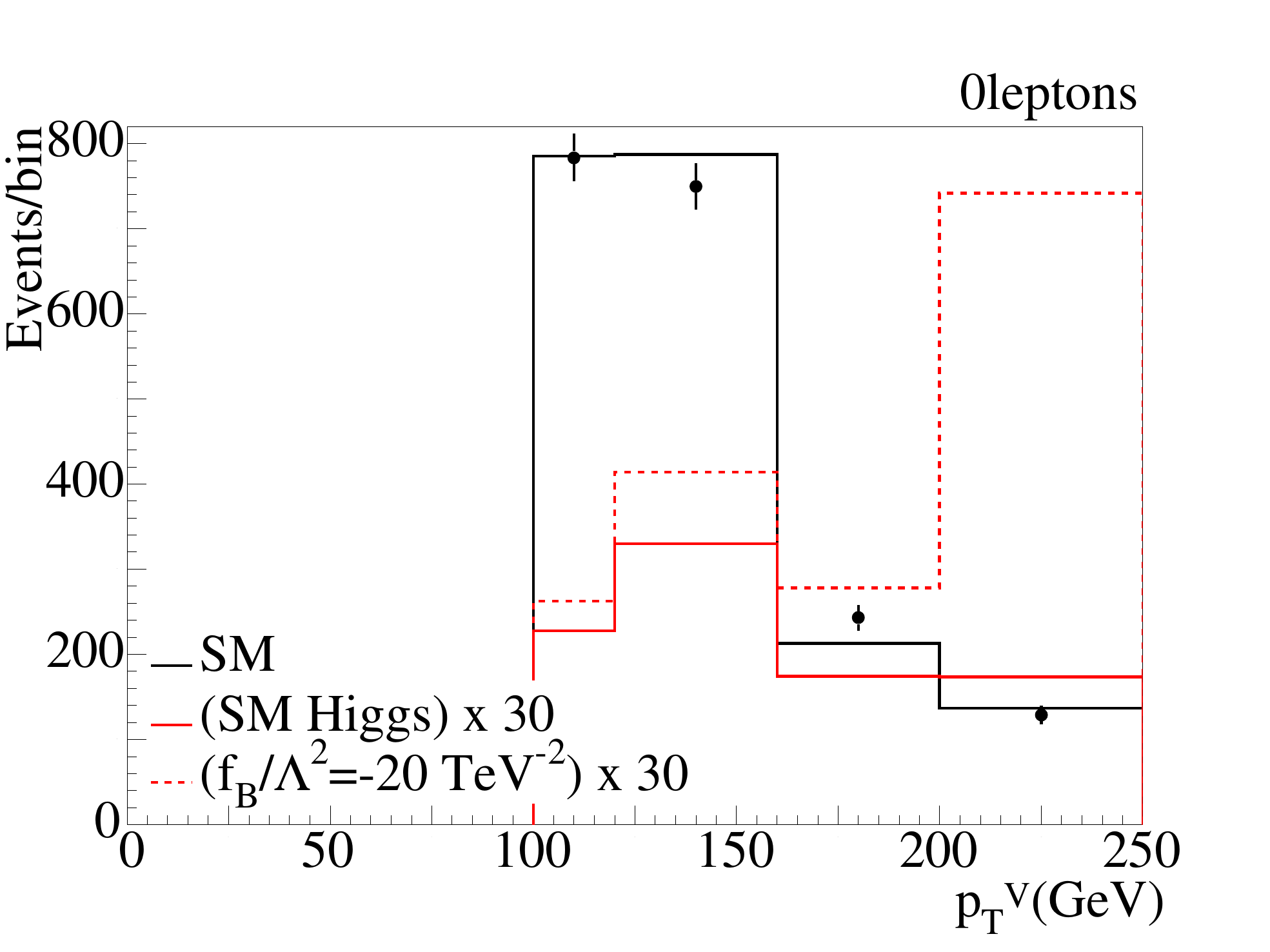}
   \includegraphics[width=0.4\textwidth]{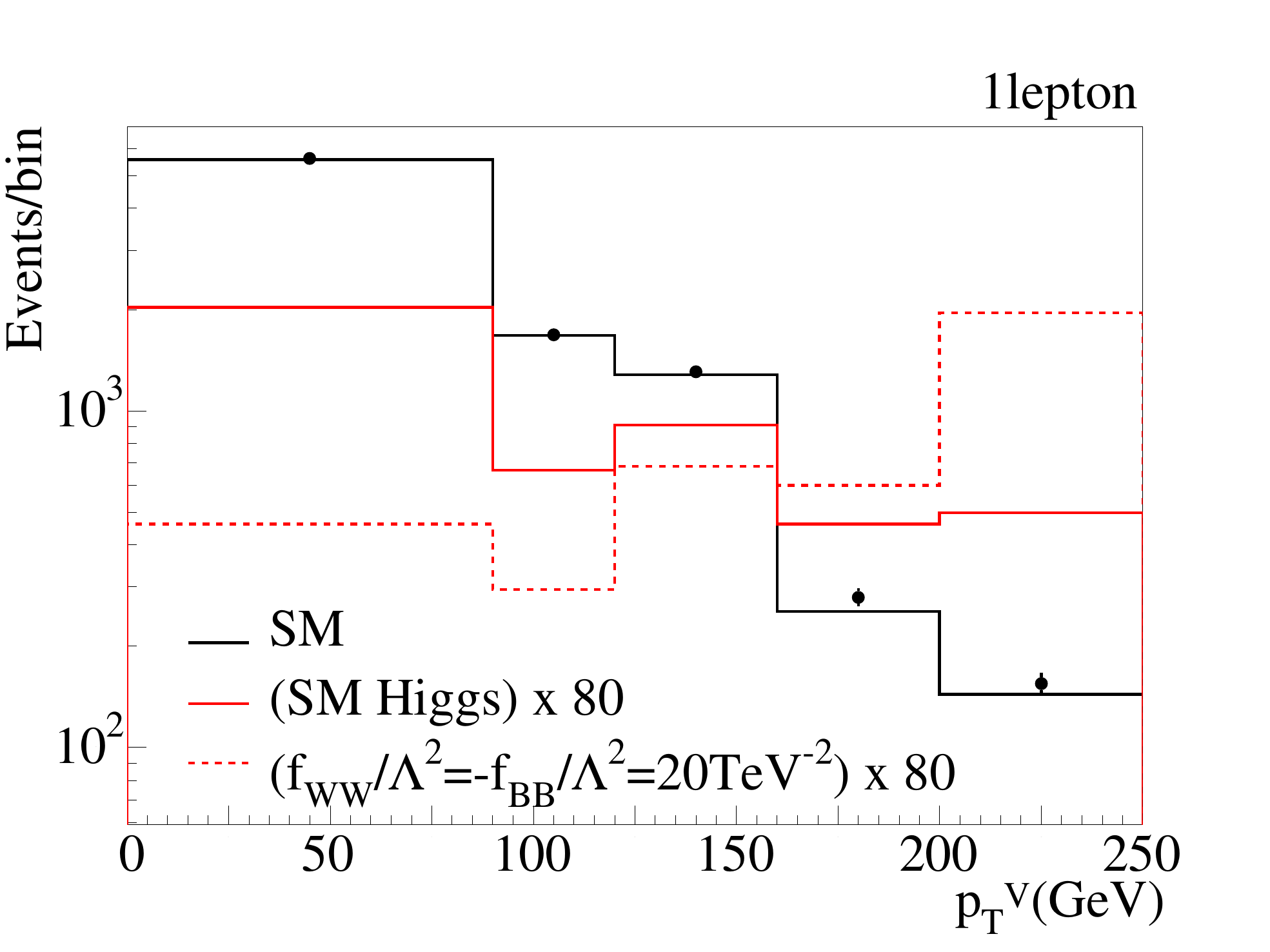}
   \includegraphics[width=0.4\textwidth]{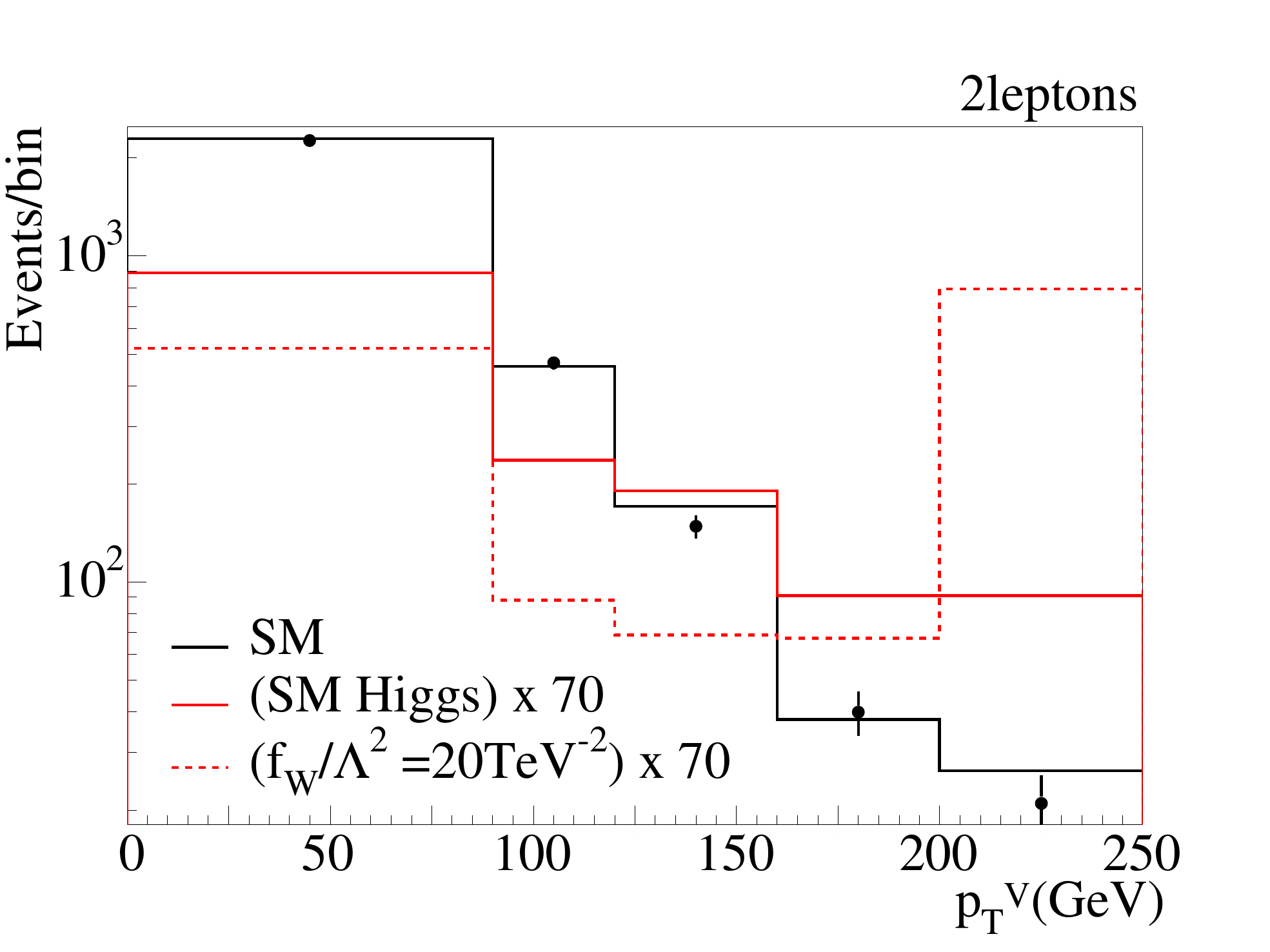}
   \includegraphics[width=0.4\textwidth]{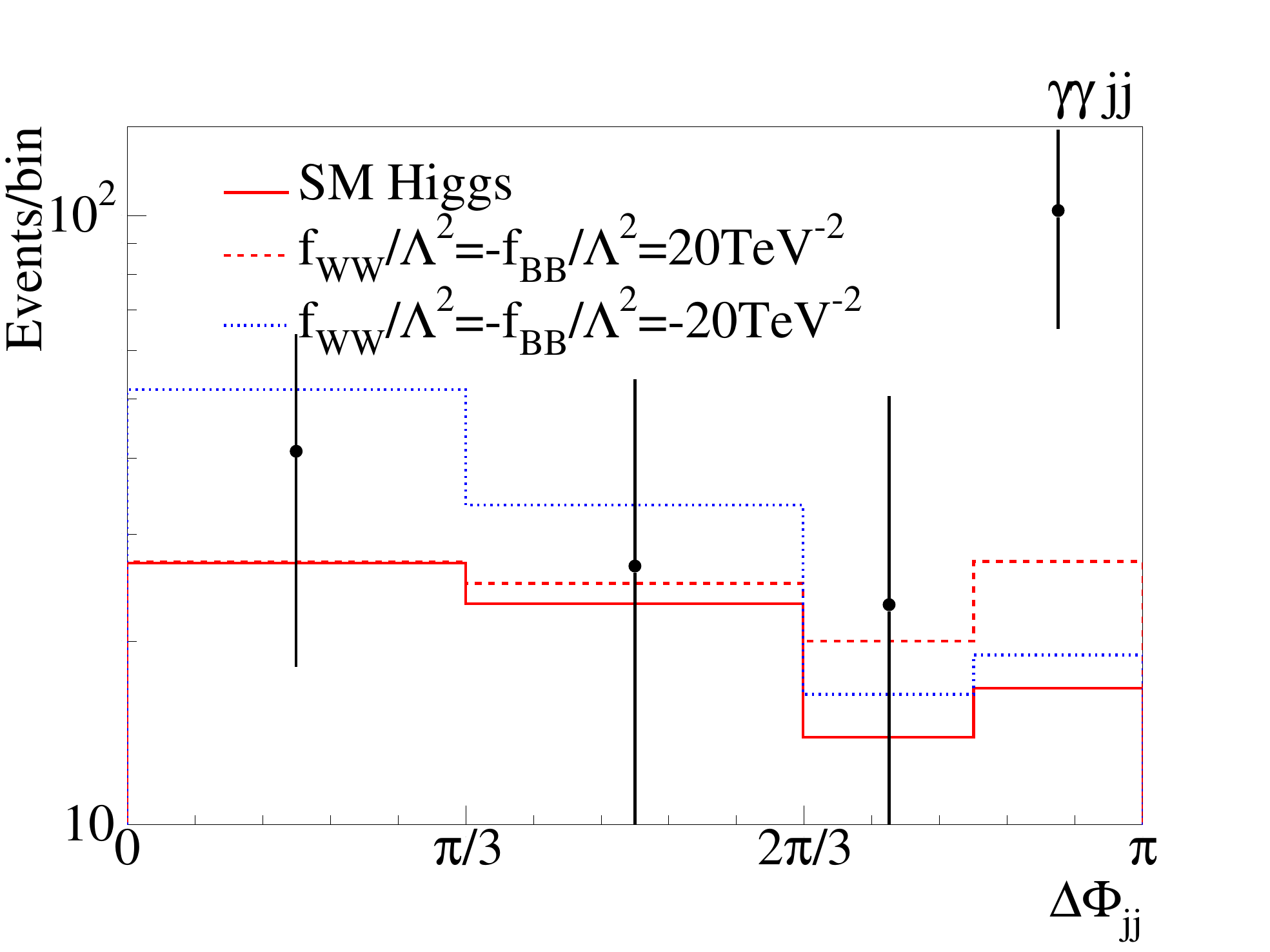}
   \caption{Upper and left lower panels: $p_T^V$ distributions from
     $VH$ production~\cite{1409.6212}. We show the SM
     Higgs-plus-background expectation (black solid), the number of
     observed events, the SM Higgs hypothesis (red solid), and the
     expectation from one dimension-6 operator (red dashed). Lower
     right: $\Delta \phi_{jj}$ distribution in Ref.~\cite{1407.4222}. We display the number of observed
     events, the SM--Higgs hypothesis (red), and the expectation from
     adding a set of dimension-6 operators (dashed red and dotted
     blue). All plots include $20.3~\ifb$ at 8~TeV. In the figure we
     neglect the effect of higher-dimensional operators on the
     branching ratios.}
 \label{fig:distributions}
 \end{figure}

We start with $VH \to V(b\bar{b})$ production. To be maximally
sensitive to $\ope_{WW}$, $\ope_{BB}$, $\ope_W$, and $\ope_B$ requires
a kinematic variable with large flow through the production vertex.
A key candidate is the transverse momentum distribution in the
hard process~\cite{1409.6212} of the cut--based experimental
analysis, which serves as a check of the measured rate, while the
measurement itself used in Sec.~\ref{sec:coup} relies on a multi-variate
analysis. The experimental search requires two jets with medium and
tight $b$-tags and defines three categories with 0, 1 and 2 leptons that receive
contributions from $HW$ and $HZ$ productions, therefore being sensitive to
different contributions of the higher-dimensional operators.  We show
these three distributions in Fig.~\ref{fig:distributions} with a selection
of dimension-6 anomalous contributions. While the
background rapidly decreases at large transverse momenta, the
main effect of the dimension-6 operators is conversely an enhancement at high
momenta, being most prominent in the last bin of the distributions.

To combine the different lepton multiplicities we use our
\textsc{FeynRules}~\cite{Christensen:2008py} implementation of
the dimension-6 operators to generate the distributions with
\textsc{MadGraph5}~\cite{Alwall:2011uj},
\textsc{Pythia}~\cite{Sjostrand:2006za}, and \textsc{PGS4}~\cite{pgs},
the latter checked with \textsc{Delphes}~\cite{delphes}. We use the SM
Higgs expectations to calibrate our setup to the distributions and
rates shown in Ref.~\cite{1409.6212}. We parametrize the kinematic
distributions as a function of the dimension-6 operators and we use
the SM background expectations and the number of measured events per
bin in Ref.~\cite{1409.6212}.

As mentioned above, the $p_T^V$ distributions shown in
Fig.~\ref{fig:distributions} correspond to a cut--based ATLAS
analysis~\cite{1409.6212}. When adding this information to the global
Run~I analysis consistently we need to be careful: first, we cannot
use the same information twice. This means we could remove the
corresponding total rates from the analysis and instead include the
binned distributions. However, the cut--based analysis is weaker than
the multi-variate analysis and they do not give the same measured
central values. This would render any estimate of the additional power
of the kinematic information impossible. Instead, we keep the
multi-variate rate information and add the kinematic information
through a set of asymmetries based on the bin content of
Fig.~\ref{fig:distributions},
\begin{equation}
A_i=\frac{\text{bin}_{i+1}- \text{bin}_i}{ \text{bin}_{i+1}+ \text{bin}_i }\; ,
\label{eq:asym}
\end{equation}
which for each leptonic channel defines three or four additional measurements.\bigskip

Our second test case is the azimuthal angle correlation in weak boson
fusion production with $H \to \gamma \gamma$~\cite{delta_phi}. Because
the measurement of $\Delta \phi_{jj}$ does not require the
reconstruction of any reference frame, its uncertainties are
reduced. Unfortunately, the corresponding distributions are not shown
in the most prominent weak boson fusion channels with decays
$H\rightarrow W^+W^-$ and $H\rightarrow \tau^+ \tau^-$. An unfolded
distribution is in contrast available for the decay $H\rightarrow
\gamma\gamma$~\cite{1407.4222}. However, due to the lack of cuts on
$m_{jj}$ and $\Delta\eta_{jj}$, the weak boson fusion mode accounts
for less than $35\%$ of all signal events, diluting consequently
the promising power of the $\Delta \phi_{jj}$ variable in this production
mechanism.  

In the present absence of a better alternative we include the above
channels in our \textsc{SFitter} analysis.  To simulate SM Higgs
production in weak boson fusion and the $VH$ channel we rely on
the same selection of tools we have used for the $p_T^V$ implementation.
To validate our calculations, we compare our SM simulations to the ATLAS result, most notably the
plots available in HEPDATA~\cite{hepdata}. Once our setup is tested
we simulate the effect of dimension-6 operators on the weak boson
fusion and $VH$ distributions. The main contribution from Higgs
production in gluon fusion is only affected by $\ope_{GG}$,
$\ope_{\Phi,2}$, and $\ope_{b,t}$, none of which change the Lorentz
structure of the hard process. We can then use the central estimate by
ATLAS, properly reweighted by the introduced shift of the relevant
operators. We use a similar reweighting to simulate the effects of the
effective operators in the di-photon decay, as none of the operators
generate a non-SM Lorentz structure for this vertex either. In the
lower right panel of Fig.~\ref{fig:distributions} we show the
$\Delta\phi_{jj}$ distribution with a selection of dimension-6 contributions.
It turns out that all $\ope_{WW}$, $\ope_{BB}$, $\ope_W$, and $\ope_B$
peak at 0 or at $\pi$~\cite{delta_phi}.

To add $\Delta\phi_{jj}$ to the global analysis we again keep the
measured total rates used in the previous analyses
and construct three additional asymmetries~\cite{delta_phi},
\begin{alignat}{5}
A_1 =&\frac{\sigma(\Delta\phi_{jj}<\frac{\pi}{3})+\sigma(\Delta\phi_{jj}>\frac{2\pi}{3})-\sigma(\frac{\pi}{3}<\Delta\phi_{jj}<\frac{2\pi}{3})}
{\sigma(\Delta\phi_{jj}<\frac{\pi}{3})+\sigma(\Delta\phi_{jj}>\frac{2\pi}{3})+\sigma(\frac{\pi}{3}<\Delta\phi_{jj}<\frac{2\pi}{3})}\;\; ,\notag \\
A_2 =&\frac{\sigma(\Delta\phi_{jj}>\frac{2\pi}{3})-\sigma(\Delta\phi_{jj}<\frac{\pi}{3})}
{\sigma(\Delta\phi_{jj}>\frac{2\pi}{3})+\sigma(\Delta\phi_{jj}<\frac{\pi}{3})}\;\; ,
\notag \\
A_3 =&\frac{\sigma(\Delta\phi_{jj}>\frac{5\pi}{6})-\sigma(\frac{2\pi}{3}<\Delta\phi_{jj}<\frac{5\pi}{6})}
{\sigma(\Delta\phi_{jj}>\frac{5\pi}{6})+\sigma(\frac{2\pi}{3}<\Delta\phi_{jj}<\frac{5\pi}{6})} \;.
\label{eq:asym2}
\end{alignat}
The first asymmetry is tailored to discriminate different production
modes and $CP$ structures~\cite{delta_phi,1407.4222}.  The second
asymmetry enhances the sensitivity to $\ope_{WW}$, $\ope_{BB}$,
$\ope_W$, and $\ope_B$ in weak boson fusion, all of which generate
non--zero values for $A_2$. Finally, the third asymmetry is orthogonal
to the other two, to not exclude any information.

\subsection{Full dimension-6 analysis}
\label{sec:eff_full}

\begin{figure}[t!]
  \includegraphics[width=0.29\textwidth]{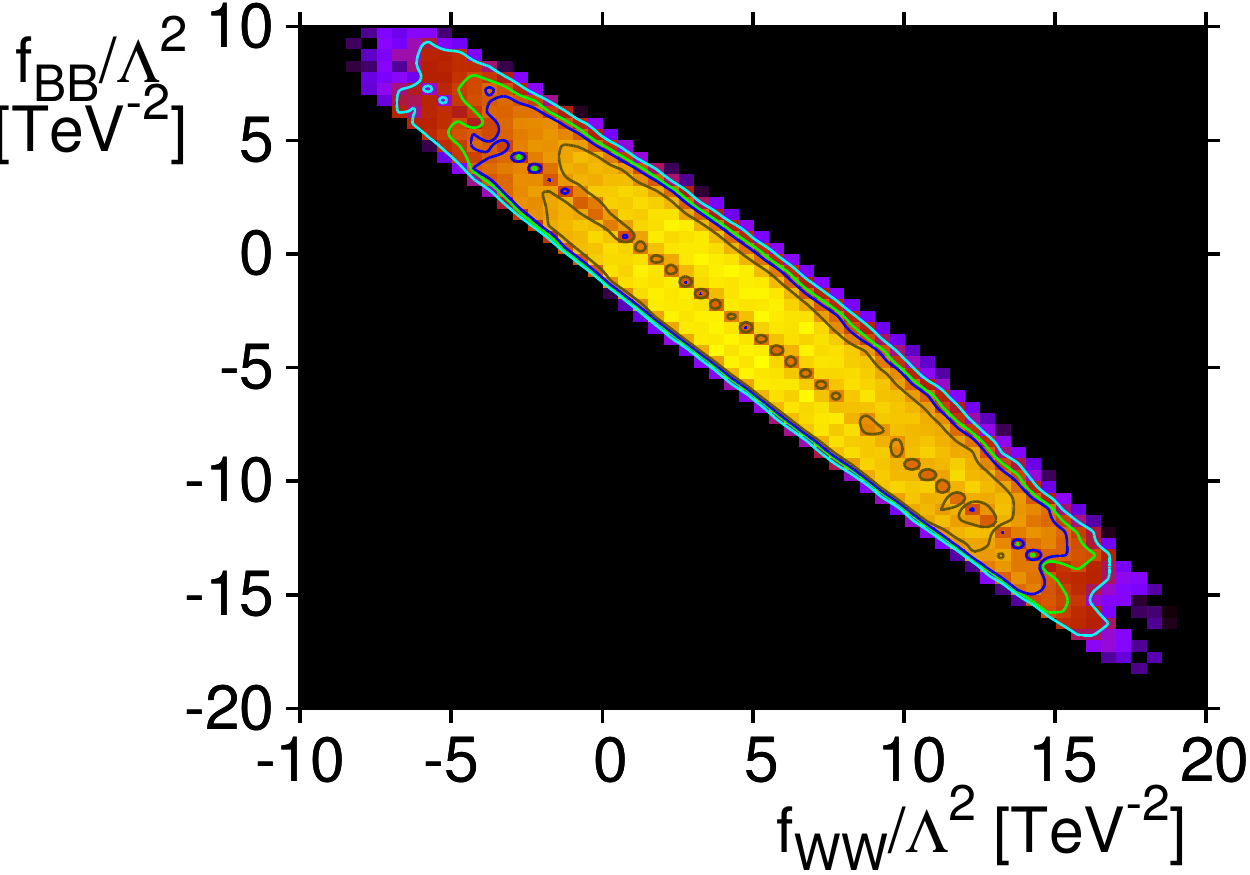} 
  \hspace*{1ex}
  \includegraphics[width=0.29\textwidth]{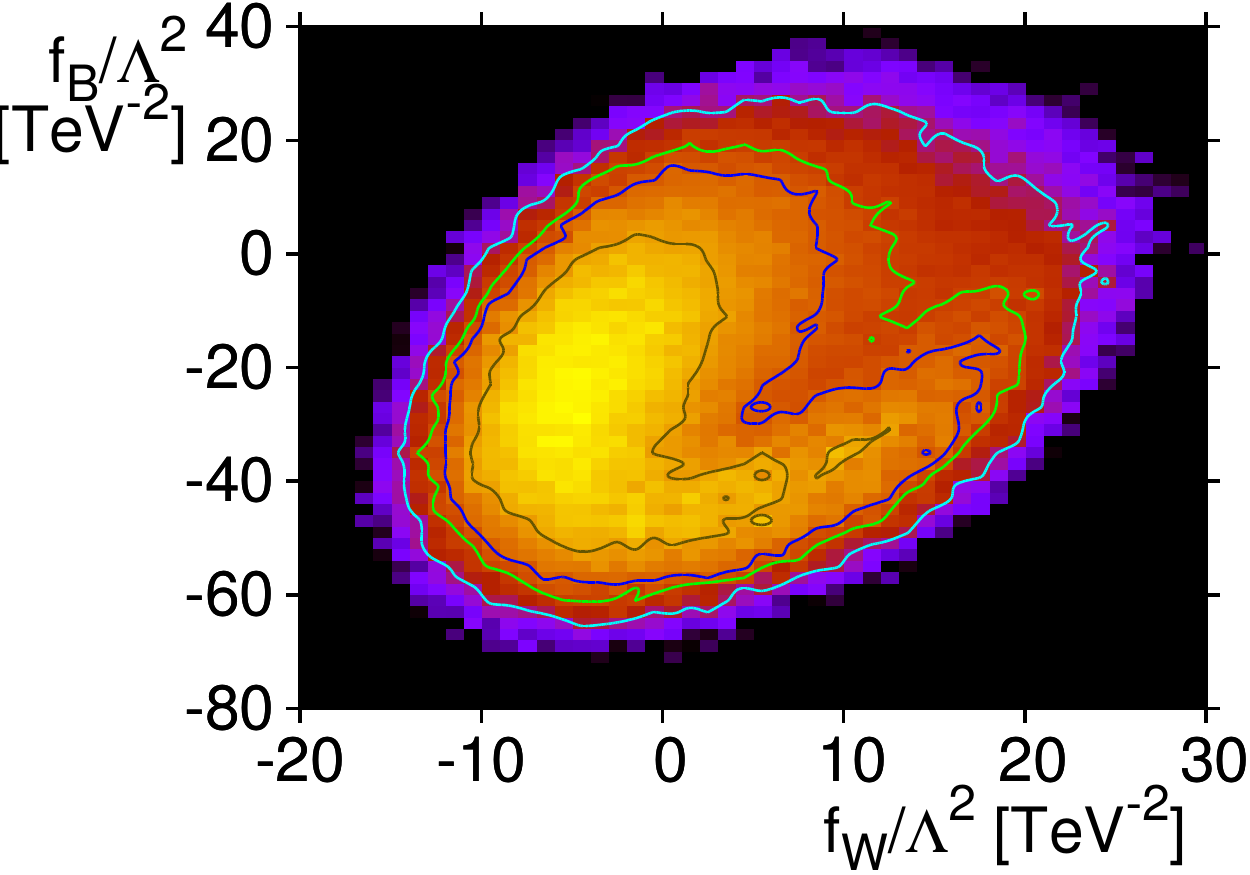} 
  \hspace*{1ex}
  \includegraphics[width=0.29\textwidth]{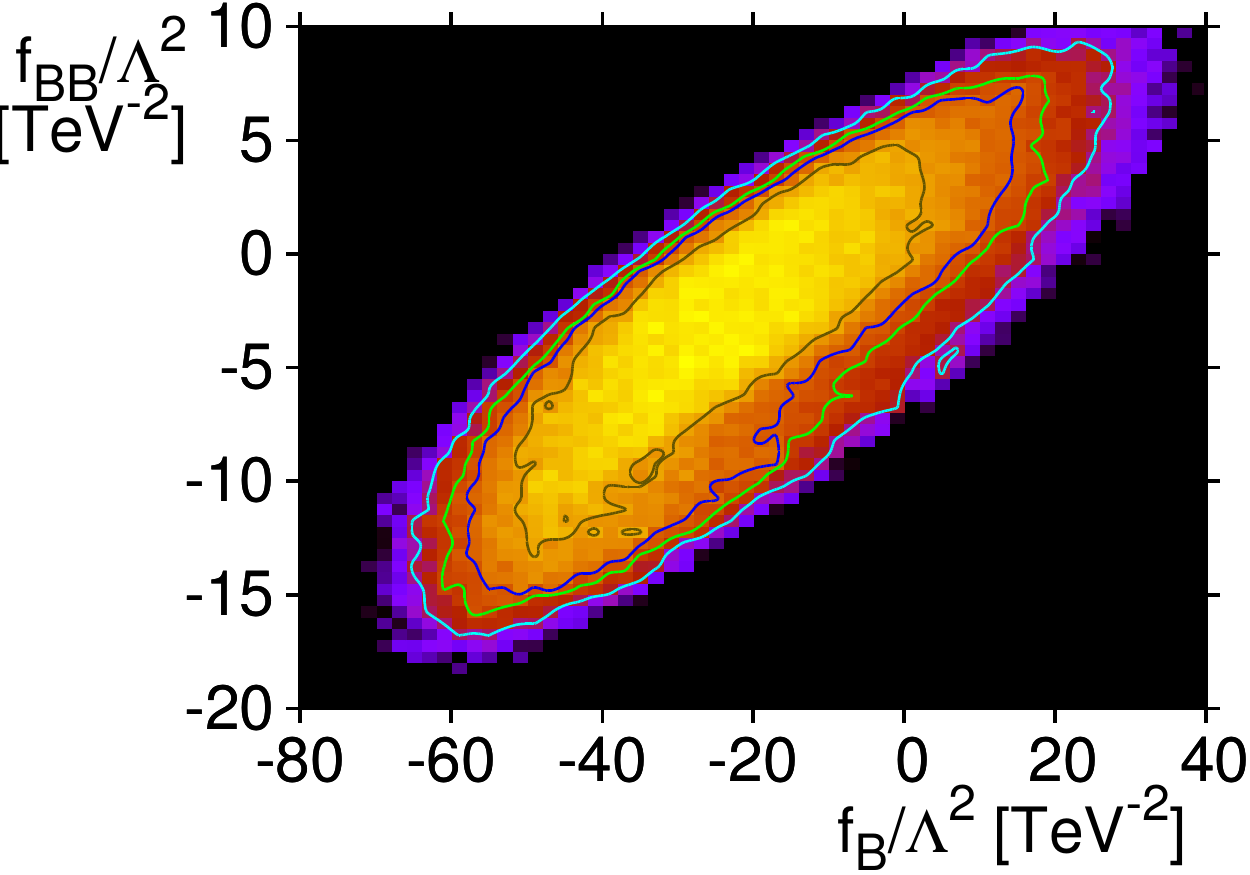}
  \hspace*{1ex}
  \raisebox{3pt}{\includegraphics[width=0.0545\textwidth]{figs/colorbox}}\\[1ex]
  \includegraphics[width=0.29\textwidth]{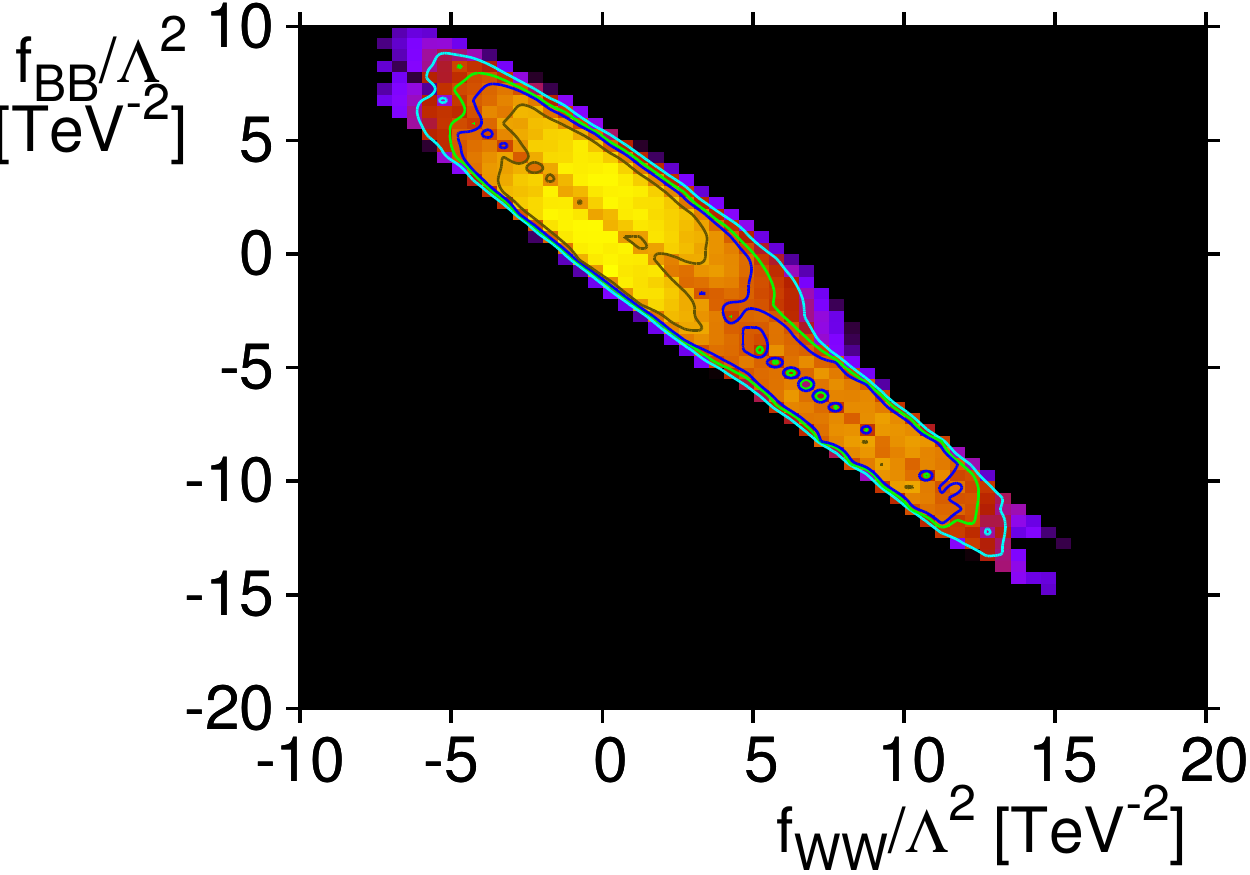} 
  \hspace*{1ex}
  \includegraphics[width=0.29\textwidth]{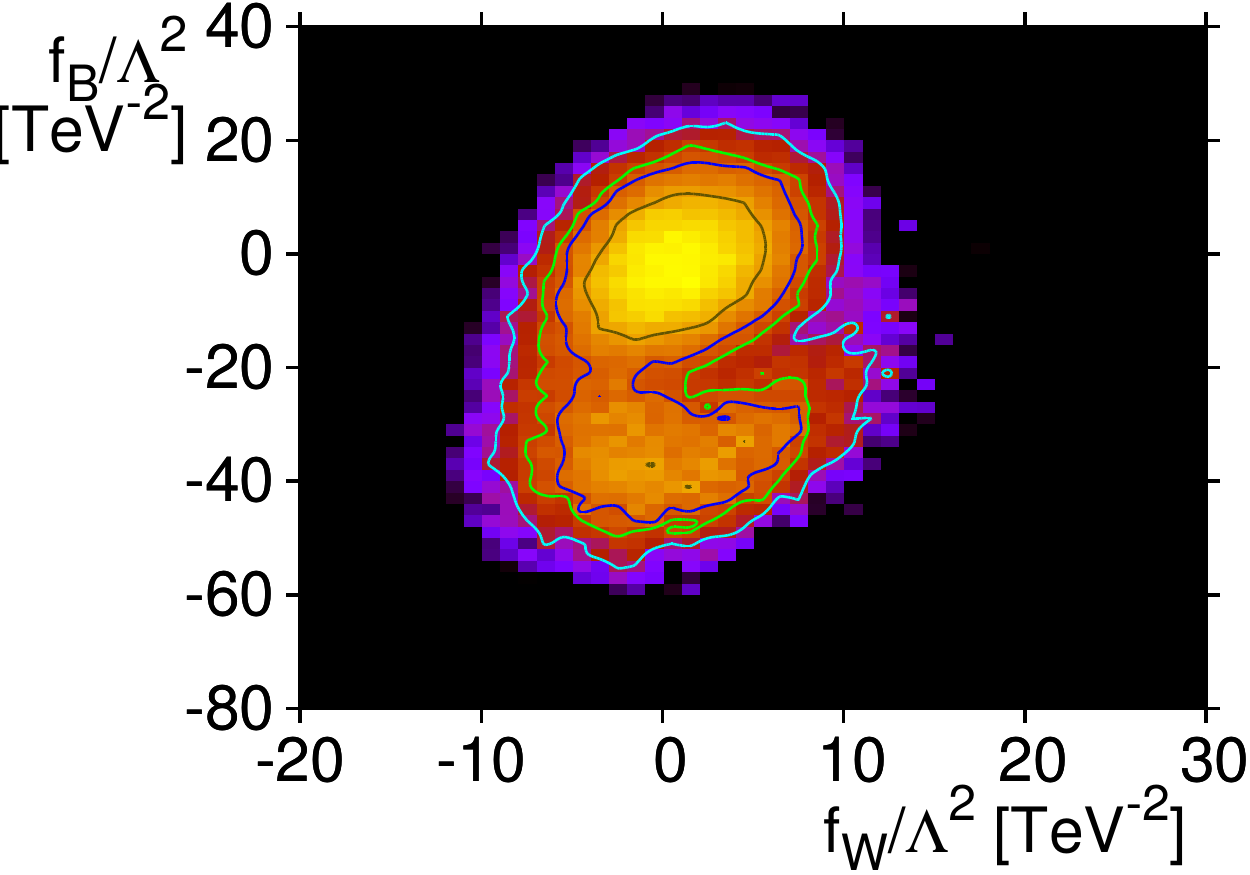} 
  \hspace*{1ex}
  \includegraphics[width=0.29\textwidth]{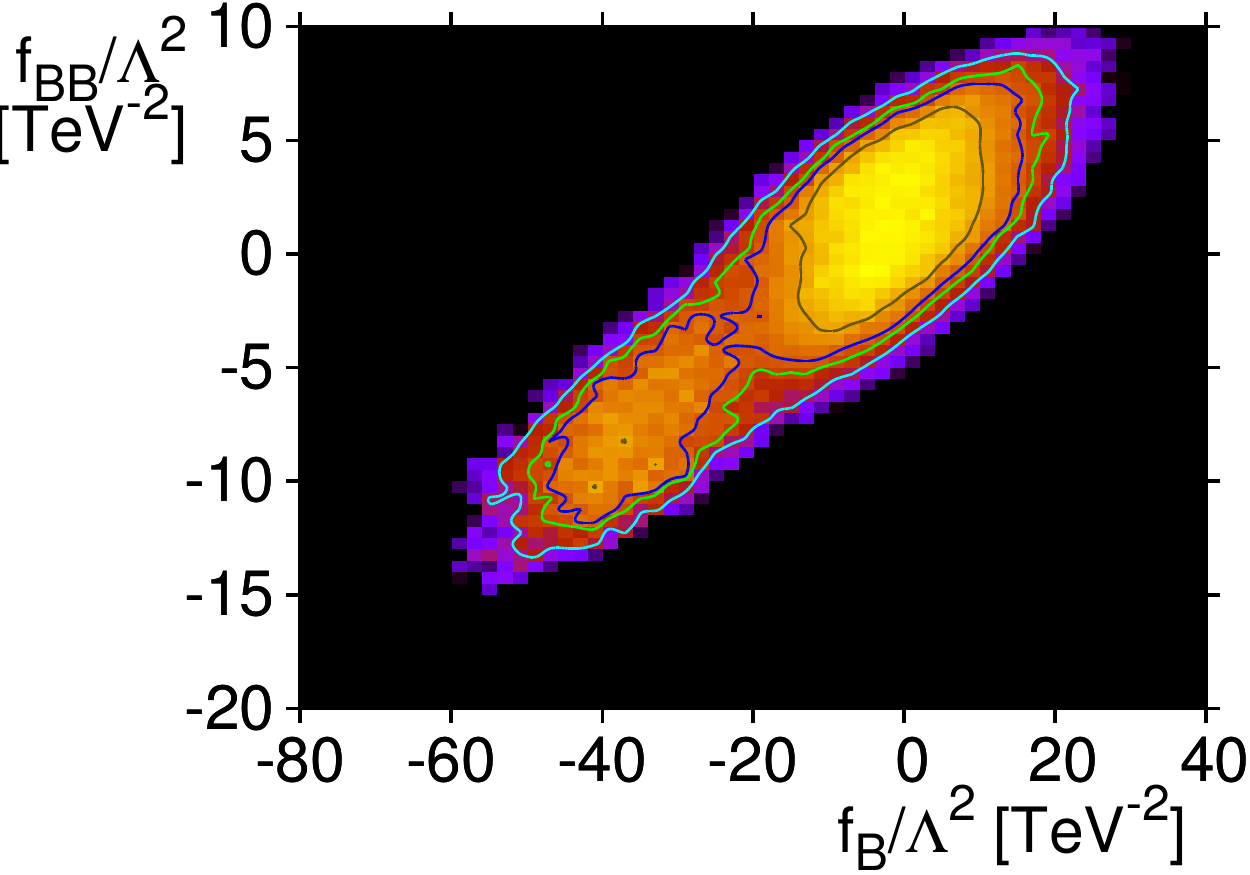}
  \phantom{\hspace*{1ex}
  \includegraphics[width=0.0545\textwidth]{figs/colorbox}}\\[1ex]
  \includegraphics[width=0.29\textwidth]{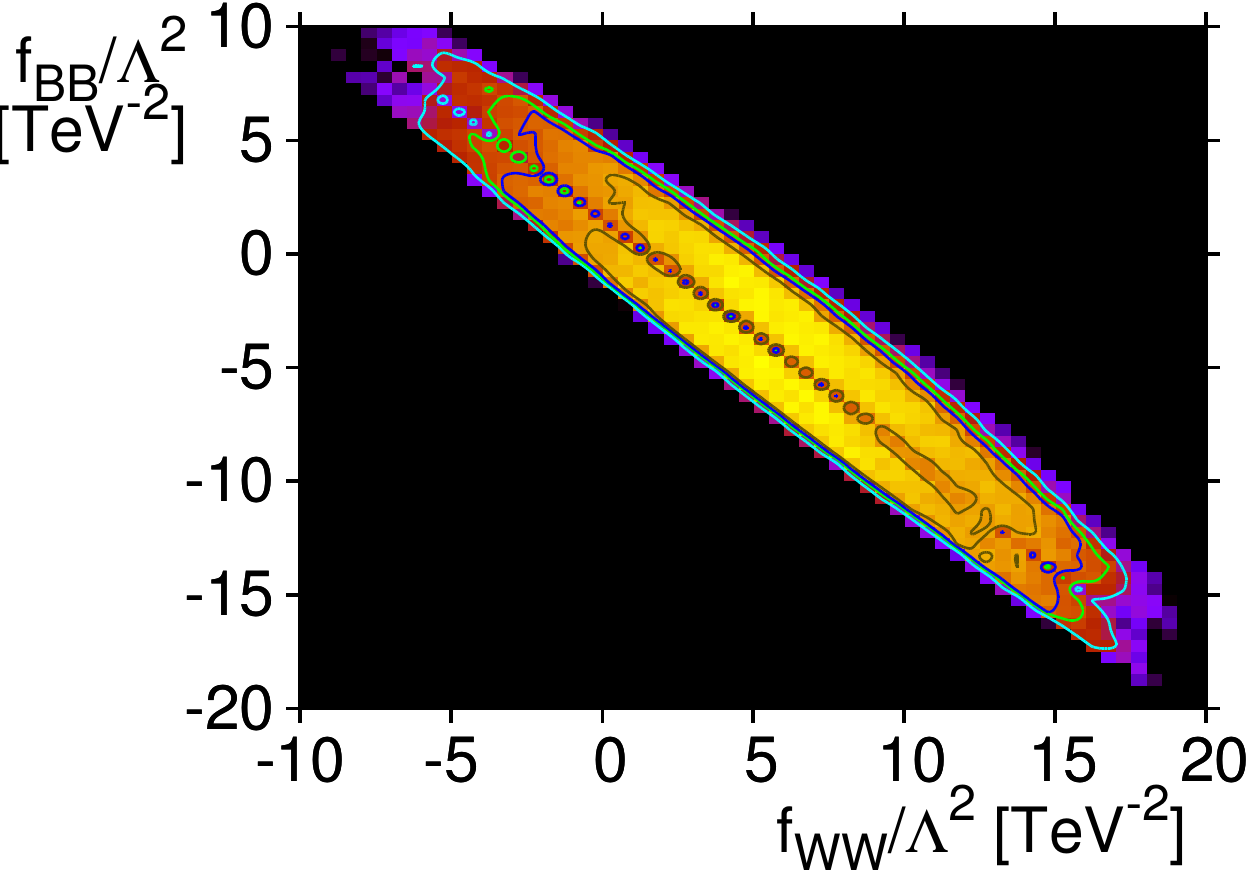}
  \hspace*{1ex}
  \includegraphics[width=0.29\textwidth]{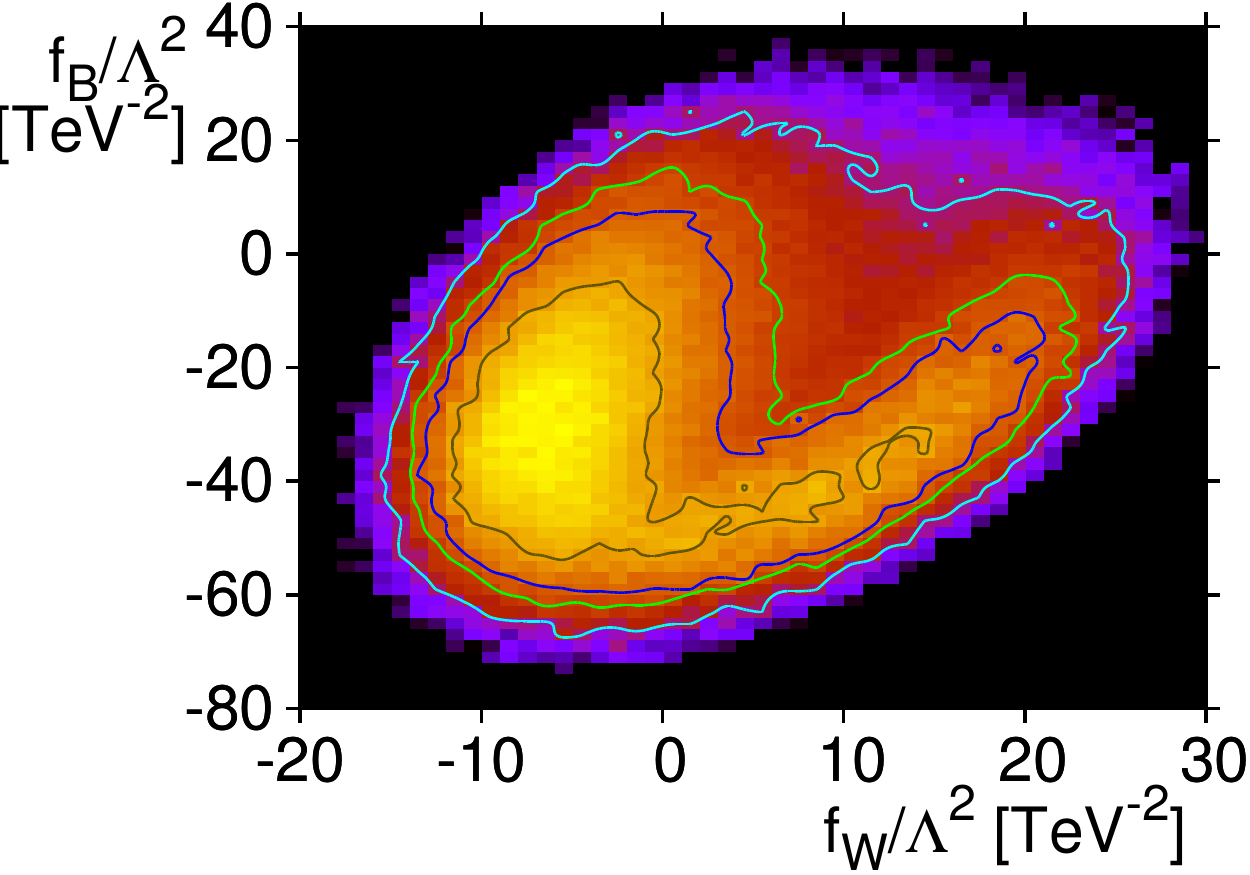}
  \hspace*{1ex}
  \includegraphics[width=0.29\textwidth]{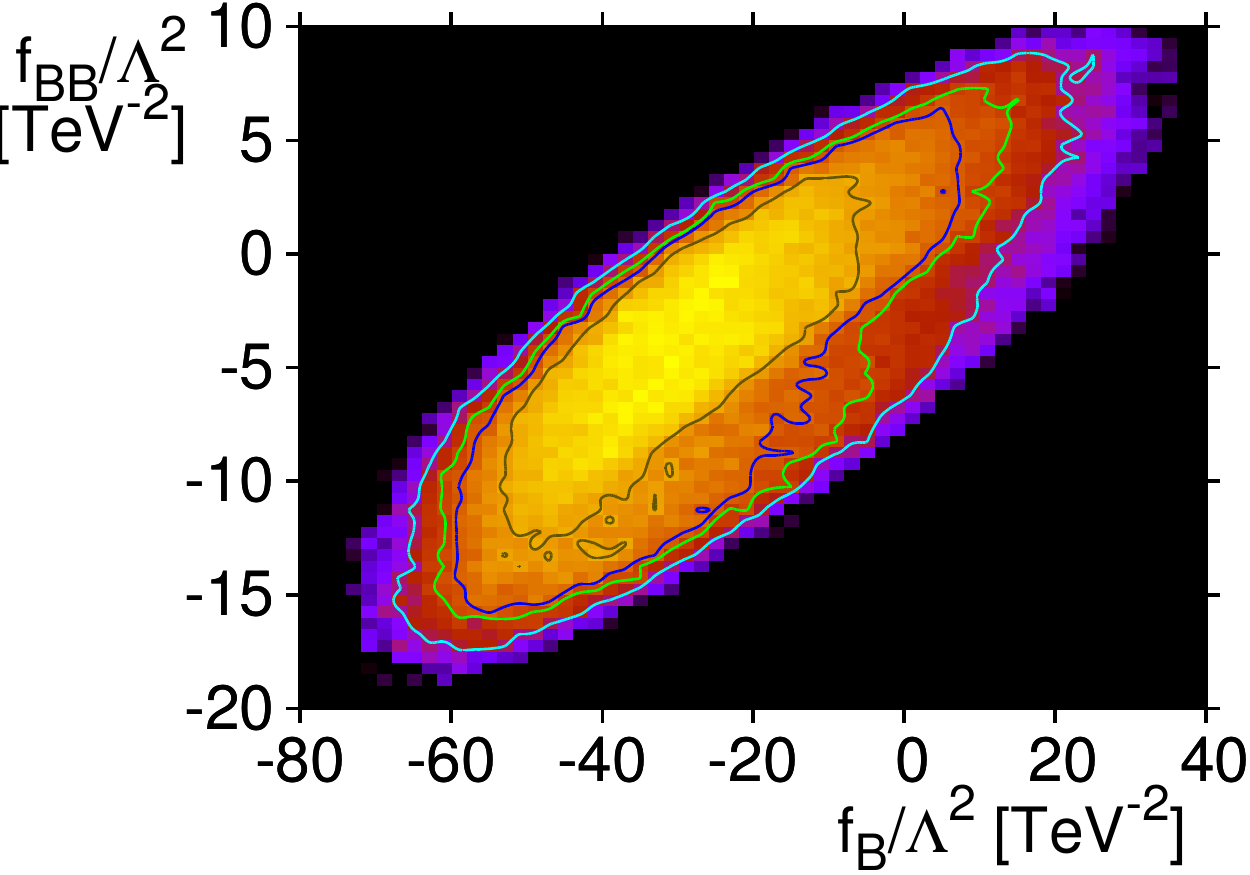}
  \phantom{\hspace*{1ex}
  \includegraphics[width=0.0545\textwidth]{figs/colorbox}}
  \caption{Correlations between different coefficients $f_x/\Lambda^2$
    (measured in $\itevx$) after including kinematic
    distributions. In the top row we add the $\Delta \phi_{jj}$
    distribution; in the second row we also include $p_T^V$ from $VH$
    production; in the bottom row we then remove the highest bin
    associated with large momentum flow through the dimension-6
    vertex. The 1-dimensional profile likelihoods of the second row
    correspond to the results shown as the blue bars in
    Fig.~\ref{fig:dim6kin}.}
\label{fig:dim6_corrkin}
\end{figure}

The final step in our higher-dimensional \textsc{SFitter} analysis is
to add these test distributions to the coupling information used
in Sec.~\ref{sec:eff_fit}. As we have discussed, we use the experimental
information shown in Fig.~\ref{fig:distributions}~\cite{1409.6212,1407.4222}
in terms of the asymmetries defined in Eqs.\eqref{eq:asym} and~\eqref{eq:asym2}. 

The main technical problem of the purely rate-based analysis of
dimension-6 operators are the correlations which make it hard to
extract 1-dimensional profile likelihoods and error bars for the
individual $f_x/\Lambda^2$, as illustrated in
Fig.~\ref{fig:dim6_corr}. We show the effect of kinematic distributions on
some critical 2-dimensional profile likelihoods in
Fig.~\ref{fig:dim6_corrkin}. In the top row we show the results after
including the $\Delta \phi_{jj}$ distribution only. Compared to the
Fig.~\ref{fig:dim6_corr} we see very small improvement, except for
a slight reduction of the secondary structure in the $\ope_{BB}$ vs
$\ope_B$ and $\ope_{W}$ vs $\ope_B$ correlations. However, this
reduced impact should not be taken as a statement about the
distinguishing power of the $\Delta \phi_{jj}$ distribution; it is
really linked to the lack of publicly available information on this
distribution, as discussed above.

In the second row of Fig.~\ref{fig:dim6_corrkin} we show the impact of
also adding the full $p_T^V$ information in $VH$ production. It
significantly improves the situation with secondary solutions, largely
removing the correlated structure for example in $\ope_W$ vs $\ope_B$.
There still exists a weak secondary minimum for example in the
$\ope_{BB}$ vs $\ope_B$ correlation, but because of its relative
weakness it will allow us to derive a more straightforward
1-dimensional profile likelihood and an associated 68\% CL error bar
for example on $f_{BB}/\Lambda^2$. The reduction of the allowed space
is notorious in the three corresponding panels in
Fig.~\ref{fig:dim6_corrkin}.  Actually, the left panel shows that
after including the $p_T^V$ and $\Delta \phi_{jj}$ distributions the fit becomes more
sensitive to $\ope_{WW}$ and $\ope_{BB}$ through their individual contributions
to the $HVV$ couplings mediating associated $VH$ production and weak
boson fusion.\bigskip

Finally, we need to check the consistency of the effective theory
approach~\cite{Biekoetter:2014jwa}. Based on Run~I data our analysis
typically probes $|f_x/\Lambda^2| \sim 10/\tev^2$. A hypothetical setting
of $f_x$ to unity would correspond to new physics scales around $\Lambda \sim
300~\gev$. According to Fig.~\ref{fig:distributions} the highest
momentum bin of the $p_T^V$ distribution starts from $p_T^V =
200~\gev$ and includes all events above this value. A conservative
approach would be to exclude this last bin, and thus the last asymmetries
defined in Eq.\eqref{eq:asym}, from the kinematic analysis. In the
bottom row of Fig.~\ref{fig:dim6_corrkin} we show the corresponding
2-dimensional correlations from this analysis. A comparison to the first row shows that
almost the entire additional information of the $p_T^V$ distribution is
encoded in the last bin. In the remainder of the discussion we will
not follow this conservative approach, so it should be noted that the
full \textsc{SFitter} analysis of the higher-dimensional operators has
to be taken with a grain of salt. On the other hand, the analysis
including kinematic distributions is mostly meant to be a proof of
principle, and the consistency of the Higgs effective theory will
clearly improve with Run~II data.\bigskip

\begin{figure}[h!]
  \centering
  \includegraphics[width=0.65\textwidth]{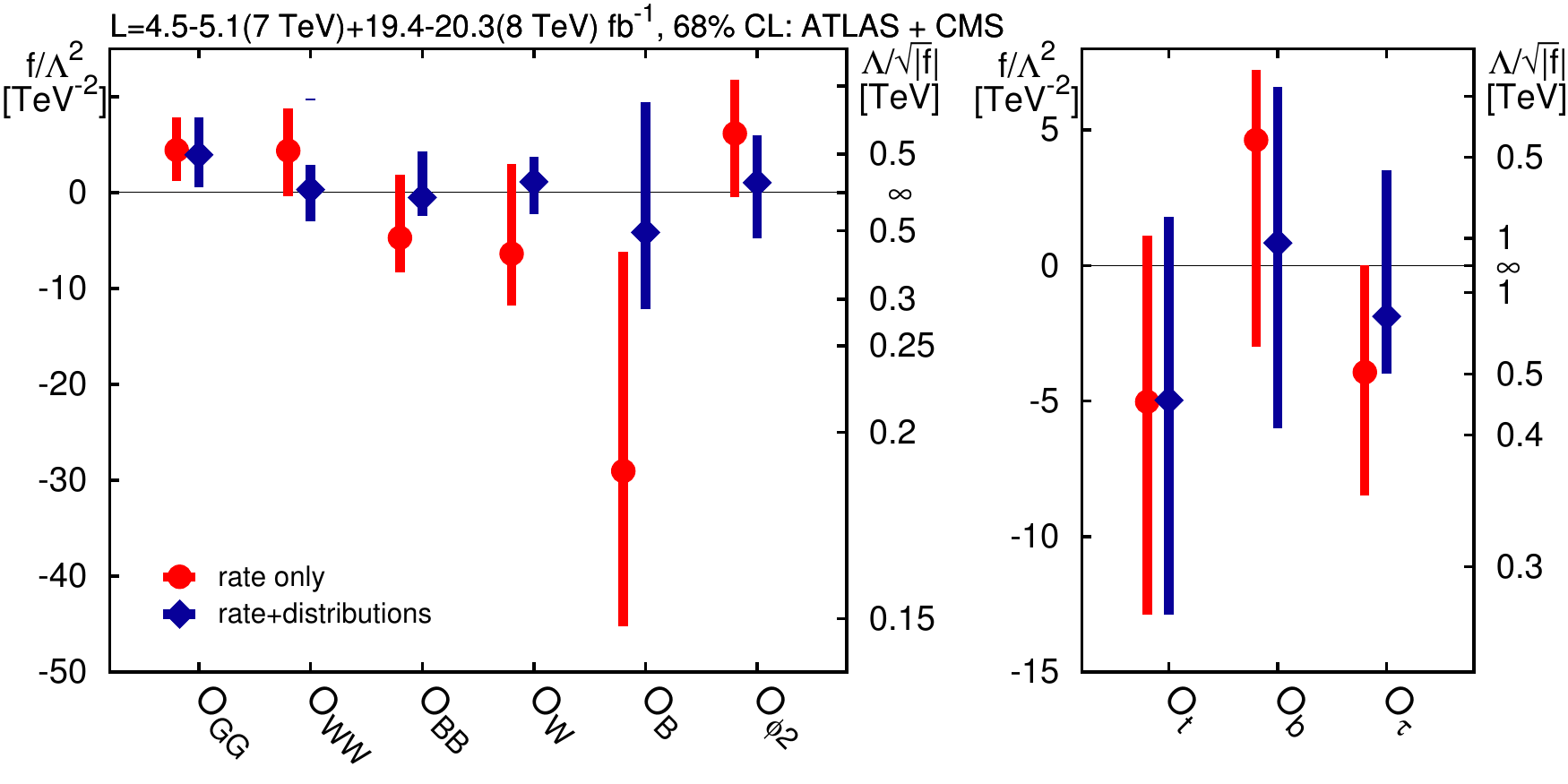}
  \caption{68\% CL error bars on the Wilson coefficients $f_x/\Lambda^2$ for the
    dimension-6 operators defined in Eq.\eqref{eq:ourleff} (left
    panel) and Eq.\eqref{eq:ourleff2} (right panel). In addition to
    total rate information we also include kinematic distributions and
    only show 68\% CL contours. For the Yukawa couplings as well as
    for $\ope_{GG}$ we limit ourselves to the SM-like solution for
  this representation.}
\label{fig:dim6kin}
\end{figure}

With this in mind we show the best fit points and the corresponding
1-dimensional 68\% CL error bars including kinematic distributions in
Fig.~\ref{fig:dim6kin}. In contrast to Fig.~\ref{fig:dim6} we do not
show secondary solutions for the signs of the Yukawa-like couplings and for
$\ope_{GG}$. We also limit ourselves to 68\% CL contours. We see that
$\ope_B$ and $\ope_W$ are the operators most affected by the addition of kinematic
distributions, closely followed by $\ope_{WW}$ and
$\ope_{BB}$. Typical energy scales probed by Run~I data are 300~GeV to
500~GeV if order one Wilson coefficients are assumed, with less significant
constraints in the fermion sector. All
coefficients are in agreement with zero, and the one to two
sigma deviations are hard to map onto individual
measurements. Including all available kinematic information visibly
stabilizes the constraint on $f_B$ and moves every single best-fit
point closer to the Standard Model prediction.

\section{Future: off-shell measurements}
\label{sec:off-shell}

ATLAS and CMS recently published a study on the contribution of Higgs
exchange to $ZZ$ production at invariant masses well above the Higgs
pole $m_{ZZ} \sim
m_H$~\cite{off_shell_th,off_shell_atlas,off_shell_cms}. Given the
small Higgs width, such a measurement would normally only show a very
moderate dependence on the Higgs mass. However, the kinematic
structure of this particular channel turns it into a sensitive
measurement. Approximately $\mathcal{O}(15\%)$ of the rate mediated by
the $s$-channel Higgs exchange lies in the off-shell regime,
$m_{4\ell}>130~\gev$. In addition the leading effect arises from the signal
interference with the continuum background.  Some representative
Feynman diagrams to this process are\medskip

\begin{center} \begin{fmffile}{feyn}
\begin{fmfgraph*}(100,60)
\fmfset{arrow_len}{2mm}
\fmfleft{f1,f2}
\fmfright{Z1,Z2}
\fmf{fermion,tension=0.8,width=0.5}{f2,v1,v2,f1}
\fmf{photon,width=0.5}{v1,Z2}
\fmf{photon,width=0.5}{v2,Z1}
\end{fmfgraph*}
\hspace*{3mm}
\begin{fmfgraph*}(100,60)
\fmfset{arrow_len}{2mm}
\fmfleft{g1,g2}
\fmfright{Z1,Z2}
\fmf{gluon,width=0.5,label.side=left}{g1,v1}
\fmf{gluon,width=0.5}{g2,v2}
\fmf{fermion,tension=0.8,label.side=left,width=0.5}{v1,v2}
\fmf{fermion,tension=1.2,width=0.5}{v2,v3}
\fmf{fermion,tension=1.2,width=0.5}{v4,v1}
\fmf{fermion,tension=0.8,width=0.5}{v3,v4}
\fmf{photon,width=0.5}{v3,Z2}
\fmf{photon,width=0.5}{v4,Z1}
\end{fmfgraph*}
\hspace*{4mm}
\begin{fmfgraph*}(100,60)
\fmfset{arrow_len}{2mm}
\fmfleft{g1,g2}
\fmfright{Z1,Z2}
\fmf{gluon,width=0.5,label.side=left}{g1,v1}
\fmf{gluon,width=0.5}{g2,v2}
\fmf{fermion,tension=.05,label.side=left,width=0.5}{v1,v2}
\fmf{fermion,tension=.8,width=0.5}{v2,v3}
\fmf{fermion,tension=.8,width=0.5}{v3,v1}
\fmf{dashes,tension=1.5,width=0.5,label.side=right}{v3,v4}
\fmf{photon,tension=0.8,width=0.5}{v4,Z1}
\fmf{photon,tension=0.8,width=0.5}{v4,Z2}
\fmfv{decor.shape=circle,decor.filled=full,decor.size=3}{v3}
\fmfv{label.side=right,label.dist=10,label=$1+\Delta_Z$}{v4}
\fmfv{decor.shape=circle,decor.filled=full,decor.size=3}{v4}
\fmffreeze
\fmf{phantom,tension=.8,width=0.5,label=$1+\Delta_t$,label.side=left,label.dist=-9}{v4,v2}
\end{fmfgraph*}
\hspace*{9mm}
\begin{fmfgraph*}(100,60)
\fmfset{arrow_len}{2mm}
\fmfleft{g1,g2}
\fmfright{Z1,Z2}
\fmf{gluon,width=0.5,label.side=left}{g1,v1}
\fmf{gluon,width=0.5}{g2,v1}
\fmfv{label.side=right,label.dist=10,label=$\Delta_g$}{v1}
\fmfv{decor.shape=circle,decor.filled=full,decor.size=3}{v1}
\fmf{dashes,tension=1.5,width=0.5}{v1,v4}
\fmfv{label.side=right,label.dist=10,label=$1+\Delta_Z$}{v4}
\fmfv{decor.shape=circle,decor.filled=full,decor.size=3}{v4}
\fmf{photon,tension=0.8,width=0.5}{v4,Z1}
\fmf{photon,tension=0.8,width=0.5}{v4,Z2}
\end{fmfgraph*}
\end{fmffile} \end{center}
\bigskip

Note that $\Delta_g$ in this representation shows a non-trivial
momentum dependence, limiting the model-independent features of the
width measurement~\cite{off_shell_no}. If the Higgs propagator in the
interference is probed far above the mass shell, it behaves like
$1/s$.  On-shell and off-shell Higgs rates then scale like
\begin{equation}
\sigma^{\text{on-shell}}_{i\rightarrow H\rightarrow f} \propto \frac{g_{i}^2(m_H) \, g_{f}^2(m_H)}{\Gamma_H} 
\qqquad \text{vs} \qqquad 
\sigma^{\text{off-shell}}_{i\rightarrow H^*\rightarrow f} \propto g_{i}^2(m_{4\ell}) \, g_{f}^2(m_{4\ell})\;\; .
\label{eq:os}
\end{equation}
where $g_i$ ($g_f$) refer to the Higgs couplings involved in the production (decay) for
the present channel.
Eventually, we will remove the assumptions about the Higgs width
described in Sec.~\ref{sec:intro_setup} from the \textsc{SFitter}
setup and instead determine the total width from the combination of
off-shell and on-shell measurements.
The Lagrangian of the underlying hypothesis reads
\begin{alignat}{5}
\lag 
= \lag_\text{SM} 
&+ \Delta_W \; g m_W H \; W^\mu W_\mu
+ \Delta_Z \; \frac{g}{2 c_w} m_Z H \; Z^\mu Z_\mu
- \sum_{\tau,b,t} \Delta_f \; 
\frac{m_f}{v} H \left( \bar{f}_R f_L + \text{h.c.} \right) \notag \\
&+  \Delta_g F_G \; \frac{H}{v} \; G_{\mu\nu}G^{\mu\nu}
+  \Delta_\gamma F_A \; \frac{H}{v} \; A_{\mu\nu}A^{\mu\nu} 
+ \text{invisible decays} 
+ \text{unobservable decays} \; .
\label{eq:lag_offshell}
\end{alignat} 
The distinction between the two terms linked to decays to non-SM
states is that `invisible decays' are reconstructable using missing
transverse momentum, while `unobservable decays' are for some other
reason not observable at the LHC, for example because of overwhelming
jet backgrounds~\cite{sfitter_orig}. Not accounting for such
unobservable decays would lead to  shifts of all $\Delta_x$ as compared 
to the analysis including these decays.\bigskip

Before we allow for a fully unconstrained Higgs width through
unobservable decay channels we combine on-shell and off-shell analysis
to probe the energy dependence of the operators
involved~\cite{taming}. On the Higgs production side, the dimension-6
operators entering the off-shell measurements are listed in
Eqs.\eqref{eq:ourleff} and~\eqref{eq:ourleff2}, namely
$\mathcal{O}_{GG}$, $\mathcal{O}_{\Phi,2}$ and $\mathcal{O}_{t,b}$.
They can be described by the two parameters $\Delta_g$ and $\Delta_t$
or equivalently $g_{Hgg}$ and $g_f$ (with a marginal contribution from
the bottom loop).  The difference between the two are top mass effects
in the kinematic structure~\cite{h_jets}. In the decay the dimension-6
operators in Eq.\eqref{eq:lhvv} lead to additional operator
structures, namely $Z_{\mu\nu}Z^\mu \partial^\nu H$ and
$HZ_{\mu\nu}Z^{\mu\nu}$ in Eq.\eqref{eq:lhvv}. None of them affects
the longitudinal $Z$-polarization~\cite{brehmer}, so they lead to
similar $m_{4\ell}$ kinematics as the SM-operator $HZ_\mu Z^\mu$. We
parametrize this Higgs decay only accounting for $\Delta_Z$ or
$g_{HZZ}^{(3)}$. Properly accounting for the continuum background we
can write the gluon fusion component to the signal as
\begin{alignat}{5}
\mathcal{M}_{gg\rightarrow ZZ}
& = (1+\Delta_Z) \left [(1+\Delta_t) \mathcal{M}_t+\Delta_g \mathcal{M}_g\right] 
    +  \mathcal{M}_c \notag \\
\frac{d\sigma}{dm_{4\ell}}
&= (1+\Delta_Z) \left[ (1+\Delta_t) \frac{d\sigma_{tc}}{dm_{4\ell}} + \Delta_g \frac{d\sigma_{gc}}{dm_{4\ell}} 
                \right] \notag \\
&+ (1+\Delta_Z)^2 \left[  (1+\Delta_t)^2 \frac{d\sigma_{tt}}{dm_{4\ell}} 
                        + (1+\Delta_t)\Delta_g \frac{d\sigma_{tg}}{dm_{4\ell}} 
                        + \Delta_g^2 \frac{d\sigma_{gg}}{dm_{4\ell}}
                  \right]
&+ \frac{d\sigma_c}{dm_{4\ell}}\; .
\label{eq:m4l}
\end{alignat}
%

\begin{figure}[t]
  \centering
  \includegraphics[width=0.38\textwidth]{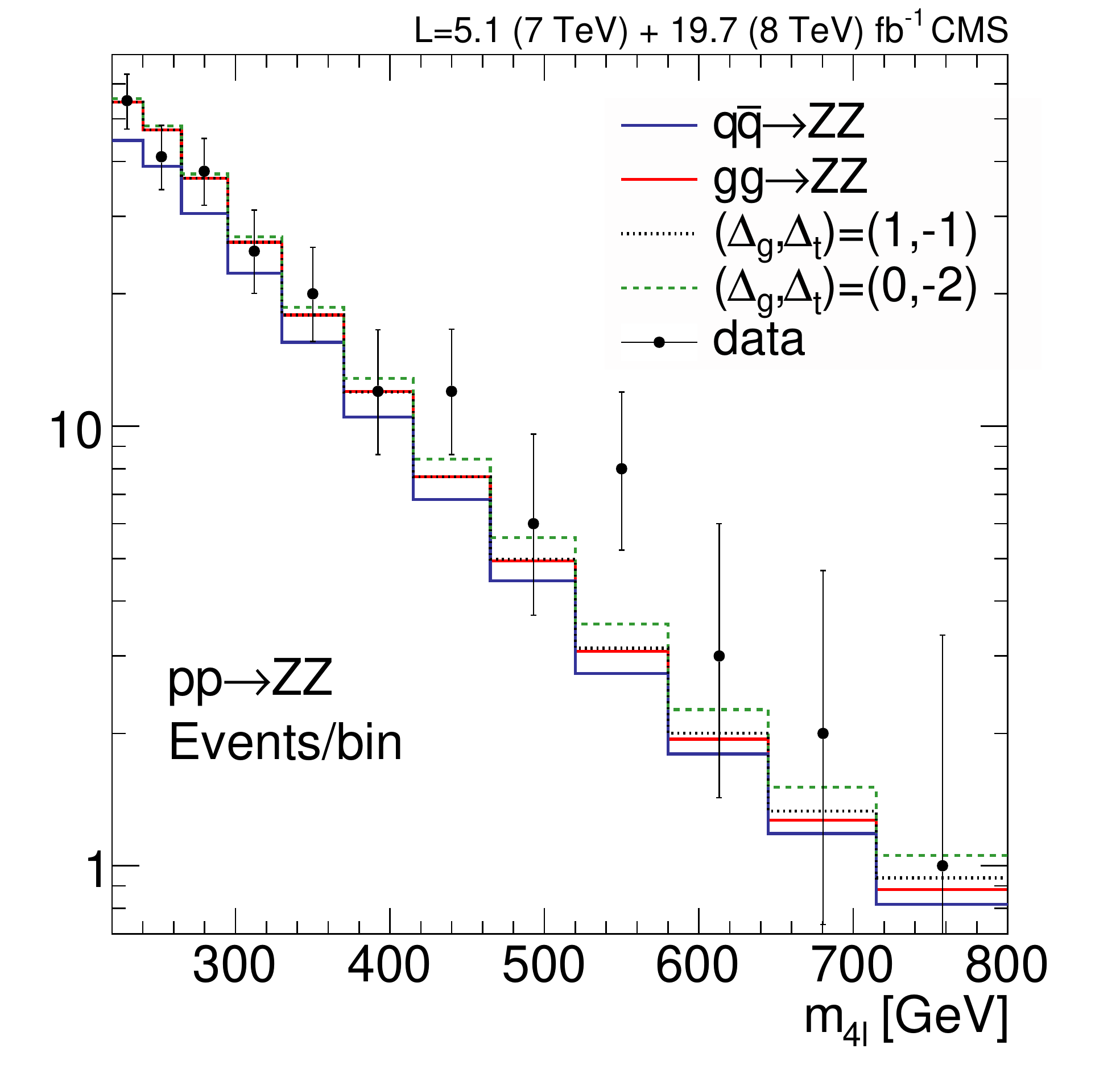}
  \hspace{0.5cm}
  \includegraphics[width=0.38\textwidth]{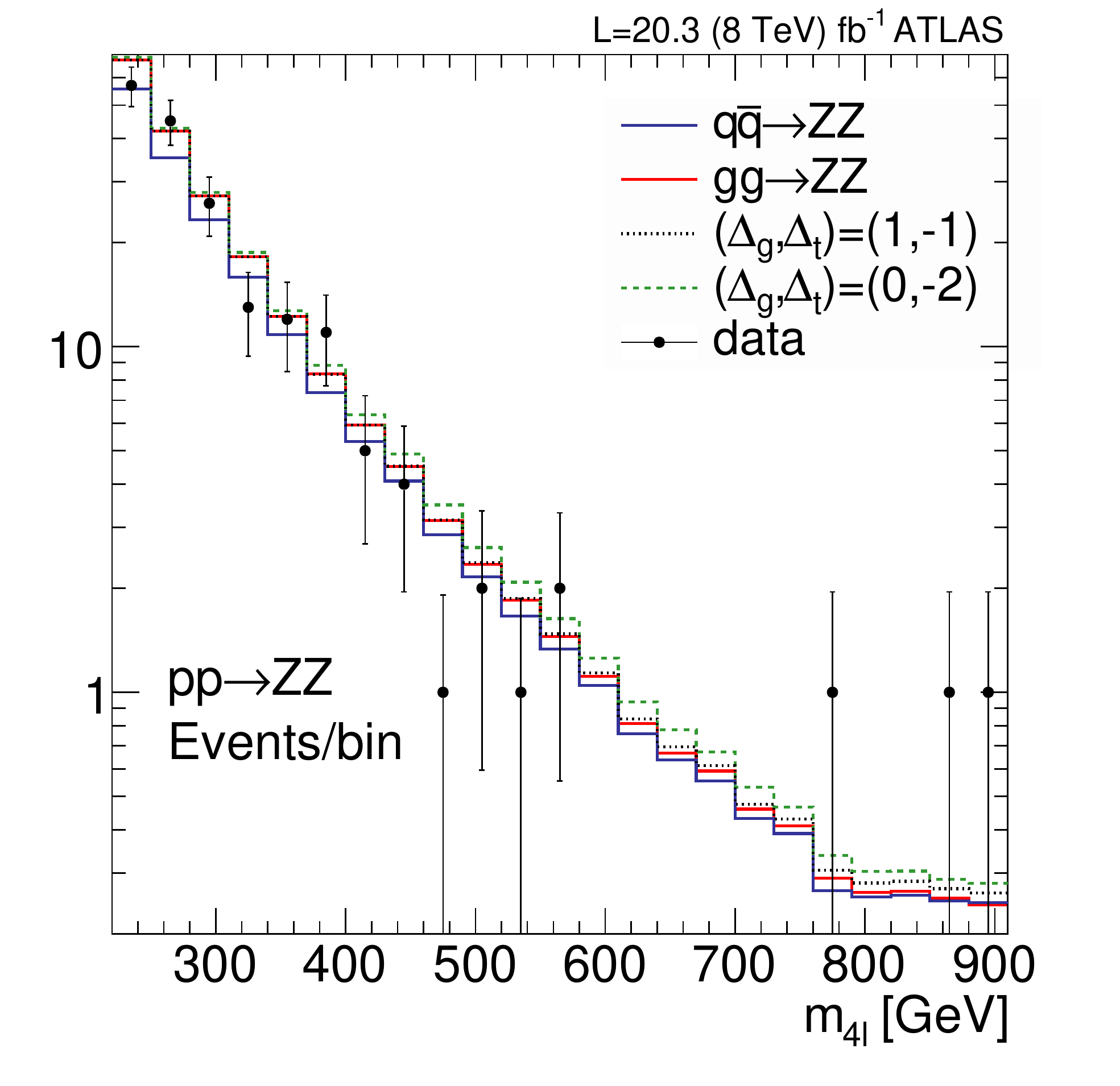}
  \caption{$m_{4\ell}$ distribution for the CMS (left) and ATLAS
    (right) analyses. The $q\bar{q}\rightarrow ZZ$ background and the
    data points are obtained from
    Refs.~\cite{off_shell_atlas,off_shell_cms}. The remaining curves
    are generated following the parametrization of Eq.\eqref{eq:m4l}.}
\label{fig:cms_off-shell}
\end{figure}

\begin{figure}[b!]
  \centering
  \includegraphics[width=0.29\textwidth]{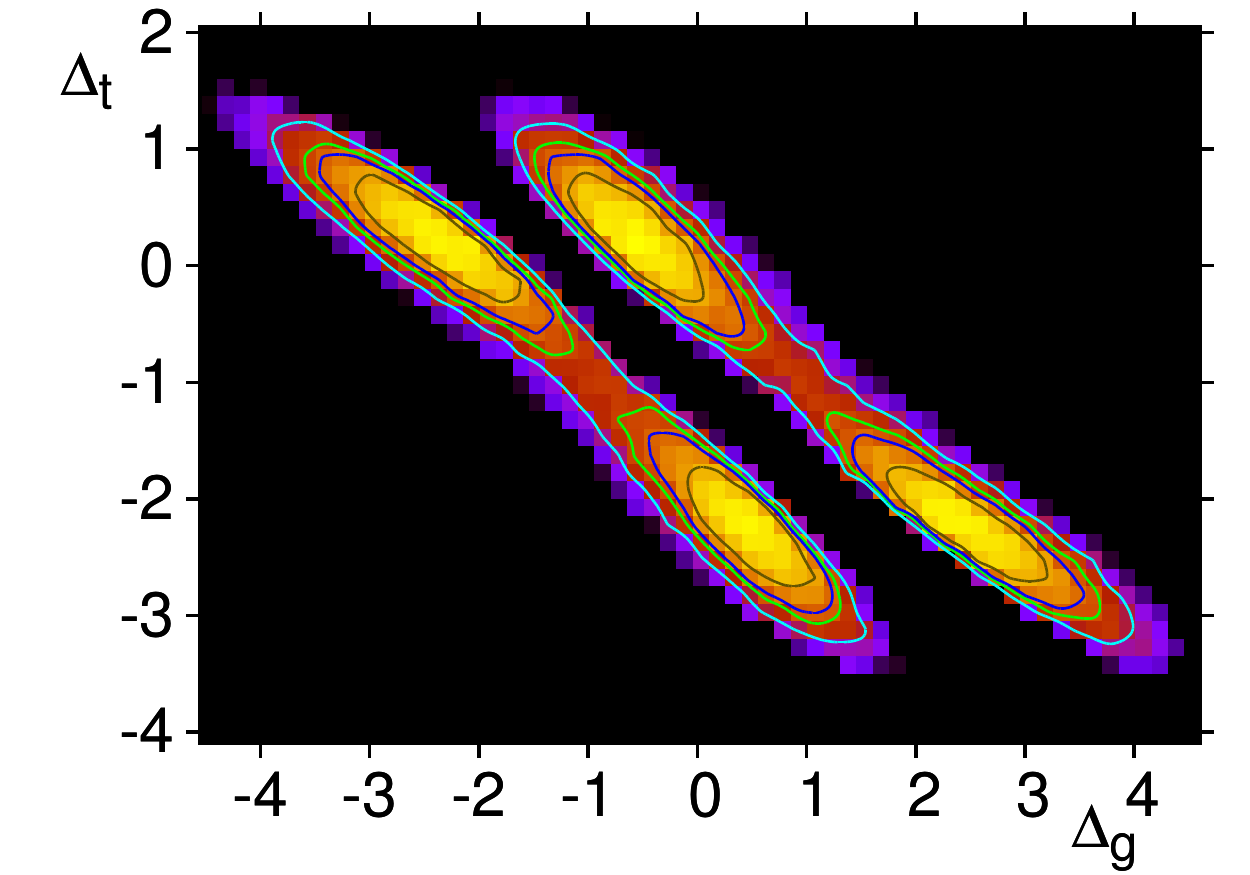}
  \hspace*{1ex}
  \includegraphics[width=0.29\textwidth]{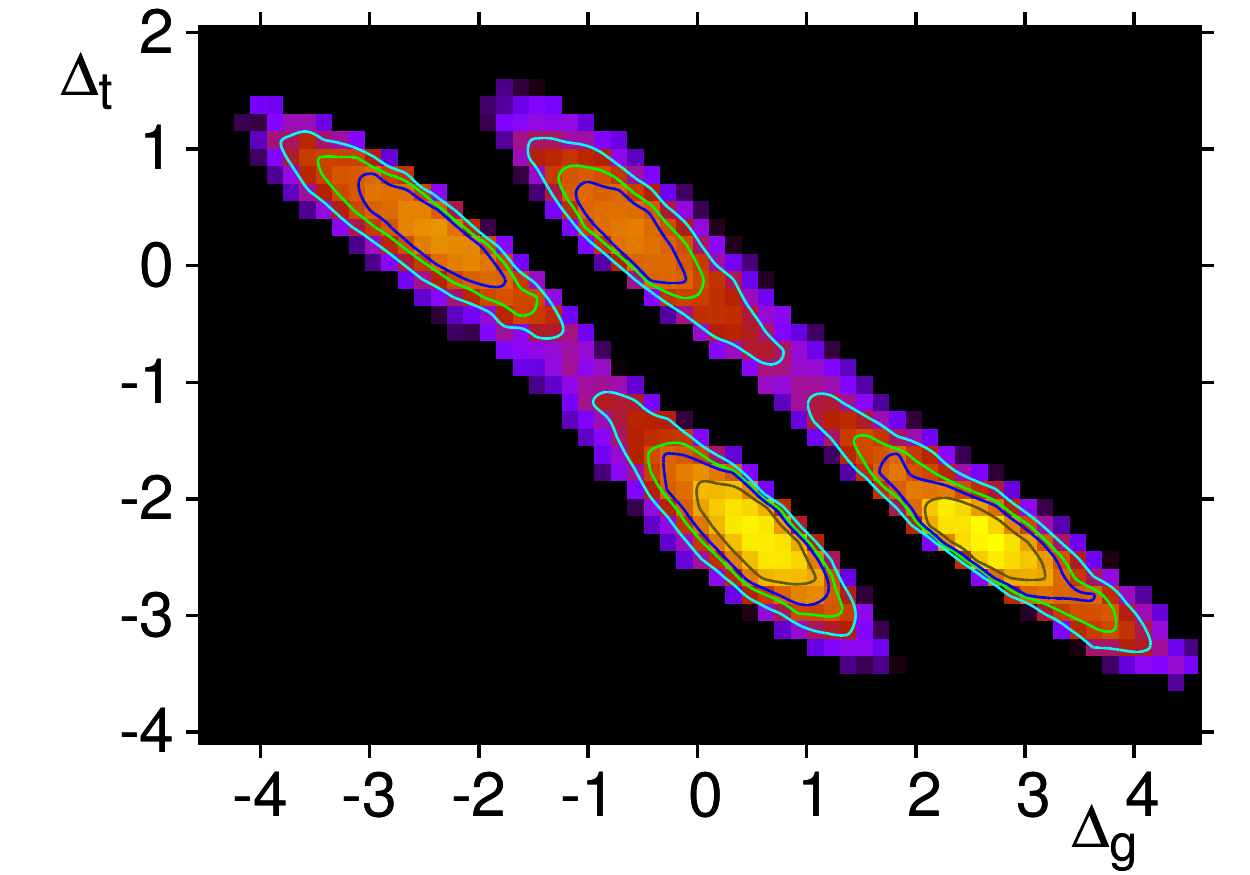}
  \hspace*{1ex}
  \raisebox{-13pt}{\includegraphics[width=0.3\textwidth]{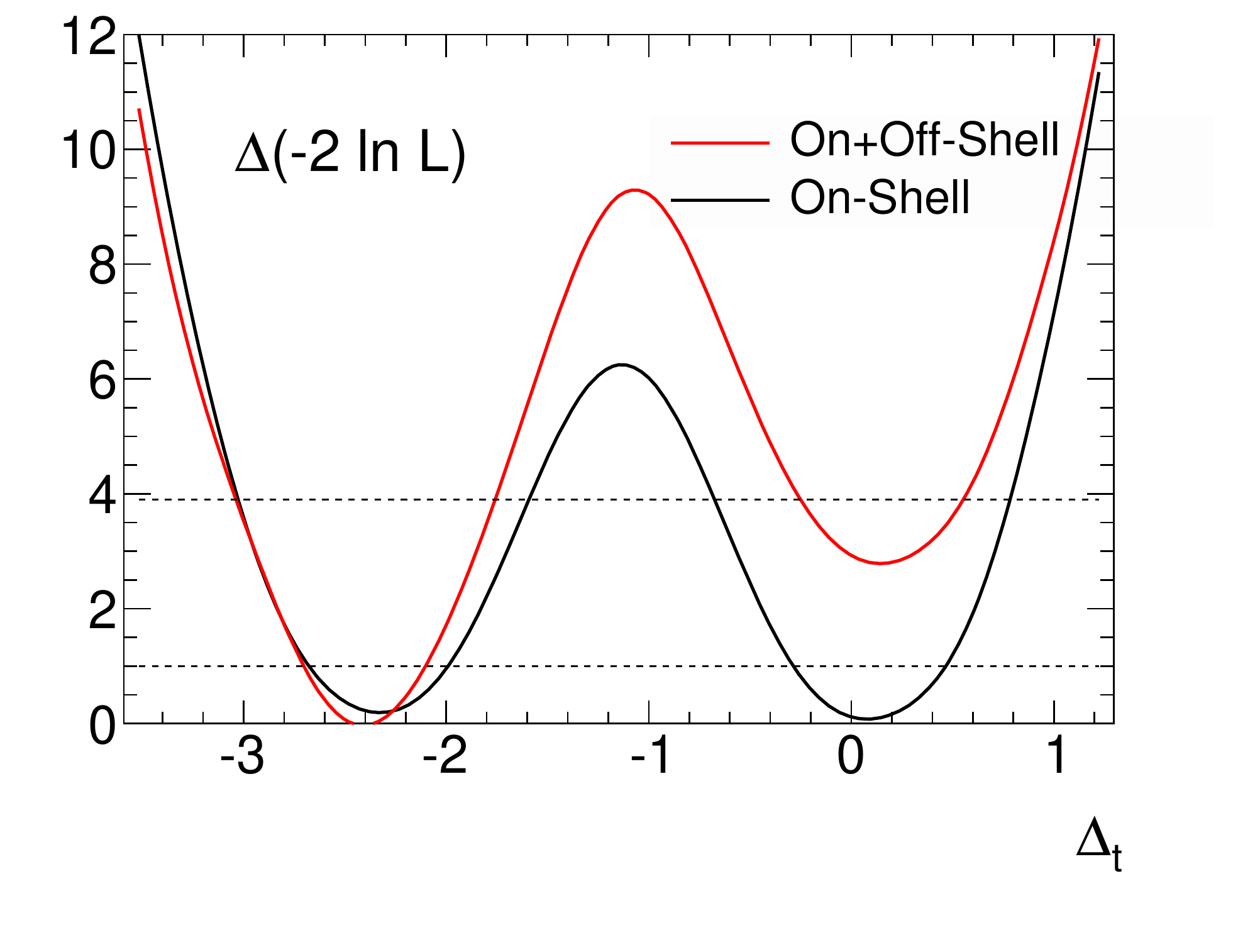}}
  \caption{Correlation between the coupling modifications $\Delta_t$
    and $\Delta_g$ without (left) and with (center) off-shell Higgs
    measurements. In the right panel we show the 1-dimensional profile
    likelihood for $\Delta_t$ with  and without off-shell measurements.}
\label{fig:dt_dg}
\end{figure}

We illustrate the top mass effects in Fig.~\ref{fig:cms_off-shell}.
The background $q\bar{q}\rightarrow ZZ$ and the data points are taken
from the experimental
publications~\cite{off_shell_atlas,off_shell_cms}. The gluon-initiated
component is generated with \textsc{Mcfm}~\cite{mcfm} following 
Eq.\eqref{eq:m4l}. QCD corrections to the gluon-induced component
are accounted via a global $K$-factor~\cite{melnikov_interference,off_shell_th}. 
We follow the ATLAS and CMS cut-flow and find full agreement with both studies.

Following Eq.\eqref{eq:m4l}, terms linear in $(1+\Delta_t)$ are
sensitive to the sign of the top Yukawa coupling~\cite{h_jets}: in the
Standard Model the off-shell interference is destructive, while a sign
change in the top Yukawa coupling increases the combined rate
significantly. In addition, we see the kinematic difference from the
missing top mass threshold and the missing logarithmic top mass
dependence for the dimension-6 operator. Similar top mass effects can
be observed in gluon fusion Higgs production with hard
jets~\cite{h_jets,boosted_higgs} and in the gluon-induced contribution
to $VH$ production~\cite{vh_topmass}. Both should eventually be
included in the Higgs couplings analysis.\bigskip

One way to exploit this feature in our Higgs couplings determination
is to include the usual coupling modifications and an invisible
branching ratio, but no unobservable width. In this case we probe the
momentum dependence of the effective Higgs-gluon coupling, linked to
its top mass dependence~\cite{h_jets}.  In Fig.~\ref{fig:dt_dg} we
present the resulting correlation between $\Delta_t$ and $\Delta_g$,
finding a significant improvement from the off-shell rate
measurement. The SM-like and flipped-sign solutions clearly
separate. In the right panel of Fig.~\ref{fig:dt_dg} we observe a
slight preference towards a negative top Yukawa coupling. It arises
from a small excess of events in the off-shell CMS data.  ATLAS sees
the opposite trend, but with considerably fewer events.\bigskip

Finally, we can allow for unobserved decays according to
Eq.\eqref{eq:lag_offshell}. This corresponds to a free total Higgs
width under the condition that the minimum width is given by the sum
of the observed partial widths.  In Fig.~\ref{fig:width}, we present
first results from a global \textsc{SFitter} analysis. In the left
panel we show the correlation between a typical coupling $\Delta_W$
and the total Higgs width. For $\Gamma_H/\Gamma_H^\text{SM} \gg 1$ the
Higgs production and decay rates scale like $g_x^4/\Gamma_H$. Indeed,
we see that the positive correlation extends to
$\Gamma_H/\Gamma_H^\text{SM} \sim 30 \sim 2.3^4$. The corresponding
value of $\Delta_W$ is then 1.3, just as expected.

In the right panels we see that the upper bound is approximately
$\Gamma_H < 9.3 \, \Gamma_H^\text{SM}$ at 68\% CL. 
While our width constraint was obtained in the EFT context, our bound is still
competitive to other analysis that account only to SM-like
interactions~\cite{off_shell_th}. The key ingredient here is the
analysis of the whole $m_{4l}$ distribution profile that probes the
possible new physics at different energy scales. A similar study via
on-shell signal strengths would clearly not be as sensitive.

\begin{figure}[t!]
  \centering
  \includegraphics[width=0.42\textwidth]{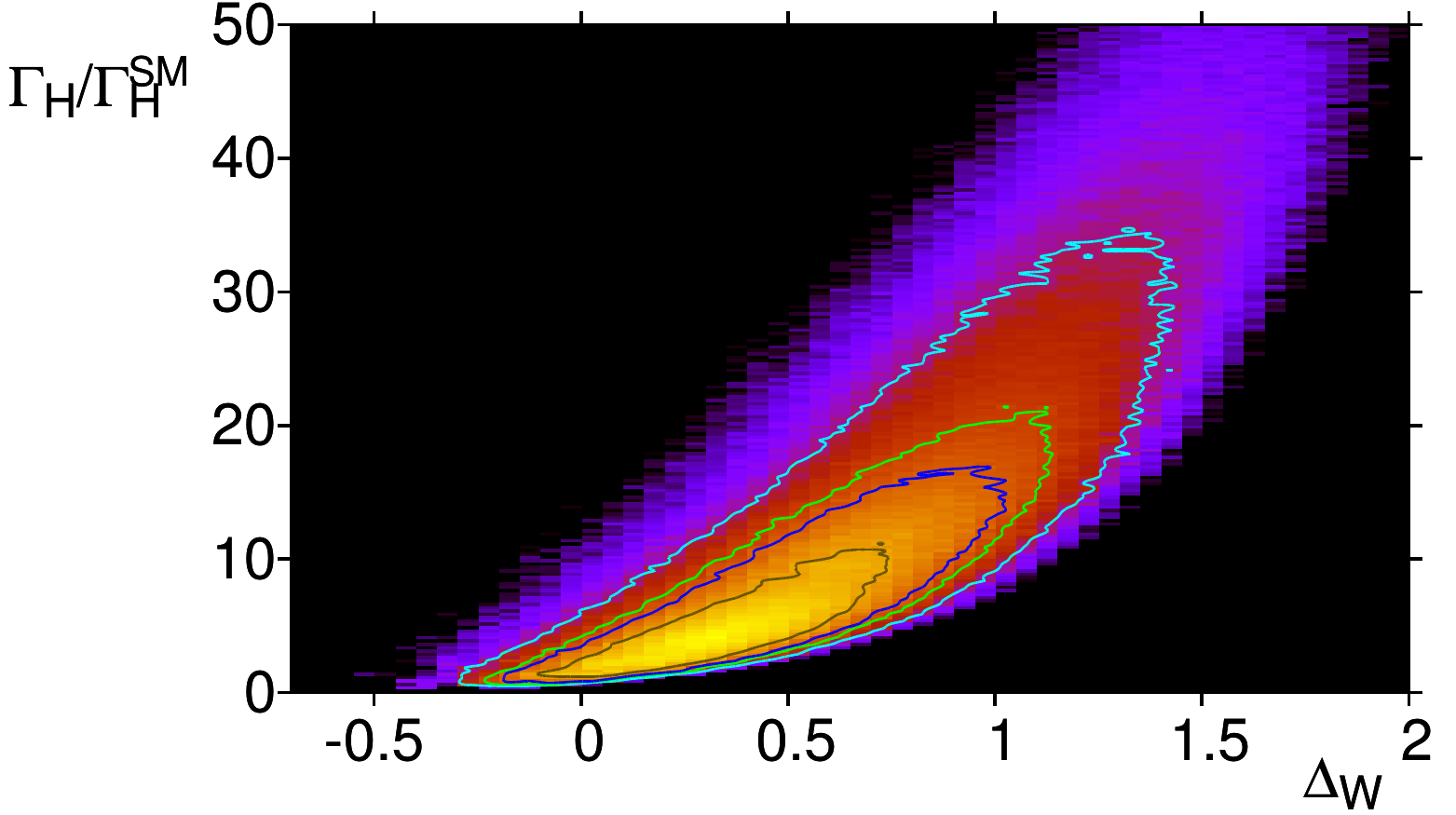}
  \hspace*{0.1\textwidth}
 \raisebox{-16pt}{\includegraphics[width=0.355\textwidth]{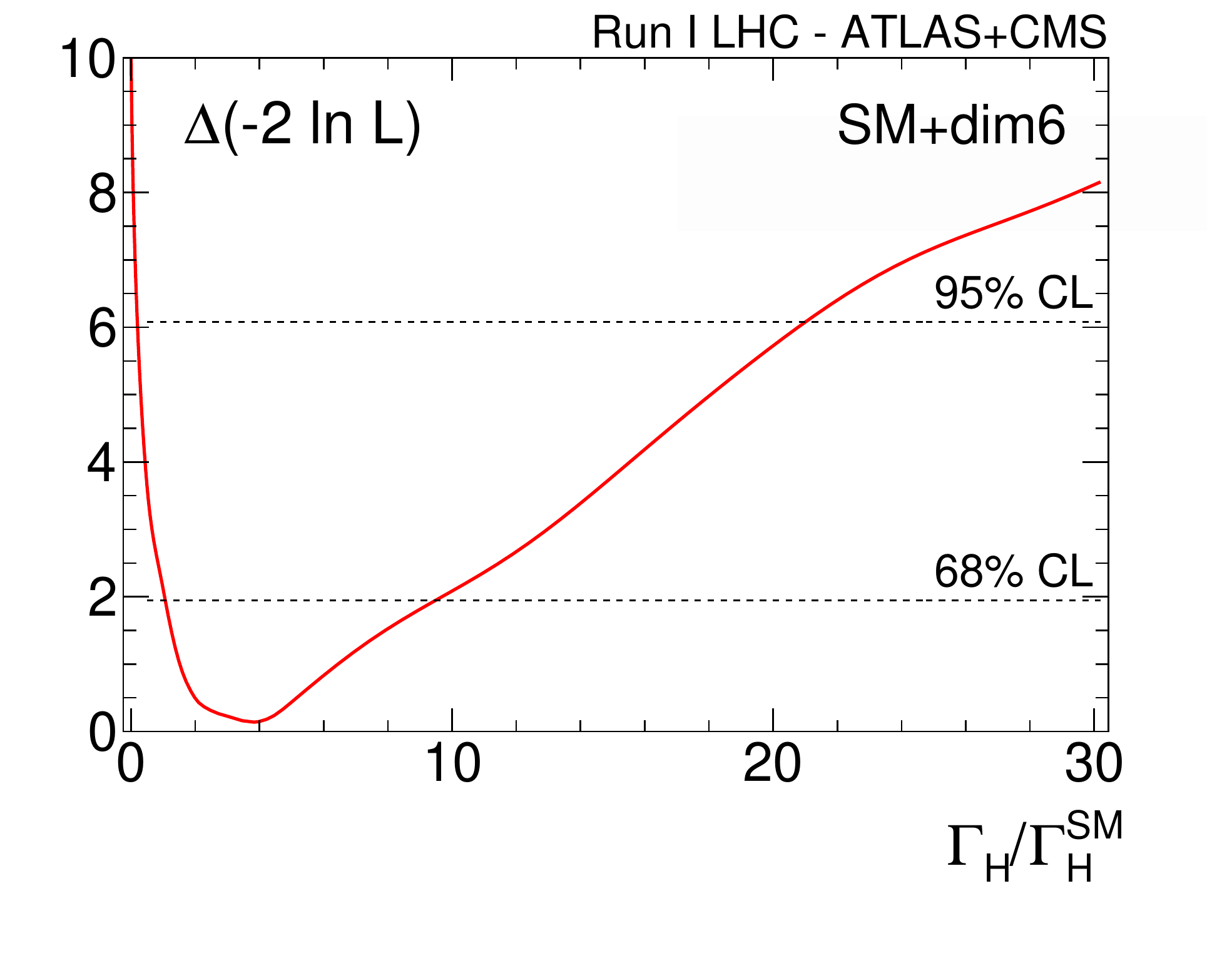}}
  \caption{Correlation between the total Higgs width and the coupling
    modifications $\Delta_W$ (left) and 1-dimensional profile
    likelihood of the total Higgs width (right).  Both, on-shell and
    off-shell Higgs rates are included.}
\label{fig:width}
\end{figure}

\section{Present and future}

In this paper we have presented a final analysis of the Run~I Higgs
measurements. For the first time, we directly compare the direct Higgs
coupling analysis with the effective Lagrangian approach. For the
coupling analysis we included independent variations of the Higgs-top
and Higgs-gluon coupling, as well as an invisible Higgs branching ratio
and eventually Higgs decays through unobserved channels.  While for
the former the current results for $t\bar{t}H$ production should be
taken with a grain of salt, we found a stable combined measurement of
$\br_\text{inv}= 0.16^{+0.07}_{-0.11}$.  Including off-shell Higgs
measurements the full couplings fit gave an upper limit on the total
Higgs width of $\Gamma_H <\ 9.3 \, \Gamma_H^\text{SM}$ at 68\%
CL.\bigskip

Theoretical uncertainties and their effect on the individual error bars of
more fundamental parameters are starting to become a crucial issue
already by the end of Run~I. We tested different assumptions on the
correlation of theoretical uncertainties, on their size, and on their
statistical treatment. As in the absence of a clear definition based
on statistical principles we at least ensured that our analysis
was conservative. Based on our findings the issue of theoretical
uncertainties and their statistical treatment needs to be carefully
considered for the upcoming Run II data.\bigskip

To include kinematic distributions the traditional coupling analysis
has to be expanded. We consider an effective field theory approach the
most natural and the most promising expansion. As long as we only
include Higgs rate information the Run~I analysis in terms of Higgs
coupling modifications $\Delta_x$ and in terms of higher-dimensional
operators $f_x/\Lambda^2$ are essentially equivalent. Once we included
measurements from the pure gauge sector this will change. Technically,
the rate-based analysis in terms of an effective field theory is more
challenging, because the underlying parameters are less directly
linked to measurements than for the Higgs coupling modifications.

We added two sample distributions from $VH$ production and from weak boson
fusion to our analysis. We found a significant stabilization of
the higher-dimensional analysis, the secondary structures were more easily
identified, and all individual error bars on the Wilson coefficients
were visibly reduced.  We end the discussion with a word of warning:
given the current precision in the dimension-6 analysis it is not
guaranteed that the effective Lagrangian expansion is within its range
of validity for all considered observables. It should then be seen as
a motivated and useful parametrization of Higgs interactions, while we
wait for the increase of precision in the future Run~II of the
LHC.\bigskip

\begin{center} \textbf{Acknowledgments} \end{center}

All of us are very grateful to Concha Gonzalez-Garcia's support during
all stages of this project. Not only would this paper look very
different without her input, the project would never even have started
without her physics contribution and her energy. Moreover, we are
grateful to the \textsc{SFitter} team with Dirk Zerwas, Markus Klute,
and R{\'e}mi Lafaye who continuously keep our project alive.

O.J.P.E. is supported in part by Conselho Nacional de Desenvolvimento
Cient\'{\i}fico e Tecnol\'ogico (CNPq) and by Funda\c{c}\~ao de Amparo
\`a Pesquisa do Estado de S\~ao Paulo (FAPESP), T.C is supported by
USA-NSF grant PHY-09-6739. We also thank support from EU grant FP7 ITN
INVISIBLES (Marie Curie Actions PITN-GA-2011-289442).

\clearpage

\appendix
\section{Numerical results}

In this appendix we give the limits presented in the figures throughout
the paper as numbers.

\begin{table}[h!]
\footnotesize
\begin{tabular}{|l|c|c|c||c|c|c|}
\cline{2-7}
 \multicolumn{1}{c|}{} 
&  \multicolumn{3}{c||}{ 5 parameter analysis} 
&\multicolumn{3}{c|}{6 parameter analysis} \\
\cline{2-7}
 \multicolumn{1}{c|}{} & Best fit & 68\% CL intervals & 95\% CL intervals
& Best fit & 68\% CL intervals & 95\% CL intervals
\\
\hline
 $\Delta_W$ & -0.113  & $(-0.23, 0.02)$  & $(-0.335, 0.145)$ & -0.185  & $(-0.35, -0.05)$  & $(-0.465, 0.095)$
\\
\hline
$\Delta_Z$ & 0.0563  & $(-0.07, 0.18)$  & $(-0.20, 0.295)$ & 0.041  & $(-0.08, 0.175)$  & $(-0.22, 0.285)$
\\
\hline
$\Delta_t$ & -0.271  & $(-0.38, -0.08)$  & $(-0.495, 0.09)$ & -0.271  & $(-0.41, -0.08)$  & $(-0.53, 0.09)$ 
\\
\hline
$\Delta_b$ & -0.291  & $(-0.535, 0.035)$  & $(-0.785, 0.36)$ & -0.304  & $(-0.57, 0.015)$  & $(-0.83, 0.355)$
\\
\hline
$\Delta_\tau$ & -0.0987 & $(-0.265, 0.095)$  & $(-0.4, 0.26)$ & -0.0826  & $(-0.265, 0.095)$  & $(-0.405, 0.29)$
\\
\hline
$\Delta_\gamma$ & ----- & ----- & ----- & 0.129  & $(0.015, 0.29)$  & $(-0.11, 0.42)$
\\
\hline\hline
$\Delta_\gamma^\text{SM+NP}$ & ----- & ----- & ----- & -0.033  & $(-0.175, 0.13)$  & $(-0.305, 0.31)$
\\
\hline
\multicolumn{1}{c|}{} 
& \multicolumn{3}{c||}{$(-2 \ln L)_\text{min}=69.2$, $(-2 \ln L)_\text{SM}=72.1$} 
& \multicolumn{3}{c|}{$(-2 \ln L)_\text{min}=68.1$, $(-2 \ln L)_\text{SM}=72.1$} \\ \cline{2-7}
\end{tabular}
\caption{Best fit values, 68\% CL and 95\% CL allowed ranges for the results of
the Higgs analysis with 5 free couplings (blue bars in Fig.~\ref{fig:delta}) and the results of the analysis
including in addition $\Delta_\gamma$ (blue bars in the left panel in Fig.~\ref{fig:delta_g}).}
\label{tab:app1}
\end{table}

\begin{table}[h!]
\footnotesize
\begin{tabular}{|l|c|c|c||c|c|c|}
\cline{2-7}
 \multicolumn{1}{c|}{} 
&  \multicolumn{3}{c||}{ 7 parameter analysis} 
&\multicolumn{3}{c|}{8 parameter analysis} \\
\cline{2-7}
 \multicolumn{1}{c|}{} & Best fit & 68\% CL intervals & 95\% CL intervals
& Best fit & 68\% CL intervals & 95\% CL intervals
\\
\hline
 $\Delta_W$ & -0.160  & $(-0.335, -0.05)$  & $(-0.46, 0.085)$ & -0.0867  & $(-0.265, 0.025)$  & $(-0.38, 0.155)$
\\
\hline
$\Delta_Z$ & 0.0559  & $(-0.07, 0.195)$  & $(-0.205, 0.305)$ & 0.158  & $(0.01, 0.28)$  & $(-0.125, 0.405)$
\\
\hline
$\Delta_t$ & 0.159  & $(-0.2, 0.46)$  & $(-0.585, 0.75)$ & 0.188  & $(-0.13, 0.57)$  & $(-0.505, 0.845)$ 
\\
\hline
$\Delta_b$ & -0.265  & $(-0.565, -0.01)$  & $(-0.82, 0.295)$ & -0.193  & $(-0.5, 0.06)$  & $(-0.77, 0.375)$
\\
\hline
$\Delta_\tau$ & -0.0492 & $(-0.25, 0.095)$  & $(-0.395, 0.28)$ & 0.0417  & $(-0.17, 0.185)$  & $(-0.33, 0.375)$
\\
\hline
$\Delta_\gamma$ & 0.226  & $(0.09, 0.40)$  & $(-0.065, 0.555)$ & 0.248  & $(0.1, 0.435)$  & $(-0.055, 0.595)$
\\
\hline
$\Delta_g$ & -0.479  & $(-0.83, -0.125)$  & $(-1, 0.37)$ & -0.430 & $(-0.855, -0.13)$  & $(-1, 0.385)$
\\
\hline
$\text{BR}_{\text{inv}}$ & ----- & ----- & ----- & 0.157 & $(0.048, 0.226)$  & $(0., 0.306)$
\\
\hline\hline
$\Delta_\gamma^\text{SM+NP}$ & -0.0191  & $(-0.17, 0.125)$  & $(-0.295, 0.285)$ & 0.0892  & $(-0.09, 0.22)$  & $(-0.22, 0.395)$
\\
\hline
$\Delta_g^\text{SM+NP}$ & 0.230  & $(-0.4, 0.115)$  & $(-0.51, 0.35)$ & -0.163  & $(-0.335, 0.04)$  & $(-0.45, 0.115)$
\\
\hline
\multicolumn{1}{c|}{} 
& \multicolumn{3}{c||}{$(-2 \ln L)_\text{min}=66.4$, $(-2 \ln L)_\text{SM}=72.1$} 
& \multicolumn{3}{c|}{$(-2 \ln L)_\text{min}=63.4$, $(-2 \ln L)_\text{SM}=72.1$} \\ \cline{2-7}
\end{tabular}
\caption{Best fit values, 68\% CL and 95\% CL allowed ranges for the results of
the analysis with 7 free couplings (blue bars in the right panel in Fig.~\ref{fig:delta_g})
and the results of the analysis including in addition $\text{BR}_{\text{inv}}$ (blue bars in Fig.~\ref{fig:delta_gi}).}
\label{tab:app2}
\end{table}

\begin{table}[h!]
\footnotesize
\begin{tabular}{|l|c|c|c||c|c|c|}
\cline{2-7}
 \multicolumn{1}{c|}{} 
&  \multicolumn{3}{c||}{ Correlated theoretical uncertainties } 
&\multicolumn{3}{c|}{N$^3$LO gluon fusion prediction} \\
\cline{2-7}
 \multicolumn{1}{c|}{} & Best fit & 68\% CL intervals & 95\% CL intervals
& Best fit & 68\% CL intervals & 95\% CL intervals
\\
\hline
 $\Delta_W$ & -0.137  & $(-0.26, -0.005)$  & $(-0.39, 0.115)$ & -0.162  & $(-0.315, -0.04)$  & $(-0.44, 0.09)$
\\
\hline
$\Delta_Z$ & 0.0814  & $(-0.02, 0.225)$  & $(-0.14, 0.335)$ & 0.0704  & $(-0.07, 0.19)$  & $(-0.205, 0.295)$
\\
\hline
$\Delta_t$ & 0.113  & $(-0.07, 0.525)$  & $(-0.315, 0.805)$ & 0.165  & $(-0.195, 0.475)$  & $(-0.565, 0.76)$ 
\\
\hline
$\Delta_b$ & -0.317  & $(-0.51, -0.01)$  & $(-0.725, 0.28)$ & -0.271  & $(-0.55, 0.005)$  & $(-0.8, 0.295)$
\\
\hline
$\Delta_\tau$ & -0.043 & $(-0.195, 0.105)$  & $(-0.32, 0.27)$ & -0.0939  & $(-0.25, 0.085)$  & $(-0.395, 0.26)$
\\
\hline
$\Delta_\gamma$ & 0.213  & $(0.105, 0.375)$  & $(-0.015, 0.52)$ & 0.241  & $(0.09, 0.395)$  & $(-0.06, 0.54)$
\\
\hline
$\Delta_g$ & -0.386  & $(-0.85, -0.215)$  & $(-1., 0.08)$ & -0.508 & $(-0.82, -0.11)$  & $(-1., 0.39)$
\\
\hline\hline
$\Delta_\gamma^\text{SM+NP}$ & 0.0088  & $(-0.115, 0.15)$  & $(-0.23, 0.29)$ & -0.00786  & $(-0.16, 0.12)$  & $(-0.28, 0.28)$
\\
\hline
$\Delta_g^\text{SM+NP}$ & -0.195  & $(-0.355, -0.055)$  & $(-0.45, 0.11)$ & -0.248  & $(0.37, -0.095)$  & $(-0.49, 0.045)$
\\
\hline
\multicolumn{1}{c|}{} 
& \multicolumn{3}{c||}{$(-2 \ln L)_\text{min}=95.6$, $(-2 \ln L)_\text{SM}=105.3$} 
& \multicolumn{3}{c|}{$(-2 \ln L)_\text{min}=71.8$, $(-2 \ln L)_\text{SM}=77.3$} \\ \cline{2-7}
\end{tabular}
\caption{Best fit values, 68\% CL and 95\% CL allowed ranges for the results of
the analysis with 7 free couplings assuming correlated theoretical
uncertainties (light blue bars in the upper--left panel in Fig.~\ref{fig:delta_err})
and the results of the analysis including N$^3$LO corrections to the Higgs gluon fusion
production rate at the LHC~\cite{n3lo} (light blue bars in lower--left panel in Fig.~\ref{fig:delta_err}).}
\label{tab:app3}
\end{table}

\begin{table}[h!]
\footnotesize
\begin{tabular}{|l|c|c|c||c|c|c|}
\cline{2-7}
 \multicolumn{1}{c|}{} 
&  \multicolumn{3}{c||}{ Passarino gluon fusion prediction} 
&\multicolumn{3}{c|}{Gaussian theoretical errors} \\
\cline{2-7}
 \multicolumn{1}{c|}{} & Best fit & 68\% CL intervals & 95\% CL intervals
& Best fit & 68\% CL intervals & 95\% CL intervals
\\
\hline
 $\Delta_W$ & -0.174  & $(-0.355, -0.06)$  & $(-0.47, 0.09)$ & -0.118  & $(-0.295, 0.01)$  & $(-0.425, 0.14)$
\\
\hline
$\Delta_Z$ & 0.0653  & $(-0.08, 0.19)$  & $(-0.21, 0.305)$ & -0.03848  & $(-0.16, 0.105)$  & $(-0.3, 0.21)$
\\
\hline
$\Delta_t$ & 0.139  & $(-0.205, 0.445)$  & $(-0.585, 0.7)$ & 0.188  & $(-0.235, 0.535)$  & $(-0.835, 0.735)$ 
\\
\hline
$\Delta_b$ & -0.291  & $(-0.58, -0.015)$  & $(-0.845, 0.31)$ & -0.151  & $(-0.48, 0.18)$  & $(-0.78, 0.585)$
\\
\hline
$\Delta_\tau$ & -0.0674 & $(-0.265, 0.09)$  & $(-0.415, 0.295)$ & -0.0700  & $(-0.255, 0.1)$  & $(-0.405, 0.305)$
\\
\hline
$\Delta_\gamma$ & 0.231  & $(0.085, 0.4)$  & $(-0.065, 0.56)$ & 0.150  & $(0., 0.315)$  & $(-0.24, 0.45)$
\\
\hline
$\Delta_g$ & -0.552  & $(-0.895, -0.205)$  & $(-1., 0.305)$ & -0.372 & $(-0.755, -0.16)$  & $(-0.96, 0.97)$
\\
\hline\hline
$\Delta_\gamma^\text{SM+NP}$ & -0.0266  & $(-0.19, 0.115)$  & $(-0.305, 0.305)$ & -0.0497  & $(-0.215, 0.09)$  & $(-0.36, 0.25)$
\\
\hline
$\Delta_g^\text{SM+NP}$ & -0.309  & $(-0.455, -0.195)$  & $(-0.55, -0.045)$ & -0.117  & $(-0.28, 0.05)$  & $(-0.415, 0.255)$
\\
\hline
\multicolumn{1}{c|}{} 
& \multicolumn{3}{c||}{$(-2 \ln L)_\text{min}=61.7$, $(-2 \ln L)_\text{SM}=71.0$} 
& \multicolumn{3}{c|}{$(-2 \ln L)_\text{min}=80.0$, $(-2 \ln L)_\text{SM}=83.0$} \\ \cline{2-7}
\end{tabular}
\caption{Best fit values, 68\% CL and 95\% CL allowed ranges for the results of
the analysis with 7 free couplings using the gluon fusion prediction from
Ref.\cite{passarino} (light blue bars in the upper--right panel in Fig.~\ref{fig:delta_err})
and the results of the analysis assuming Gaussian theoretical uncertainties
(light blue bars in lower--right panel in Fig.~\ref{fig:delta_err}).}
\label{tab:app4}
\end{table}

\begin{table}[h!]
\footnotesize
\begin{tabular}{|l|c|c|c||c|c|c|}
\cline{2-7}
 \multicolumn{1}{c|}{} 
&  \multicolumn{3}{c||}{Rate-based analysis} 
&\multicolumn{3}{c|}{Analysis with kinematic distributions} \\
\cline{2-7}
 \multicolumn{1}{c|}{} & Best fit & 68\% CL intervals & 95\% CL intervals
& Best fit & 68\% CL intervals & 95\% CL intervals
\\
\hline
 $f_{GG}/\Lambda^2$ \footnotesize{(TeV$^{-2}$)} & -19.6 &
$(-24, 16.8)$ &
$(-27.9, -12.9)$  & -24.39  & $(-27, -18.9)$  & $(-30, -14.1)$ 
\\
& -4.79 & $(-7.2, 0.6)$ & $(-9.3, 9.3)$ & -3.75  & $(-7.2, -1.2)$ & $(-9.6, 9.6)$
\\
& 4.42 & $(1.2, 7.2)$ & $(13.5, 28.8)$ & 3.94  & $(0.6, 7.8)$  & $(15.3, 30)$
\\
& 19.35 & $(16.8, 25.1)$ & & 24.18  & $(19.2, 27)$ &
\\
\hline
$f_{WW}/\Lambda^2$ \footnotesize{(TeV$^{-2}$)} & 4.36 & $(-0.4, 9.5)$ & $(-3.7, 13.7)$ &
0.296 & $(-2.95, 2.9)$, $(9.65, 9.8)$ & $(-4.3, 4.4)$, $(15.3, 30)$
\\
\hline
$f_{BB}/\Lambda^2$ \footnotesize{(TeV$^{-2}$)} & -4.72 & $(-8.6, 2.8)$ & $(-13.4, 6.1)$ & -0.518 & $(-2.45, 4.3)$ & $(-10, 7)$
\\
\hline
$f_{\phi,2}/\Lambda^2$ \footnotesize{(TeV$^{-2}$)} & 6.15 & $(0, 12.5)$ & $(-5, 18)$ & 1.03 & $(-4.75, 6)$ & $(-8, 10)$, $(12.25, 19.75)$
\\
\hline
$f_{W}/\Lambda^2$ \footnotesize{(TeV$^{-2}$)} & -6.38 & $(-12.4, 7)$ & $(-13, 22.5)$ & 1.12 & $(-2.25, 3.75)$ & $(-5.75, 7.25)$
\\
\hline
$f_{B}/\Lambda^2$ \footnotesize{(TeV$^{-2}$)} & -29.04 & $(-45.2, -5.6)$ & $(-55.4, 13)$ & -4.16 & $(-12.2, 9.4)$ & $(-45.2, 14.8)$
\\
\hline
$f_{b}/\Lambda^2$ \footnotesize{(TeV$^{-2}$)} & 4.63 & $(-2.1, 7.8)$ & $(-9.6, 10.8)$ & 0.83 & $(-6, 6.6)$ & $(-13.2, 19.2)$
\\
& 33.7 & $(24.9, 47.3)$ & $(13.2, 61.2)$ & 46.4 & $(33, 56.4)$ & $(24, 67.8)$
\\
\hline
$f_{\tau}/\Lambda^2$ \footnotesize{(TeV$^{-2}$)} & -3.94 & $(-9, 2)$ & $(-12, 4.5)$ & -1.88 & $(-4, 3.5)$ & $(-9, 7)$
\\
 & 41.7 & $(35, 51)$ & $(24, 56.5)$ & 43.9 & $(39, 53.5)$ & $(23.5, 60)$
\\
\hline
$f_{t}/\Lambda^2$ \footnotesize{(TeV$^{-2}$)} & -5.03 & $(-12.8, 1.6)$ & $(-21.3, 10.9)$ & -4.97 & $(-12.9, 1.8)$ & $(-18.5, 10.2)$
\\
 & 43.4 & $(34, 54.4)$ & $(24.9, 63.4)$ & 53.37 & $(41, 59.9)$ & $(24.2, 66.9)$
\\
\hline
 \multicolumn{1}{c|}{} 
& \multicolumn{3}{c||}{$(-2 \ln L)_\text{min}=66.7$, $(-2 \ln L)_\text{SM}=72.1$} 
& \multicolumn{3}{c|}{$(-2 \ln L)_\text{min}=88.4$, $(-2 \ln L)_\text{SM}=91.9$} \\ \cline{2-7}
\end{tabular}
\caption{Best fit values, 68\% CL and 95\% CL allowed ranges for the results of
the effective Lagrangian rate-based analysis (blue bars in Fig.~\ref{fig:dim6}) and
the results when including in addition kinematic distributions (blue bars in Fig.~\ref{fig:dim6kin}).}
\label{tab:app5}
\end{table}

\clearpage


\end{document}